\documentclass[twocolumn]{aastex62}
\usepackage{natbib,rotating}

\newcommand{\cMpch}{$h^{-1}$~cMpc}
\newcommand{\dperp}{\langle d_{\perp} \rangle}

\newcommand{\msol}{{\rm M}_{\odot}}
\newcommand{\mss}{\langle d_{\perp} \rangle}

\received{2019 November 13}
\revised{2020 February 5}
\accepted{2020 February 11}

\shorttitle{LATIS}
\shortauthors{Newman et al.}

\begin{document}

\title{LATIS: The Ly$\alpha$ Tomography IMACS Survey}

\correspondingauthor{Andrew B. Newman}
\email{anewman@carnegiescience.edu}

\author[0000-0001-7769-8660]{Andrew B. Newman}
\author{Gwen C. Rudie}
\affiliation{The Observatories of the Carnegie Institution for Science, 813 Santa Barbara St., Pasadena, CA 91101, USA}
\author{Guillermo A. Blanc}
\affiliation{The Observatories of the Carnegie Institution for Science, 813 Santa Barbara St., Pasadena, CA 91101, USA}
\affiliation{Departamento de Astronom\'ia, Universidad de Chile, Camino del Observatorio 1515, Las Condes, Santiago, Chile}
\author{Daniel D. Kelson}
\affiliation{The Observatories of the Carnegie Institution for Science, 813 Santa Barbara St., Pasadena, CA 91101, USA}
\author{Sunny Rhoades}
\affiliation{Department of Physics and Astronomy, Pomona College, Claremont, CA 91711, USA}
\affiliation{The Observatories of the Carnegie Institution for Science, 813 Santa Barbara St., Pasadena, CA 91101, USA}
\author{Tyson Hare}
\affiliation{The Observatories of the Carnegie Institution for Science, 813 Santa Barbara St., Pasadena, CA 91101, USA}
\author{Victoria P\'{e}rez}
\affiliation{Departamento de Astronom\'ia, Universidad de Chile, Camino del Observatorio 1515, Las Condes, Santiago, Chile}
\author{Andrew J. Benson}
\author{Alan Dressler}
\affiliation{The Observatories of the Carnegie Institution for Science, 813 Santa Barbara St., Pasadena, CA 91101, USA}
\author{Valentino Gonzalez}
\affiliation{Departamento de Astronom\'ia, Universidad de Chile, Camino del Observatorio 1515, Las Condes, Santiago, Chile}
\affiliation{Centro de Astrofisica y Tecnologias Afines (CATA), Camino del Observatorio 1515, Las Condes, Santiago 7591245, Chile}
\author{Juna A. Kollmeier}
\author{Nicholas P. Konidaris}
\author{John S. Mulchaey}
\author{Michael Rauch}
\affiliation{The Observatories of the Carnegie Institution for Science, 813 Santa Barbara St., Pasadena, CA 91101, USA}
\author{Olivier Le F\`{e}vre}
\affiliation{Aix Marseille University, CNRS, CNES, LAM (Laboratoire d'Astrophysique de Marseille), UMR 7326, 13388 Marseille, France}
\author{Brian C. Lemaux}
\affiliation{Department of Physics, University of California, Davis, One Shields Ave., Davis, CA 95616, USA}
\author{Olga Cucciati}
\affiliation{INAF - Osservatorio di Astrofisica e Scienza dello Spazio di Bologna, via Gobetti, 93/3, I-40129, Bologna, Italy}

\author{Simon J. Lilly}
\affiliation{Department of Physics, ETH Z{\"u}rich, CH-8093 Z{\"u}rich, Switzerland}

\begin{abstract}
We introduce LATIS, the Ly$\alpha$ Tomography IMACS Survey, a spectroscopic survey at Magellan designed to map the $z=2.2$-2.8 intergalactic medium (IGM) in three dimensions by observing the Ly$\alpha$ forest in the spectra of galaxies and QSOs. Within an area of 1.7 deg${}^2$, we will observe approximately half of $\gtrsim L^*$ galaxies at $z=2.2$-3.2 for typically 12 hours, providing a dense network of sightlines piercing the IGM with an average transverse separation of 2.5~$h^{-1}$ comoving Mpc (1 physical Mpc). At these scales, the opacity of the IGM is expected to be closely related to the dark matter density, and LATIS will therefore map the density field in the $z\sim2.5$ universe at $\sim$Mpc resolution over the largest volume to date. Ultimately LATIS will produce approximately $3800$ spectra of $z=2.2$-3.2 galaxies that probe the IGM within a volume of $4\times10^6h^{-3}$ Mpc${}^3$, large enough to contain a representative sample of structures from protoclusters to large voids. Observations are already complete over one-third of the survey area. In this paper, we describe the survey design and execution. We present the largest IGM tomographic maps at comparable resolution yet made. We show that the recovered matter overdensities are broadly consistent with cosmological expectations based on realistic mock surveys, that they correspond to galaxy overdensities, and that we can recover structures identified using other tracers. LATIS is conducted in Canada--France--Hawaii Telescope Legacy Survey fields, including COSMOS. Coupling the LATIS tomographic maps with the rich data sets collected in these fields will enable novel studies of environment-dependent galaxy evolution and the galaxy-IGM connection at cosmic noon.\end{abstract}


\section{Introduction} \label{sec:intro}

The central goal of the study of galaxy evolution is to understand how the main physical characteristics of galaxies and their diversity arise from their initial conditions and the actions of many physical processes. Although it is clearly a simplification, many studies have distinguished processes that are primarily internal versus external, and a major focus of galaxy evolution studies has been to gauge the influence of these categories by correlating galaxy properties with two proxies: the mass of a galaxy or its dark matter halo, and the density of the environment measured on some larger scale. Virtually all galaxy properties are correlated with mass at all observed epochs. In the local universe, environment or local density is also clearly correlated with some galaxy properties \citep[e.g.,][]{Dressler80,Postman84,Kauffmann04,Peng10}, and such correlations have clearly been in place since at least $z \sim 1$ \citep[e.g.,][]{Dressler97,Cooper06,Patel09,Muzzin12,Hahn15,Darvish16}. This connection to environment seems to be closest for properties related to a galaxy's star formation history \citep[e.g.,][]{Bamford09,Blanton09,Lemaux19,Tomczak19}.  Measuring the evolution of environmental trends is key to understanding their origins, which are a mixture of physical processes that are sensitive to local density or halo mass (e.g., ram pressure stripping, starvation, galaxy interactions) along with differences in assembly history (e.g., earlier collapse of halos within large-scale overdensities). Yet at earlier epochs $z \gtrsim 1.5$, observations that probe the relation between galaxy properties and the environment are much less definitive (see review by \citealt{Overzier16}). 

A serious impediment is the difficulty of quantifying galaxy environments and mapping large-scale structures at these redshifts. Massive overdensities at $z \gtrsim 2$ are expected to be diffuse, with a modest density contrast spread over $\sim20$ arcmin \citep{Chiang13}. Galaxy density can be used as an indicator of environment, but spectroscopic surveys at these redshifts cover smaller volumes with poorer sampling than at $z \lesssim 1$. Although photometric redshifts can be used to trace galaxy density, particularly when a subset of sources have spectroscopic redshifts, their decreasing accuracy and precision begin to degrade environmental measures beyond $z\sim1$ \citep[e.g.,][]{Darvish17}. Observations of an intragroup or intracluster medium push the sensitivity limits of present X-ray and CMB observatories and will miss massive structures at $z \gtrsim 2$ that have not yet developed a hot atmosphere.

The state of the study of protoclusters, the progenitors at $z \gtrsim 1.5$ of today's massive galaxy clusters, provides an illustrative example.
Present samples of early clusters and protoclusters are heterogeneously selected and likely quite diverse. Some have been identified as a by-product of a general spectroscopic survey \citep[e.g.,][]{Steidel05,Diener13,Cucciati14,Lemaux14,Lemaux18,Kelson20}. Others have been identified by searching for overdensities of red-sequence galaxies \citep{Andreon09,Newman14}, Ly$\alpha$ emitters \citep{Chiang15}, or dusty starbursts \citep{Clements14,Casey15}. Others were found by surveying the neighborhood of radio galaxies thought to signpost overdensities \citep{Pentericci00,Kurk04,Galametz10,Hatch11,Wylezalek13,Noirot18}. These methods can all detect high-redshift structures, but many depend on the presence of particular (often rare) galaxy types, which could bias studies of galaxy evolution in these structures. Furthermore, masses of unvirialized structures are important to connect to theory but are challenging to estimate. \citet{Overzier16} surveyed the literature and compiled a set of just 21 protoclusters that were confirmed at $z=2$-3 with measurements suggesting they will evolve into a halo exceeding $10^{14}~\msol$ at $z=0$.

A promising complementary technique for measuring galaxy environments and detecting large-scale structures at $z \simeq 2$-3 is to map the intergalactic medium (IGM). Fluorescent Ly$\alpha$ emission from IGM filaments has begun to be detected in the centers of protoclusters \citep{Umehata19}. At more typical locations in the IGM, the surface brightness of this emission falls below the sensitivity limits of current facilities, but the hydrogen gas can be detected through the ``forest'' of Ly$\alpha$ absorption that it produces. The Ly$\alpha$ forest arises from trace amounts of \ion{H}{1} in photoionized gas that is within a factor of $\sim10$ of mean density. On scales larger than roughly the Jeans length ($\simeq 100$~comoving kpc; \citealt{Gnedin98,Kulkarni15}), the distribution of \ion{H}{1} follows that of the dark matter. There is a long history of studying structure formation using the Ly$\alpha$ forest observed in the spectra of quasars (see reviews by \citealt{Rauch98,McQuinn16}). Quasar observations probe the matter distribution only along a single sightline. If a bundle of sightlines piercing the same volume is observed, the three-dimensional (3D) matter distribution can be reconstructed \citep{Pichon01,Caucci08}, a technique that has become known as IGM or Ly$\alpha$ forest tomography.

The resolution achievable in such a reconstruction depends on the density of sightlines that are observed. With a sufficiently high density, multiple sightlines will probe the distribution and kinematics of \ion{H}{1} and metals in the circumgalactic gas surrounding individual galaxies (scales of $\sim300$~kpc), enabling the flow of gas between galaxies and their gaseous halos to be studied in unprecedented detail \citep{Theuns06,Steidel09,Evans12,Newman19,Rudie19}. However, that project requires spectroscopy of very faint sources with moderate spectral resolution and relatively high signal-to-noise ratios, which must await 30-m-class telescopes. \citet{Lee14} pointed out that if the goal is instead to map the IGM with a resolution of a few comoving Mpc (cMpc), then the observational requirements are greatly reduced and become practical with current facilities.

On these larger scales of $\gtrsim 3$~\cMpch, the mean Ly$\alpha$ opacity is expected to be well-correlated with the matter density \citep{McDonald02,Kollmeier03,Cai16} and is observed to correlate with the galaxy density \citep{Adelberger03}. Measuring this opacity does not require identifying individual Ly$\alpha$ absorption lines, only spatially coherent flux decrements within the Ly$\alpha$ forest, which can be measured in fairly noisy spectra. \citet{Stark15b,Stark15a} performed realistic mock surveys in cosmological simulations and showed that IGM tomography can effectively detect and estimate the masses and sizes of protoclusters and large voids at $z \sim 2.5$, as along as the mean transverse separation between the sightlines is $\dperp \lesssim 3$~\cMpch. This requirement corresponds to a sightline density of $>550$ deg${}^{-2}$, which is 20-$60\times$ higher than the peak effective density of quasar sightlines in the BOSS or DESI surveys, respectively \citep{Ozbek16}. Despite their sparsity, these quasar surveys can be used to locate some very extended overdensities, as the MAMMOTH survey has shown \citep{Cai16,Cai17}, but such samples are quite incomplete \citep{Miller19}.

Reaching higher source densities requires moving beyond quasars and observing the Ly$\alpha$ forest in the spectra of galaxies as faint as $g \sim 24.5$~mag. Such observations were first implemented in the COSMOS Ly$\alpha$ Mapping and Tomography Observations (CLAMATO) survey \citep{Lee14b}. The CLAMATO map now covers an area of 0.16 deg$^2$ spanning $z=2.05$-2.55 with a resolution set by $\dperp = 2.5$~\cMpch~\citep{Lee18}. This pioneering survey convincingly demonstrated the power of Ly$\alpha$ tomography in several applications, including a study of a protocluster at $z=2.44$ with a tomographic mass of $(1.1 \pm 0.6) \times 10^{14} h^{-1} \msol$ \citep{Lee16} and the identification of a sample of voids \citep{Krolewski18}. Extending this technique over a larger volume could enable the discovery and characterization of statistical samples of large-scale structures. Furthermore, \citet{Lee16b} showed that a larger $\sim1$ deg$^2$ survey could effectively map the topology of the cosmic web (voids, filaments, sheets, and nodes), enabling a new measure of the environments of high-redshift galaxies that may be equally or more useful than the local density.

Motivated by the results of these studies, we have begun the Ly$\alpha$ Tomography IMACS Survey (LATIS) using the Inamori-Magellan Areal Camera and Spectrograph (IMACS; \citealt{Dressler11}) at the Magellan Baade telescope. The goal of LATIS is to map a representative volume of the distant universe ($z = 2.2$-2.8) by densely sampling the Ly$\alpha$ forest in a network of Lyman-break galaxies having a mean separation of $\dperp = 2.5-3$~\cMpch~(1 physical Mpc). LATIS will ultimately cover 1.7 deg${}^2$, corresponding to a volume of $4 \times 10^6 h^{-3}$~cMpc${}^3$, in three of the Canada--France--Hawaii Telescope Legacy Survey (CFHTLS) Deep fields, including COSMOS. The large volume of LATIS is key to producing representative samples of large structures, including protoclusters and large voids, while also minimizing edge effects that can limit tomographic maps when the survey footprint is small. For instance, we expect to detect and characterize $\sim$24 massive protoclusters with present-day masses exceeding $10^{14.5} h^{-1} \msol$. This sample is comparable in number to the compilation by \citet{Overzier16}, but  homogeneously selected. Equally important, Ly$\alpha$ tomography identifies structures independently of their galaxy populations and provides an estimate of their total mass. The LATIS maps will provide a novel measure of Mpc-scale environments of galaxies in well-observed extragalactic fields, enabling new studies of environment-dependent galaxy evolution and the galaxy-IGM connection at cosmic noon.

LATIS observations are now complete over one-third of the survey area. In this paper, in order to help inform future tomographic surveys, we first describe the design and implementation of LATIS (Sections 2-5). We then describe our methods for categorizing and analyzing the spectra of 2596 galaxies (Sections 6-7) and for constructing maps of the IGM opacity covering an area of 0.58~deg$^2$ and a redshift range $z=2.2$-2.8 (Section 8). These are already the largest tomographic maps with Mpc-scale resolution. We characterize and validate the LATIS maps using mock surveys (Section 8) and by demonstrating correlations with the galaxy distribution, with structures previously identified via other tracers, and with the CLAMATO maps in their region of overlap (Section 9). Finally we discuss the complementarity of IGM tomography with other environmental metrics and future plans (Section 10). Readers who are primarily interested in the Ly$\alpha$ tomography methods and maps rather than the implementation of the spectroscopic survey may wish to begin in Section~\ref{sec:specmodeling}.

Throughout the paper we use a flat $\Lambda$CDM cosmology with $\Omega_m = 0.307$ and $h = 0.677$ \citep{Planck15}.

\section{Survey Design Overview} \label{sec:overview}

Before describing the implementation of LATIS, we will first review the parameters that drove our main design decisions.

\emph{Area:} As motivated in the Introduction, a wide area is necessary to identify a statistical sample of structures. \citet{Stark15a} studied the performance of Ly$\alpha$ tomography for detecting protoclusters in simulated observations. They defined protoclusters as the progenitors of $z=0$ clusters that exceed a given mass. When the mean sightline separation is $\dperp  < 3$~\cMpch, they estimated that $\gtrsim 60\%$ of protoclusters that will have masses $\log M_{z=0} / (h^{-1} \msol) > 14.5$ are recovered at $z$=2.5. The present number density of clusters in this mass range is $1.05 \times 10^{-5}$ $h^3$ cMpc${}^{-3}$ \citep{Angulo12,Murray13}. Therefore, over the redshift range $z \approx 2.2$-2.75 where we expect to reach $\dperp < 3$~\cMpch~(see Section~\ref{sec:sldens}), we can expect to detect roughly 14 protoclusters per deg${}^2$. We consider that studying the galaxy populations in protoclusters requires a minimum sample of $\simeq 20$. This requires surveying $\sim$1.4~deg${}^2$, which sets an overall minimum scale. We plan to observe 12 IMACS ``footprints'' (the instrument field of view) that will cover 1.7~deg${}^2$ in total.

\begin{figure*}
\centering
\includegraphics[width=0.43\linewidth]{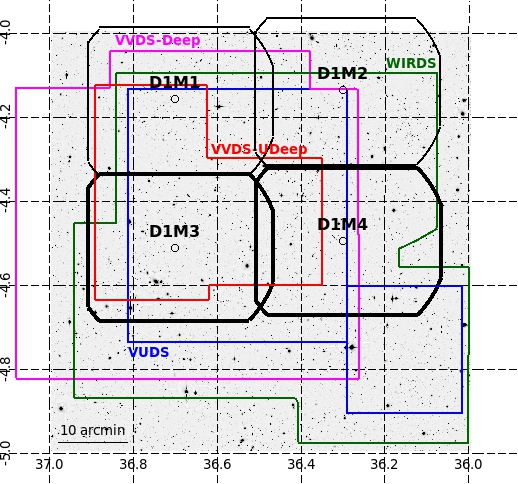}
\includegraphics[width=0.43\linewidth]{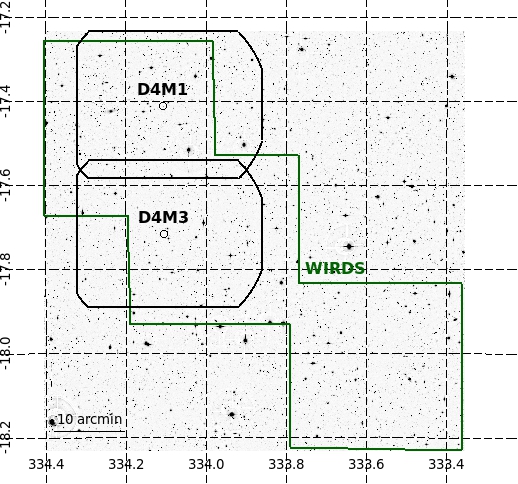} \\
\includegraphics[width=0.65\linewidth]{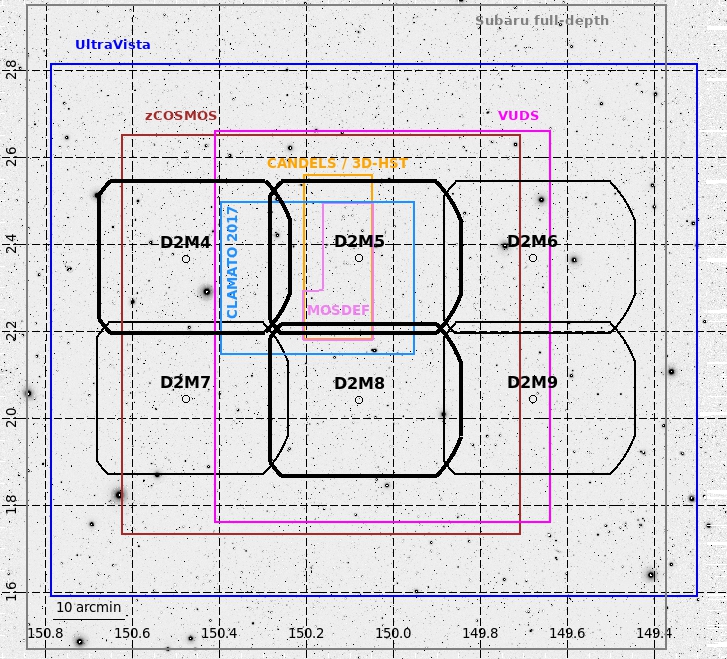}
\caption{The planned positions of the 12 IMACS footprints that comprise LATIS (black outlines) are overlaid on $r$ band images of each field and labeled. Thicker black outlines show the 5 footprints containing the observations used in this paper. The top left and right panels show the CFHTLS D1 and D4 fields, respectively, and the bottom panel shows D2/COSMOS. The footprints of various others surveys listed in Sections~\ref{sec:overview} and \ref{sec:photcatalogs} are overlaid for reference. Axes show the R.A.~and Decl.~in degrees.\label{fig:fieldmaps}}
\end{figure*}

\emph{Survey fields:} This area will be divided among three of the CFHTLS fields. Half of the survey will be conducted in the D2/COSMOS field, and the remainder will be divided between D1 and D4. Figure~\ref{fig:fieldmaps} shows the fiducial layout of the survey area, although the final configuration is flexible to accommodate telescope scheduling constraints. (The layout is discussed further in Section~\ref{sec:tiling}.) We selected the CFHTLS fields for three reasons. First, the CFHTLS provides deep, homogeneous optical imaging over the necessary area, including the $u^*$ filter that is critical for selecting $z=2$-3 galaxies. Second, all fields except D3 are visible from Las Campanas and span a range of right ascension that permits flexible scheduling from August through April. Third, the fields are well observed and benefit from a legacy of deep imaging and spectroscopy. For example, public near-infrared imaging from the UltraVISTA \citep{McCracken12} and WIRDS \citep{Bielby12} surveys covers most of the LATIS area, spectroscopy from the zCOSMOS \citep{Lilly09} and VIMOS Ultra-Deep Surveys \citep{LeFevre15} covers much of D1 and COSMOS, and space-based imaging from the \emph{Hubble} \citep{Koekemoer07,Mowla19}, \emph{Spitzer} \citep{Sanders07}, and \emph{Chandra} \citep{Civano16} telescopes cover the COSMOS field. 

\emph{Instrument:} IMACS is well suited for LATIS due to the wide field of 0.5~deg of its f/2 camera. To increase multiplexing, we purchased a custom bandpass filter that transmits 383-591 nm (Section~\ref{sec:filtergrism}) and enables 2-3 ranks of slits to be ``stacked'' in the dispersion direction. We can observe targets over 0.15 deg${}^2$ with full spectral coverage over this bandpass. To improve sensitivity at blue wavelengths, we designed and purchased a new grism blazed at 460~nm (Section~\ref{sec:filtergrism}).  In order for the spectral resolution to not degrade the resolution of tomographic maps more than 10\%, $\sigma_{\rm inst}$ should be at least $2\times$ smaller than the transverse smoothing scale $\sigma_{\rm trans} \sim \dperp$ expressed in velocity, which translates to a resolving power $R \gtrsim 800$. Our custom grism delivers an average $R = 880$ in the Ly$\alpha$ forest. IMACS is among the most efficient instruments worldwide for conducting LATIS. A simple metric of mapping speed is $\Omega \times D^2 \times e$, where $\Omega$ is the field of view in deg${}^2$, $D$ is the telescope diameter in meters, and $e$ is the throughput of the instrument and telescope. We estimate that Magellan/IMACS, VLT/VIMOS (now decommissioned) and Keck/LRIS (600/4000 grism) have survey speeds of 1.0, 1.3, and 0.5, respectively.

\emph{Target density:} Besides the volume, a critical parameter for tomographic surveys is the areal density $n$ of sightlines, or equivalently the mean transverse sightline separation $\mss = n^{-1/2}$. With our IMACS configuration, we observe $\sim 270$ targets per mask. By using two masks within each footprint, we can therefore observe $\sim 3600$ targets per deg${}^2$. About half of the photometric targets are ultimately useful for tomographic mapping (Section~\ref{sec:purity}), providing a total sightline density of 1800~deg${}^{-2}$. However, an individual sightline does not probe the entire redshift range of our reconstruction. We aim to reconstruct $z=2.2$-2.8, with the low cutoff set by the blue sensitivity of IMACS and the high cutoff set by the falling density of suitably bright galaxies. But a sightline typically spans $\Delta z \approx 0.3$ in its Ly$\alpha$ forest before confusion with Ly$\beta$ absorption begins, and we therefore expect a mean sightline density of $1800 \times 0.3/0.6 \approx 900$~deg${}^{-2}$ piercing a given $z_{\rm Ly\alpha}$. This is an upper limit, since some sightlines will have a Ly$\alpha$ forest that extends outside the reconstruction volume, but this rough calculation shows that we can expect LATIS to achieve a sightline separation in the range $\langle d_{\perp} \rangle \sim 2.5$-3~\cMpch~($n \sim550$-800~deg${}^{-2}$) that has been shown to adequate for the detection and characterization of large structures \citep{Lee14,Stark15a}.

\section{Target Selection} \label{sec:targets}

Our selection of targets is motivated by two goals: first, to achieve the highest practical signal-to-noise ratio in the tomographic map, and second, to maintain a well-defined selection function so that the properties of galaxies in different environments can be robustly characterized. There is some tension between these goals. For example, a color selection with higher purity, coupled with a bias against lower-surface brightness or blended sources, might be more effective for delivering tomographic sightlines, but it would introduce complex biases in the galaxy population that is selected. We therefore limited our selection to relatively simple and inclusive color criteria, supplemented by public databases of spectroscopic redshifts for a minority of targets. 

\subsection{Photometric Catalogs}
\label{sec:photcatalogs}

In the D1 and D4 fields, the basis of our photometric catalogs is the final release (T0007) of the CFHTLS.\footnote{\url{http://terapix.calet.org/terapix.iap.fr/cplt/T0007/doc/T0007-doc.html}} We use the catalogs produced from $u^*griyz$ stacks that are sigma-clipped means of the 85\% best seeing images. The depth in $r$ is 25.6 AB~mag (85\% completeness for point sources), which is 0.8~mag fainter than our flux-limited selection described below. We use fluxes measured within $2\farcs2$ diameter apertures, corrected for Galactic extinction and for the light outside of the aperture as estimated using bright point sources. 

In the D2/COSMOS field, we instead use the \citet{Ilbert09} catalog of $I < 25$ sources covering 2~deg${}^2$ with 30-band photometry. Using this catalog enables a potential future extension of the survey beyond the central 1 deg${}^2$ covered by the CFHTLS. Since the Ly$\alpha$ forest is most easily observed in rest-UV-bright galaxies, we preferred the optical selection in this catalog to the near-infrared selection used in the more recent \citet{Laigle16} catalog.

We cross-matched these catalogs to publicly available databases of spectroscopic redshifts, including VVDS \citep{LeFevre13}, VUDS DR1 \citep{LeFevre15,Tasca17}, MOSDEF \citep{Kriek15}, DEIMOS 10K \citep{Hasinger18}, 3D-HST \citep{Brammer12,Momcheva15}, ZFIRE \citep{Nanayakkara16}, FMOS-COSMOS \citep{Silverman15}, CLAMATO \citep{Lee18}, MilliQuas \citep{Flesch15}, VIPERS \citep{Scodeggio18}, and the G10/COSMOS catalog \citep{Davies15} which includes the zCOSMOS-Bright \citep{Lilly09} and PRIMUS surveys \citep{Cool13}.\footnote{Later we will place galaxies from the full VUDS and zCOSMOS-Deep data sets in our tomographic maps; however, these catalogs were not used to inform targeting before semester 2019B, which include all observations used in this paper.}

\subsection{Selecting LBGs}\label{sec:selecting_lbgs}

\begin{figure*}
\includegraphics[width=\linewidth]{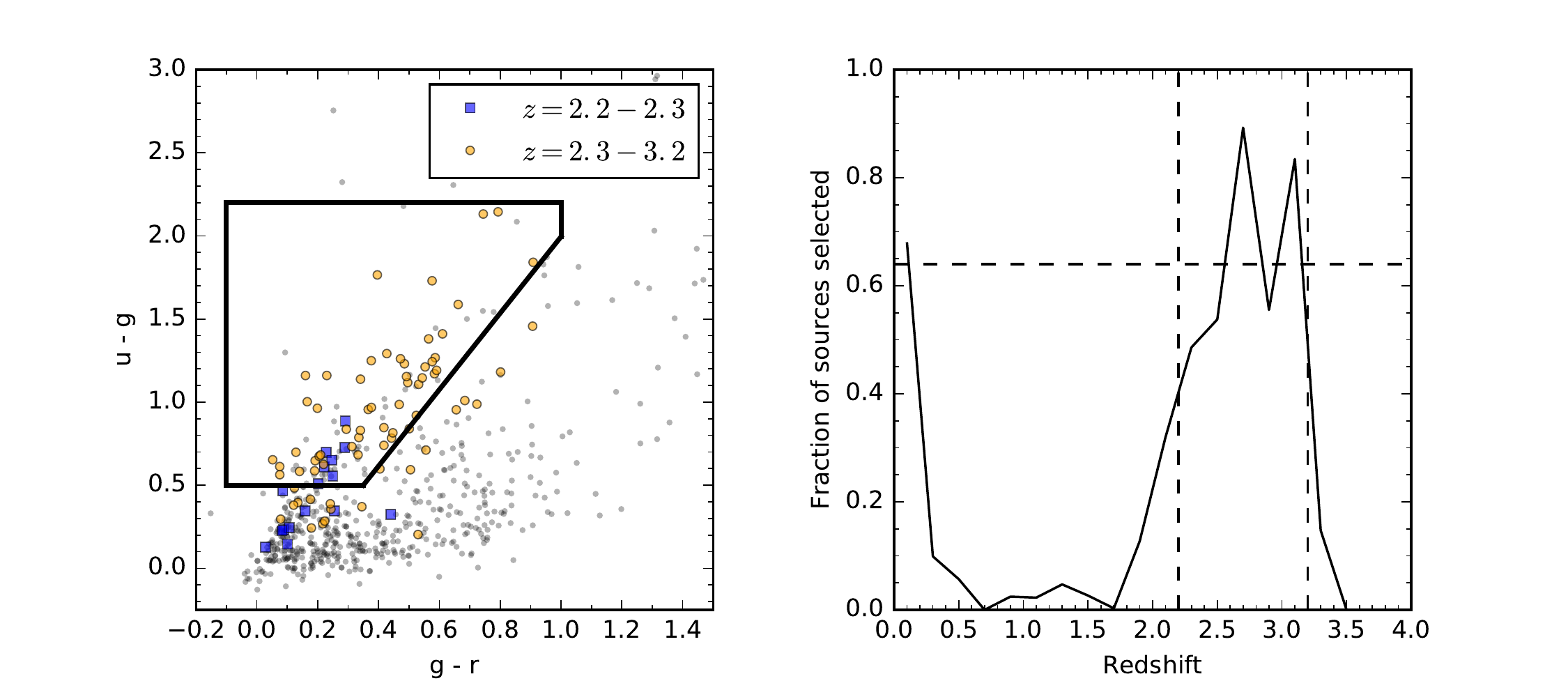}
\caption{\emph{Left:} $ugr$ colors of galaxies in the flux-limited VVDS-UltraDeep survey with $23 < i < 24.75$. Galaxies in the redshift range of interest ($z = 2.2$-3.2) are colored, while those outside it are shown in gray. The thick line encloses the selection box used to select LATIS targets (in conjunction with photometric redshifts in the COSMOS field). \emph{Right:} The fraction of sources that lie within the selection box in the left panel as a function of redshift. Here galaxies are weighted according to the VVDS selection function described by \citet{Cucciati12} and \citet{LeFevre13,LeFevre13b}. The target redshift range is enclosed by the vertical lines, with the mean completeness of 64\% indicated.\label{fig:ugr}}
\end{figure*}

Obtaining a high density of sightlines requires an efficient color-based selection of galaxies in the desired redshift range. Two approaches are widely used to select Lyman-break galaxies (LBGs): $ugr$ colors and photometric redshifts. (We refer to UV-bright, high-redshift galaxies as LBGs generically, irrespective of their exact redshift.)

The quality of photometric redshifts is highly dependent on the number of filters used and their wavelength sampling, which is not uniform over the LATIS fields. A particular problem is that near-infrared photometry only partly covers the D1 and D4 fields. With only optical photometry, there is a significant degeneracy in the photometric redshifts for high-$z$ sources due to ambiguity between the Balmer and Lyman breaks. In order to maintain a consistent selection function within each field, we decided to adopt a $ugr$ selection in all fields and to supplement this with a photometric redshift selection within the COSMOS field, which contains the best-tested and most highly constrained photometric redshifts.

\subsubsection{Color Selections and Completeness}
\label{sec:colorselection}

Our goal is to devise an efficient color selection for galaxies in the redshift range $z=2.2$-3.2. The lower limit is driven by the limited sensitivity of IMACS at $\lambda < 390$~nm, i.e., $z_{\rm Ly\alpha} < 2.2$. (Although a galaxy must have $z > 2.28$ in order to observe absorption at $z_{\rm Ly\alpha} = 2.2$ in a usable region of its spectrum, we also want to include galaxies at $z=2.2$-2.28 in order to study their positions and properties within the IGM map.) Beyond the upper limit of $z=3.2$, the utility of sightlines diminishes as less than about half of the Ly$\alpha$ forest lies within the intended tomographic volume from $z=2.2$-2.8. The sample selection is much less sensitive to the high-$z$ cutoff, since there are few sufficiently bright galaxies at $z \gtrsim 3$.

A $ugr$ color selection has been widely and effectively used to identify $z \approx 2$-3 sources, and the color limits can be tuned to select redshifts of interest \citep[e.g.,][]{Adelberger04}. Ideally, the bounds of the color selection are derived from a flux-limited sample of galaxies with spectroscopic redshifts. Fortunately, the VVDS-UltraDeep survey falls within the CFHTLS D1 field and contains a flux-limited sample with $i=23$-24.75 and sufficiently deep exposures to achieve a spectroscopic success rate of $\gtrsim 80\%$ for $z \approx 2$-3 sources \citep{LeFevre13}. The left panel of Figure~\ref{fig:ugr} shows the distribution of VVDS-UltraDeep sources in $ugr$ space, with the galaxies in our target range $z_{\rm spec}=2.2$-3.2 colored. Based on this color distribution, we defined the selection box outlined in black: $0.5 < u-g < 2.2$ and $-0.1 < g-r < 1.0$ and $u-g > 0.50 + 2.3(g-r-0.35)$. 

The upper limit of $u-g$ sets the upper redshift limit; as mentioned before, the sample is not very sensitive to this limit since the density of available targets is low. The lower limit of $u-g$ sets the lower redshift limit. Toward bluer $u-g$ colors, the number of $z < 2.2$ interlopers increases rapidly, so there is a trade-off between completeness and purity, particularly for the $z=2.2$-2.3 sources highlighted in blue in Figure~\ref{fig:ugr}. The $u-g > 0.5$ limit was chosen since it selects about half of $z \sim 2.2$ sources. The notch in the upper right corner of the selection box helps to avoid part of the stellar locus when the color selection is applied at brighter magnitudes.

The right panel of Figure~\ref{fig:ugr} shows the completeness of this $ugr$ selection relative to the VVDS-UltraDeep sample.  The color selection identifies 64\% of galaxies within our target redshift range of $z=2.2$-3.2. The main contaminants are galaxies slightly below $z=2.2$ and low-$z$ interlopers with $z \lesssim 0.3$. 

Our target selection differs in the COSMOS field in two respects. First, the \citet{Ilbert09} catalog contains photometry with different filters than the CFHTLS catalogs, particularly the $u$ band. In order to use the $ugr$ selection that we calibrated in the D1 field, we apply a conversion to the \citet{Ilbert09} $u-g$ and $g-r$ colors. The conversion was derived by comparing the colors of galaxies in the two catalogs that lie in the color selection box in Figure~\ref{fig:ugr}: $\Delta (u-g) = 0.09$, $\Delta (g-r) = 0.10 (g-r)_{\rm COSMOS}^2 - 0.31 (g-r)_{\rm COSMOS} + 0.03$, and $\Delta r = - 0.04$, where $\Delta$ is CFHTLS-COSMOS.

Second, we supplement the $ugr$ color selection in the COSMOS field by adding galaxies with $2.2 < z_{\rm phot} < 3.2$. We take $z_{\rm phot}$ from the COSMOS2015 catalog \citep{Laigle16}. For any objects not present in this NIR-selected catalog, we use the \citet{Ilbert09} $z_{\rm phot}$ instead. Although we cannot assess the completeness of this $z_{\rm phot}$ selection against the VVDS, we find that among sources selected by either the $ugr$ or the $z_{\rm phot}$ selection, only 15\% are not $ugr$-selected in the magnitude range $23.5 < r < 24.8$ motivated below. Thus the $z_{\rm phot}$ selection does not add many targets, but as we will see in Section~\ref{sec:purity}, the $z_{\rm phot}$ selection has a significantly higher purity, especially at brighter fluxes, and so is useful for prioritizing targets.

\subsubsection{Flux Limits}
\label{sec:fluxlimits}

The magnitude range is constrained by dual considerations. First, we must achieve a sightline density adequate for tomography. Second, we favor brighter photometric candidates, since the signal-to-noise ratio in their Ly$\alpha$ forest will be higher, but only as long as the fraction of low-$z$ interlopers is not prohibitive.

We will discuss the purity of the LATIS selection in Section~\ref{sec:purity}, but based on the VVDS-UltraDeep sample shown in Figure~\ref{fig:ugr}, we anticipate that $\approx50$\% of $ugr$-selected sources around $r \approx 24$ fall in the target range $z = 2.2$-3.2, and that this purity declines rapidly at brighter fluxes and becomes very small for $r < 23$ sources, which are dominated by interlopers. For our main target selection, we include sources with $r > 23$ and prioritize those with $r > 23.5$ (see Section~\ref{sec:priorities}). We also prepare separate ``bright target masks'' used in poorer weather conditions that consist of color-selected $r=22$-23.5 sources (Section~\ref{sec:brighttargetmasks}). 

As discussed in Section~\ref{sec:overview}, we must observe 3600 targets per deg${}^2$, or about twice the sampling density of an individual IMACS mask. This requires observing sources at least as faint as $r = 24.3$, which is a lower limit, since not all targets can be accommodated on two slit masks, and we will not be able to measure a redshift from every spectrum. A second consideration is that we would like to use the LATIS spectra not only for the construction of the tomographic map, but also to investigate the properties of galaxies as a function of their local density derived from the map. For this purpose, we would like to incorporate galaxies at least as faint as $L_{\rm UV}^*$ out to $z=2.8$, which corresponds to $r = 24.5$ based on the \citet{Reddy08} luminosity function. 

Based on these considerations, we have defined the magnitude range for the highest-priority targets as $r=23.5$-24.4, but we also include brighter and fainter sources in the range $r = 23.0$-24.8 at lower priority.

\subsection{Selecting QSOs}\label{sec:selecting_qsos}

\begin{figure*}
\centering
\includegraphics[width=\linewidth]{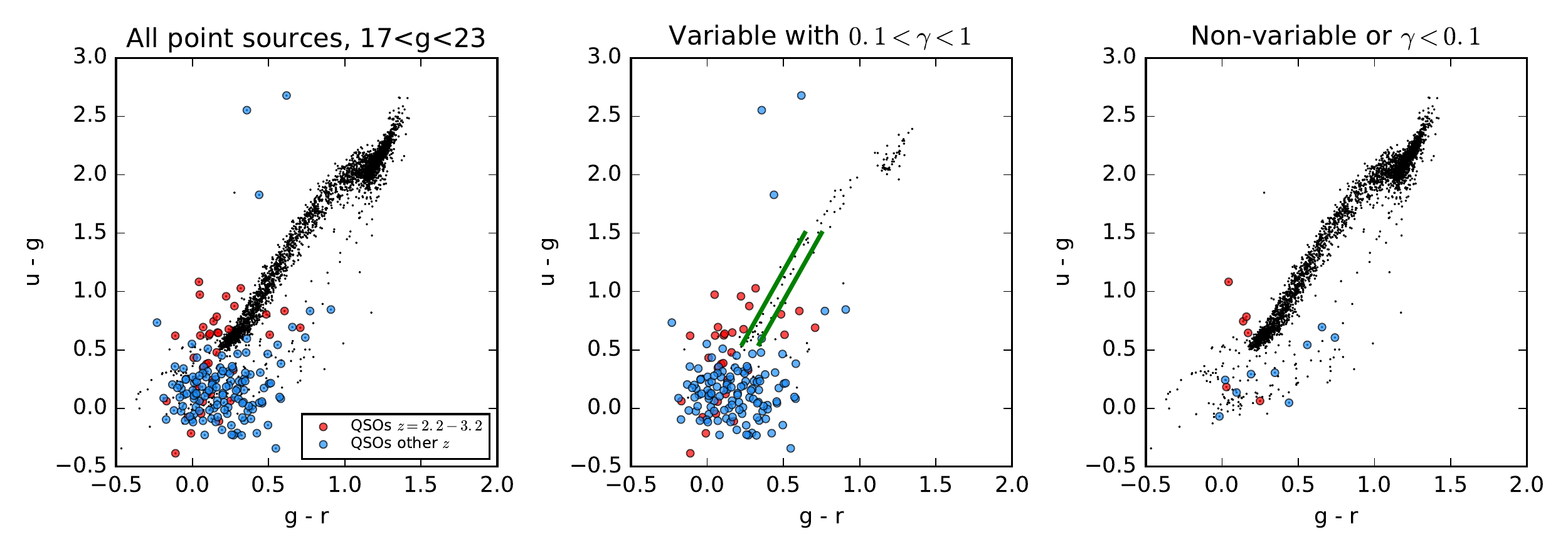}
\caption{\emph{Left:} Colors of point sources with $g=17$-23 (black) are compared to those of broad-line QSOs from the MilliQuas catalog at $z=2.2$-3.2 (red) and at other redshifts (blue). \emph{Middle:} Only sources from the left panel that are variable with a power law index $0.1 < \gamma < 1$ (see Figure~\ref{fig:qso_variability}) are plotted, showing the greatly reduced contamination from stars while including most known QSOs. The green bands enclose the residual locus of variable stars, which are excluded from the QSO selection. \emph{Right:} Only sources that do not pass the variability and $\gamma$ cuts are plotted. Note that only a few quasars are missed.\label{fig:qso_stellar_locus}}
\end{figure*}

Ultimately QSOs contribute only 2\% of the sightlines in our tomographic reconstructions. Although their inclusion is not likely to make a major improvement in the map quality, they are worth observing in LATIS because they provide high-fidelity probes of \ion{H}{1} and metals along sightlines that may pierce regions of particular interest (e.g., a protocluster). However, given their low numbers, our QSO selection must maintain an acceptable level of purity. Selecting QSOs in our target redshift range $z=2.2$-3.2 based on their colors alone is difficult. The left panel of Figure~\ref{fig:qso_stellar_locus} shows the distribution of point sources in the CFHTLS D2 catalog in $ugr$ space. Colored circles identify known broad-line QSOs from the MilliQuas catalog, and red circles indicate those at $z=2.2$-3.2. These overlap the stellar locus considerably, which would introduce an unacceptable contamination rate if not mitigated.

\begin{figure*}
\centering
\includegraphics[width=0.7\linewidth]{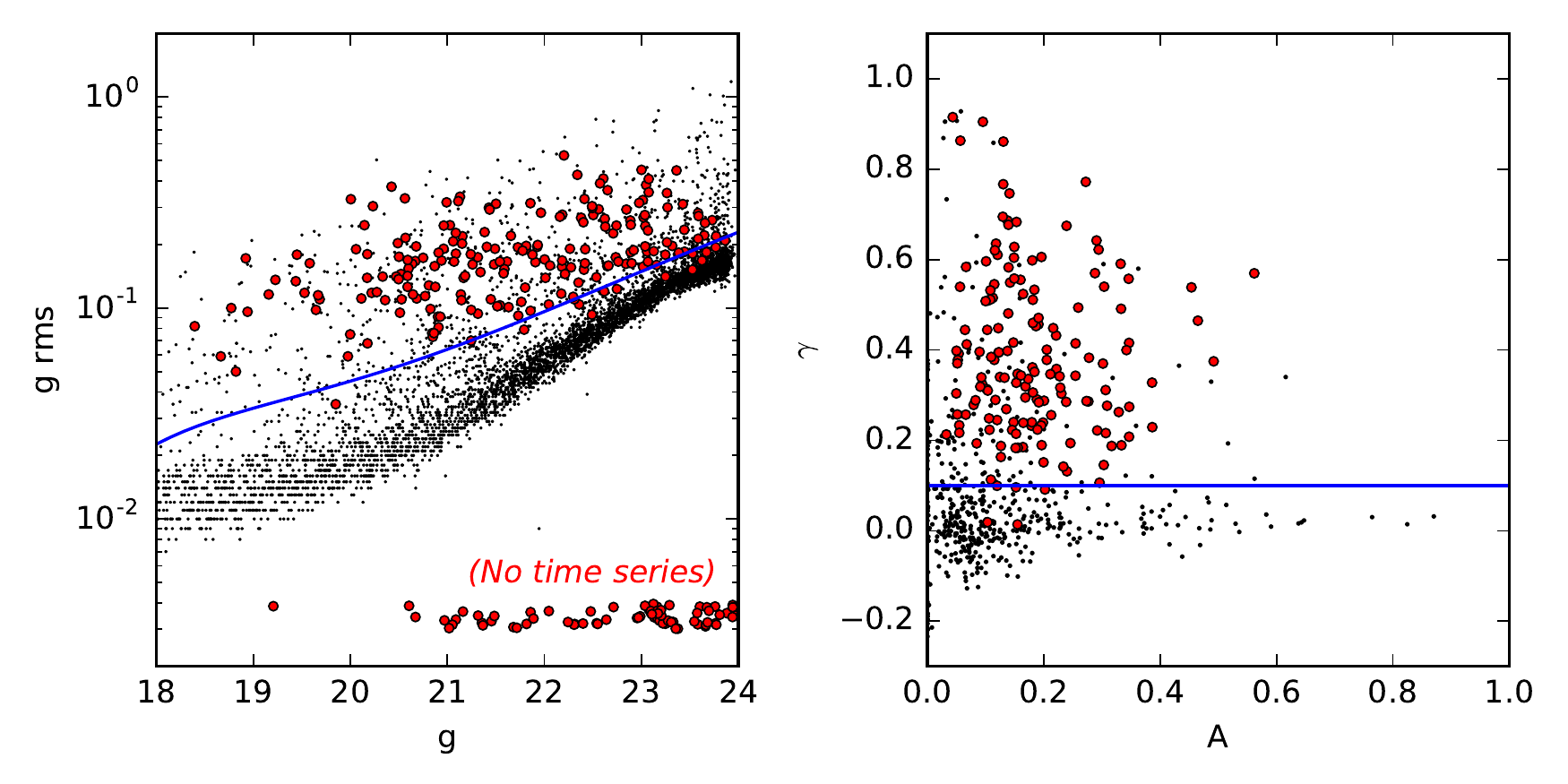}
\caption{\emph{Left:} The rms variability of point sources in the CFHTLS D2 (COSMOS) field (black points) is compared to the known QSOs (red). Sources above the blue line (see text), which includes almost all of the QSOs, are identified as variable. Known QSOs that are absent from the time series catalog are plotted at the bottom at an arbitrary abscissa. \emph{Right:} The normalization $A$ and slope $\gamma$ of a power law fit to the variability structure function of each variable source (i.e., those above the blue line in the left panel). Symbols have the same meaning as the left panel. A cut of $0.1 < \gamma < 1$ identifies nearly all known QSOs while eliminating many variable stars.\label{fig:qso_variability}}
\end{figure*}

Photometric variability provides one way to distinguish QSOs from stars. The CFHTLS fields were observed regularly over a decade, providing a time baseline for monitoring. Time series photometry for point sources in the CFHTLS-Deep fields with $17.5 < g < 24$ have been constructed by \citet{Gwyn12}.\footnote{ \url{http://www.cadc-ccda.hia-iha.nrc-cnrc.gc.ca/en/megapipe/cfhtls/dfspt.html}} The left panel of Figure~\ref{fig:qso_variability} shows the rms $g$-band magnitude variations for point sources in the D2 field, with known QSOs from the MilliQuas catalog identified as red circles. It is immediately apparent that virtually all of the QSOs are variable with fluctuations of tenths of magnitudes. We select variable sources as those lying above the blue curve in the left panel of Figure~\ref{fig:qso_variability}, which delineates the region where the rms exceeds the mode by $4\sigma$. At $g=23$-24, many of the known QSOs are not present in the time series catalog because they are not point-like, so we confine our subsequent variability analysis and QSO selection to $g < 23$ sources. 

This variability selection reduces contamination but still includes variable stars. We follow \citet[][see also \citealt{DESI16}]{Palanque11} and compute the structure function of the variable sources. We fit a power law $A(\Delta t)^{\gamma}$ to the magnitude difference $\Delta m$ as a function of the time lag $\Delta t$ in years. The right panel of Figure~\ref{fig:qso_variability} shows that QSOs are clearly distinct from the bulk of the variable sources in their distribution of $\gamma$, reflecting the fact that their magnitude differences tend to increase with the time separation. 

Based on this analysis, we make an initial identification of QSO candidates as variable $g=17.5$-23 point sources with $0.1 < \gamma < 1$. The middle and right panels of Figure~\ref{fig:qso_stellar_locus} show that this selection dramatically reduces contamination by stars while rejecting only a small fraction of QSOs. To further reduce the residual contamination by variable stars, we exclude sources along a narrow strip (enclosed by green lines in Figure~\ref{fig:qso_variability}, middle panel) aligned with the peak density of the stellar locus.

\begin{figure}
\centering
\includegraphics[width=\linewidth]{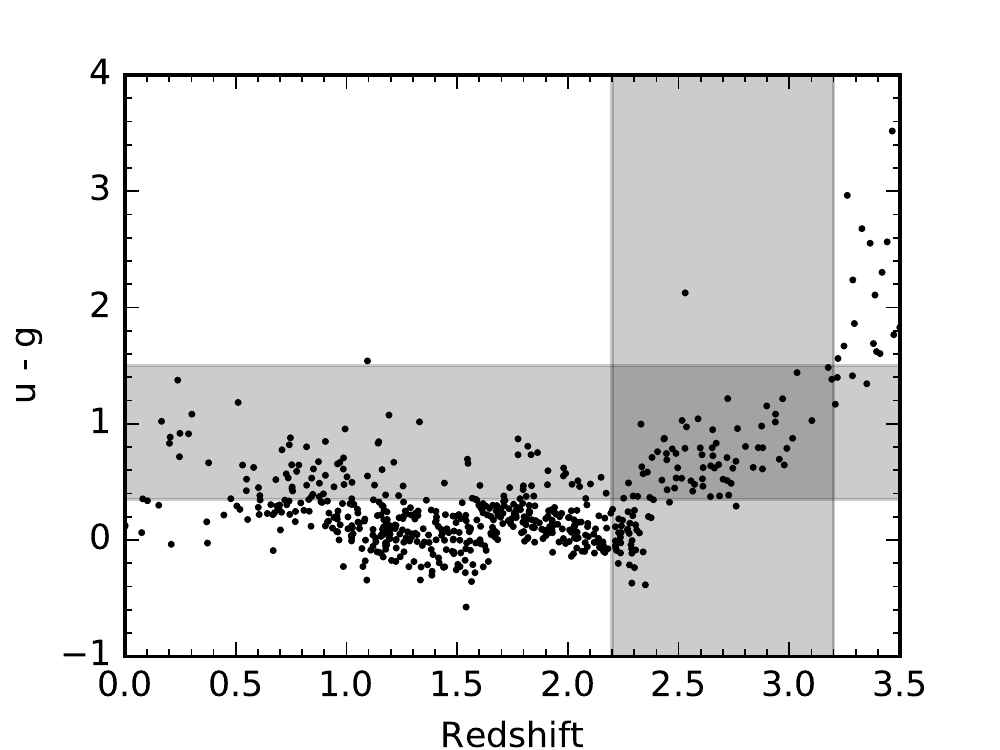}
\caption{The relationship between redshift and $u-g$ color for 624 known QSOs in the CFHTLS-Deep fields with $g < 23$. To identify the $z=2.2$-3.2 population (vertical band) while maintaining an acceptable level of contamination by lower-redshift sources, we select QSO candidates with $0.35 < u-g < 1.5$ (horizontal band).\label{fig:quasar_z}}
\end{figure}

We now must further restrict the QSO candidates to those likely to lie in the target redshift range $z=2.2$-3.2. Figure~\ref{fig:quasar_z} shows the relationship between $u-g$ color and redshift for the known QSOs in the CFHTLS-Deep fields. To select QSOs in the target range while minimizing contamination from lower redshifts, we require $0.35 < u-g < 1.5$.\footnote{In the COSMOS field, where we transform the colors from the \citet{Ilbert09} catalog as described in Section~\ref{sec:colorselection}, we find that a slightly different cut of $0.5 < u-g < 1.5$ performs better.} This cut should remove most QSOs at $z\approx1$-2, but we expect some contamination from $z\lesssim1$ QSOs.

Of the 112 QSOs at $z=2.2$-3.2 in the MilliQuas catalog and CFHTLS-Deep fields, 101 are variable, 94 also pass the $\gamma$ cut, and 57 also pass the color criteria, for a completeness of 51\%. Most of the missed targets are at the front of the volume, with $z < 2.4$, and must be excluded since their $u-g$ colors are indistinguishable from the bulk of the QSO sample at $z \approx 1$-2. Among the $z > 2.4$ QSOs in the sample, which are the most useful for tomography, this method selects 78\%. In the D2/COSMOS field, 83\% of the QSO photometric candidates have a literature spectroscopic redshift. However, in the D1 and D4 fields the fraction is only 48\% and 3\%, respectively, so our variability selection method takes on greater importance.

\section{Observational Setup and Mask Design} \label{sec:obs}

With the selection of LBG and QSO targets defined, we now describe how targets are prioritized and assembled into IMACS masks.

\subsection{Bandpass Filter and Grism} \label{sec:filtergrism}

\begin{figure}
\includegraphics[width=\linewidth]{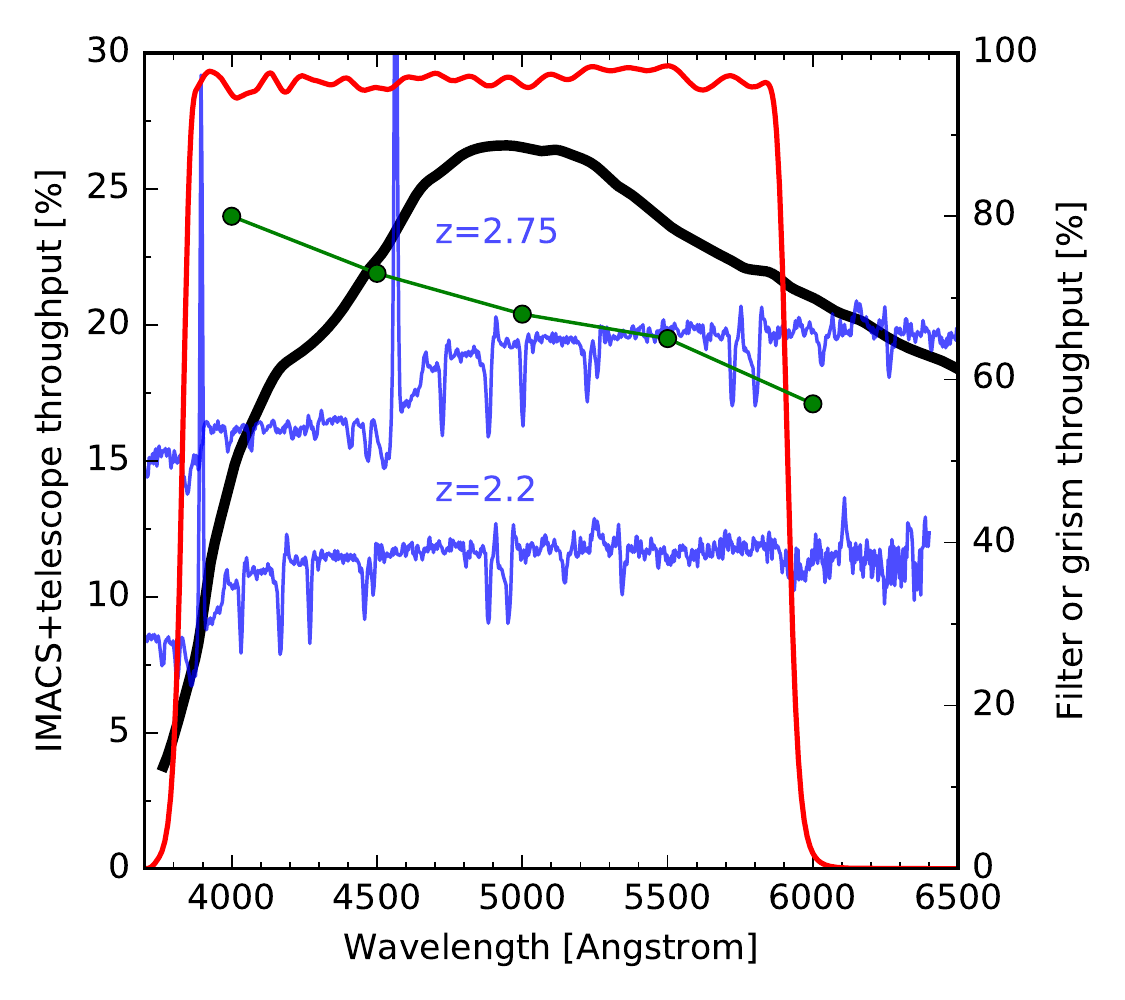}
\caption{The total measured throughput of IMACS and the telescope (black line, left axis). The throughput of the custom filter (red line) and grism (green points), as measured by the vendors, are also shown (right axis). Spectra of Lyman break galaxies at $z=2.2$ and 2.75 \citep{Shapley03} are shown for reference.\label{fig:filter_grism}}
\end{figure}

To increase multiplexing, we conduct observations through a custom bandpass filter. The blue cutoff was motivated by the 390 nm design limit of the IMACS f/2 camera \citep{Dressler11}. The red cutoff was motivated by our desire to observe the strongest interstellar lines, including \ion{C}{4} $\lambda\lambda$ 1549,1551, in the spectra of galaxies out to $z \simeq 2.75$, which we anticipated as roughly the useful limit of the tomographic map. This motivates a red cutoff near 589~nm, which additionally serves to isolate the darkest part of the night sky spectrum. The filter was fabricated by Asahi Spectra on Ohara PBL25Y glass. The measusured half-power points are 383 nm and 591 nm; the transmission is $>95$\% (average 97\%) over the range 387-586~nm.

To improve the sensitivity of IMACS at blue wavelengths, we moved the more blue-sensitive detector mosaic to the f/2 focus in December 2017. We also designed and purchased a new grism. Based on the spectral resolution considerations outlined in Section~\ref{sec:overview}, we selected from the Richardson Grating Lab (RGL) catalog a grating with 400 grooves mm${}^{-1}$ and a nominal first-order blaze wavelength of 460~nm. The grating was replicated by RGL onto a BK7 prism that has an anti-reflection coating on the input side. In the mean seeing of $0\farcs7$, the typical image size is $1\farcs1$ (the galaxies are semi-resolved and IMACS contributes some broadening). For such objects, the grism provides an average resolution of $R=880$ in the Ly$\alpha$ forest, ranging from $R=830$-920 over $z_{\rm Ly\alpha} = 2.2$-2.8. The absolute first-order diffraction efficiency, as measured by RGL, is shown by the green line in Figure~\ref{fig:filter_grism}. Due to a manufacturing error, the grating dispersion is not precisely aligned with the symmetry plane of the prism. The effect of this is to shift the spectra orthogonally to the dispersion, which results in a minor loss of 5\% of targets that are shifted off the detector mosaic.

Figure~\ref{fig:filter_grism} (black curve) shows the throughput of the instrument and telescope measured in April 2019. The throughput increases from 9-24\% over the range 390-460~nm, i.e., $z_{\rm Ly\alpha} \approx 2.2-2.8$.

\subsection{Mask Design and Target Prioritization}
\label{sec:priorities}

Targets are selected from three sources: LBG candidates based on the criteria  in Section~\ref{sec:selecting_lbgs}, QSO candidates based on the criteria in Section~\ref{sec:selecting_qsos}, and LBGs or QSOs with prior spectroscopic redshifts from the literature. 
 Masks were designed using the {\tt maskgen} software, which accounts for our filter bandpass and allows multiple ranks of slits. We used a slit width of $1\farcs2$. Slits are $6''$ long by default, but we extended the boundaries when necessary to ensure that a length of at least $3\farcs5$ is free of sources and useful for sky subtraction. {\tt maskgen} can resolve slit conflicts using user-provided numerical priorities, or alternatively it can attempt to maximize the number of slits. Although these modes may suffice for general galaxy surveys, for tomography the distribution of sightlines is also important. We therefore used a custom procedure, described below, in which we run {\tt maskgen} in several stages to prioritize targets while also evening out the sightline distribution. Since this is most easily accomplished among targets with similar priority, we introduce targets with progressively lower priorities in subsequent stages. 
 
 The highest priorities are assigned to known QSOs and QSO candidates. We then add LBG candidates in stages. We first consider $z_{\rm phot}$-selected or $z_{\rm spec}$-selected targets in the magnitude range $r=23.0$-24.4 (Section~\ref{sec:fluxlimits}). The $z_{\rm phot}$-selected sources are considered before the $ugr$-selected sources since, as we will show in Section~\ref{sec:purity}, they have a higher purity. Among sources with $z_{\rm phot} = 2.2$-3.2, we attempt to concentrate the redshift distribution slightly to maximize the Ly$\alpha$ forest pathlength within the tomography volume from $z=2.2$-2.8. We do this by drawing a random subset of the LBG photometric candidates with a probability $W(z_{\rm phot})$ that is unity over $z_{\rm phot} = 2.3$-3.0 and ramps linearly to 0 over $z_{\rm phot} = 2.2$-2.3 and 3.0-3.2. 

Using this initial subset of highest-priority targets, we generate a target list for {\tt maskgen} and produce a mask. We then attempt to redistribute the targets more uniformly throughout the IMACS footprint using a simple Monte Carlo procedure. Targets are initially prioritized randomly. We first randomly select a target for which the local density of assigned slits is particularly low. We swap its priority with a second target in the same region of the mask that has a higher local density of slits. We run {\tt maskgen} and measure the rms separation between a random point in the field and the nearest slit. If the priority swap has decreased this metric, we consider the spatial distribution to have improved and keep the swap. Iterating the procedure produces a somewhat more uniform target distribution.

In the second stage, we fix the slits already assigned, and we add $ugr$-selected targets in the magnitude range $r=23.5$-24.4. (Note that the bright limit is fainter for $ugr$-selected sources since, as we will see, their purity declines rapidly at $r < 23.5$.) We again selected a subsample of these targets following a priority $W(u-g)$ that is unity over $u-g=0.8$-1.5 and linearly ramps to zero over $u-g=0.5$-0.8 and 1.5-2.2. As for the $z_{\rm phot}$-selected galaxies, this is an attempt to slightly taper the ends of the redshift distribution. We again attempt to even out the sightline distribution as described above. 

In the third stage, we revisit all $z_{\rm phot}$-selected targets (without any subsampling) and consider the full magnitude range $r=23.0$-24.8. Slits already assigned are fixed, and additional slits are allocated according to a  priority based on the sum of $W(z_{\rm phot})$, $W(r)$, and $W(n_{\rm slit})$. Here $W(r)$ is unity over the range $r=23$-24.4 and declines linearly to 0 over $r=24.4$-24.8 to deprioritize faint sources. The $W(n_{\rm slit})$ term prioritizes galaxies in sparsely populated regions of the mask. 

In the fourth stage, we consider all $ugr$-selected sources over the full magnitude range. The procedure is the same as for the $z_{\rm phot}$-selected sources, except that the $W(r)$ term ramps from 0 to 1 over the range $r=23.0$-23.5, rather than remaining at unity, due to the lower purity of bright $ugr$-selected galaxies (Section~\ref{sec:purity}). At the end of the fourth stage, 280-310 slits are assigned on the final mask. Typically 8\% of these slits are not observable, usually because the spectrum falls into a gap between detectors or is shifted off the mosaic by the grism defect described in Section~\ref{sec:filtergrism}, which leaves 270 usable slits on average. Masks for the CFHTLS D1 and D4 fields are constructed similarly, but the above procedure is simplified since there is no $z_{\rm phot}$ selection.

These slits comprise the first of two ``target sets'' for the footprint. Masks for the second target set are constructed similarly, with two main differences. First, in order to enable studies of the inner circumgalactic medium with LATIS, we prioritize a small number (typically $\sim 3$-10) of candidates within $6''$ of a galaxy with a redshift $z=2.2$-3.2 determined from the first target set observations or a literature source. Second, sources from the first target set are repeated only where no other targets are available.

Although the two target sets largely correspond with two masks in each footprint, this is not true in detail. We attempt to improve purity by initially observing a slitmask for $\simeq 1/3$ of the total exposure time. We can then identify $\sim10$-20\% of targets as being outside the range $z=2.2$-2.8, and we generate a new mask by deleting these slits and repeating the third and fourth stages described above to add new targets. The remainder of the exposure is then spent on this improved mask.

\subsection{Bright Target Masks}
\label{sec:brighttargetmasks}

In addition to the main survey masks described in the previous subsection, we also constructed masks consisting of brighter LBG candidates in the magnitude range $r=22$-23.5. Although few of these are genuine high-redshift galaxies, they can be observed when the conditions are not suitable for the main masks, and they allow us to place the very brightest LBGs and AGN within the tomographic maps. We construct the bright target masks in two tiers by feeding {\tt maskgen} a prioritized list. In the first stage, we include the $z_{\rm phot}$-selected targets in COSMOS, while in the second stage, we add the $ugr$-selected targets. In both tiers, we prioritize fainter candidates given their much lower rate of contamination. 

\subsection{Field Tiling}
\label{sec:tiling}

Figure~\ref{fig:fieldmaps} shows a fiducial layout of footprints for the survey. Currently we have obtained full or partial observations in the D1M3, D1M4, D2M4, D2M5, D2M8 (full), D1M1, D1M2, and D4M3 (partial) footprints. The overall positioning of the footprints is designed to maximize overlap with the external surveys shown in the figure. We also chose footprints that are a subset of a complete tiling of each field, in order to allow for the possibility of future observations over a larger area. (This also accounts for the non-sequential numbering of the footprints shown.) Small shifts from a uniformly spaced tiling are needed to allow the guide probes to reach suitable guiding and Shack-Hartmann stars.

Fields are separated by $24'$ in R.A. because at field radii $R > 12'$, IMACS suffers from some vignetting and degraded image quality. Our tiling scheme ensures that much of the $R > 12'$ region is covered by two footprints, which allows targets in the vignetted overlap region to have twice the exposure time. Accounting for our wavelength coverage constraints, the addressable field of view is a circle with $R = 15'$ truncated by two lines of constant declination separated by $21'$, and also lines of constant right ascension located $14.1'$ west and $12.3'$ east of center. The east-west truncations reflect the detector mosaic boundary, and they are asymmetric because of the lateral shift of the spectral traces described in Section~\ref{sec:filtergrism}. Each footprint covers 0.15 deg${}^2$.

\section{Completed Observations and Data Reduction}

\subsection{Observations}\label{sec:observations}

Over 28.5 operable nights from December 2017 to April 2019, we conducted LATIS observations in all of the footprints listed in Section~\ref{sec:tiling}. At least one target set has received the full planned exposure in the D1M3, D1M4, D2M4, D2M5, and D2M8 footprints (outlined in bold in Figure~\ref{fig:fieldmaps}), and the remainder of the paper will focus on these data, although we have partial observations in other footprints. We have fully observed both of the main target sets in all of these fields except D1M4, where only one is complete. In addition, we have observed bright target masks in D1M4 and D2M4.

The total exposure time that a galaxy receives varies according to several factors, e.g., weather conditions, duplication on multiple target sets or footprints, or removal from a target set following identification as an interloper. The median exposure time is 12.2 hours, or 14.2 hours for those galaxies we will ultimately use for tomography. This exposure time was intended to produce a typical signal-to-noise ratio of roughly $\sim2$ \AA${}^{-1}$ in the Ly$\alpha$ forest. This limit was in turn motivated by \citet{McQuinn11}, who showed that gains in measuring the flux correlation function using a quasar survey begin to diminish at higher signal-to-noise ratios, as the noise in the spectra becomes smaller than the amplitude of IGM fluctuations for a wide range of scales $\gtrsim 2$~cMpc.

In order to minimize the effect of read noise, we operate IMACS in $1 \times 2$ binning, i.e., with $0\farcs2$ pixels and a dispersion of 1.8~\AA~($0\farcs4$) per pixel, and use the slow read mode coupled with 45~min exposures. Wavelength calibration is obtained using helium and mercury lamps that illuminate the flat field screen at the telescope pupil. To obtain adequate counts at blue wavelengths, we use exposures of the twilight sky for flat fielding. The Magellan Baade telescope is equipped with an atmospheric dispersion corrector, which removes chromatic differential atmospheric refraction (DAR). Due to the wide field of view, achromatic DAR (i.e., a gradient in scale) can be appreciable. We calculate the typical hour angle for a planned observing sequence and design the mask using the DAR capability of {\tt maskgen}. For observations of a mask over its full arc, we design two masks for use east and west of the meridian. This strategy should reduce the DAR-induced offsets between images and slits to $\lesssim 0\farcs2$.

\subsection{Data Reduction}

The data were reduced using a series of Python scripts designed to process IMACS observations in a highly automated way. For a given mask, a fiducial mapping from the focal plane to the detector is first refined using direct images of the slitmask. Lines are then identified in the arc lamp spectra, and a two-dimensional polynomial is fit to the global wavelength solution on each of the 8 detectors. Twilight flats are reduced by modeling and dividing out the sky spectrum. The slit functions, which encode the variation in throughput along a slit, are factored from the pixel-to-pixel variations in the flat. We generally take twilight flats at a series of gravity angles and then reduce each science frame with the closest matching flat. 
For each science exposure, we subtract bias using the overscan region before using cross-correlations to estimate small residual flexure between the flat and science exposure. These shifts are applied to the slit functions, which are then divided from the science frame along with the pixel flat. Sky subtraction is performed in two phases using bspline techniques \citep{Kelson03}. The first pass is used to roughly remove the sky emission and locate the targets. The portion of the slit within $0\farcs7$ of the target position is then masked, thereby isolating the sky flux for the second pass. For each galaxy on a given mask, the spectra are then rectified, normalized to a common flux level, and averaged using inverse-variance weighting with outlier rejection. A one-dimensional spectrum is then optimally extracted \citep{Horne86}. Noise spectra based on standard CCD statistics are propagated throughout.

Galaxies are usually observed on multiple masks. For each galaxy in the survey, we then optimally combine all of the extracted spectra. Flux calibration to $f_{\nu}$ is performed based on twilight observations of white dwarfs in the X-Shooter standards library \citep{Moehler14}. Spectra can be contaminated in several ways, most commonly by overlapping the zero-th order spectra of other slits. We use automated methods to identify many of these contaminated regions, which are also flagged during our visual inspection of the spectra (Section~\ref{sec:specclass}).

\section{Spectrum Analysis and Sample Statistics}

With the data now reduced, we turn to our methods for visually inspecting and classifying the 2895 spectra distributed over 11 target sets in the 5 footprints listed in Section~\ref{sec:observations}. We will first review the classifications, sampling rate, and purity for the 2596 galaxies observed in the 9 main target sets. We will then consider the 299 targets that have been observed only on a bright target mask (Section~\ref{sec:brighttargetmasks}), since these  have very distinct statistics.

\subsection{Initial Spectral Classification and Redshifts}
\label{sec:specclass}

We developed an interactive GUI to examine the 1D and 2D spectra of every target. For each target, we attempted to identify the spectrum and measure an approximate initial redshift by comparing to the \citet{Shapley03} LBG composite spectrum and a set of SDSS templates that include low-redshift galaxies, stars, and QSOs.\footnote{\url{https://classic.sdss.org/dr2/algorithms/spectemplates/index.html}} These initial redshifts serve only as starting points for the refined versions based on an expanded template library that we will describe in Section~\ref{sec:specmodeling}. We assigned a redshift quality {\tt zqual} as follows:

\begin{itemize}
\item {\tt zqual} = 0: No redshift could be assigned (12.3\% of spectra)
\item {\tt zqual} = 1: Only a single emission line was identified and assumed to be Ly$\alpha$ (1.2\%)
\item {\tt zqual} = 2: Low-confidence guess, not suitable for most analyses (9.0\%)
\item {\tt zqual} = 3: High-confidence redshift, multiple lines and a well-modeled spectrum (19.8\%)
\item {\tt zqual} = 4: Certain redshift, high signal-to-noise spectrum with numerous lines identified (57.8\%)
\end{itemize}

Throughout the remainder of the paper, we consider reliable redshifts as those with ${\tt zqual}$ = 3 or 4, which comprise 78\% of the spectra. Since our exposure times are driven by requirements in the Ly$\alpha$ forest, the region of the spectrum redward of Ly$\alpha$ achieves a rather high signal-to-noise ratio of 3.3 pixel${}^{-1}$ on average, which accounts for the high fraction of high-confidence redshifts. The interactive tool also allows us to flag QSOs and AGN.

\begin{figure}
\centering
\includegraphics[width=\linewidth]{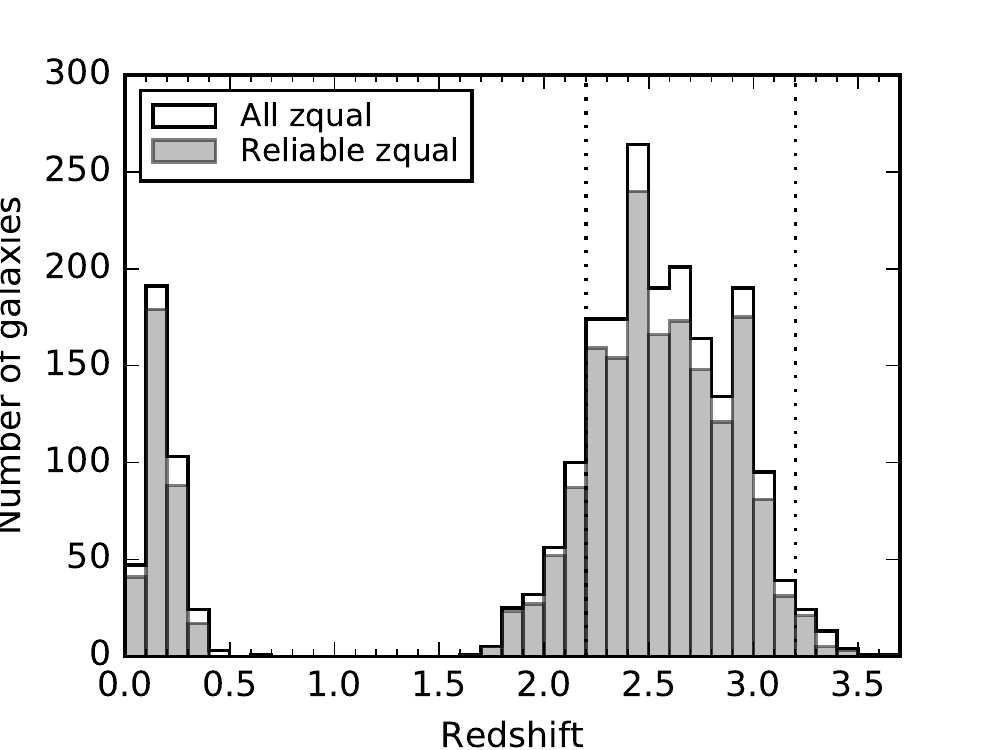}
\caption{Redshift distribution of galaxies and QSOs. The subset with reliable redshifts ({\tt zqual} = 3 or 4) are shown in the filled histogram. Note the concentration of sources in the target range $z=2.2$-3.2 (dotted lines).\label{fig:zdist}}
\end{figure}

\begin{deluxetable}{lcc}
\tablecolumns{3}
\tablewidth{\linewidth}
\tablecaption{Inventory of main target sets\label{tab:numbers}}
\tablehead{
\colhead{Type} & \colhead{Observed to date} & \colhead{Full LATIS} }
\startdata
All targets & 2596 & 6920 \\
$z=2.2$-3.2 galaxies/QSOs & 1593 & 4250 \\
With {\tt zqual} $\geq 3$ & 1425 & 3800 \\
Within tomographic area & 1268 & 3800 \\
Used for tomography & 1071 & 3210
\enddata
\tablecomments{The right column shows extrapolations to the full LATIS survey. Each row is a subset of the last. Note that 1 of the 9 main target sets that has been observed falls outside of the tomographic reconstruction in this paper (Section~\ref{sec:mapmethod}).}
\end{deluxetable}

The distribution of redshifts is shown in Figure~\ref{fig:zdist}. It is clear that the sources are indeed concentrated in the target range $z=2.2$-3.2 (dotted lines), as we will quantify below. The main identifiable contaminant is a population of low-mass galaxies primarily at $z \lesssim 0.4$. Some galaxies at $z \approx 0.5$-1.5 are probably also present, but their redshifts would be hard to identify given the bandpass of our filter. Table~\ref{tab:numbers} shows the numbers of galaxies observed to date and extrapolated to the full LATIS survey. We note that 97\% of the $z=2.2$-3.2 targets are LBGs while only 3\% are QSOs.

\begin{figure*}
\centering
\includegraphics[width=\linewidth]{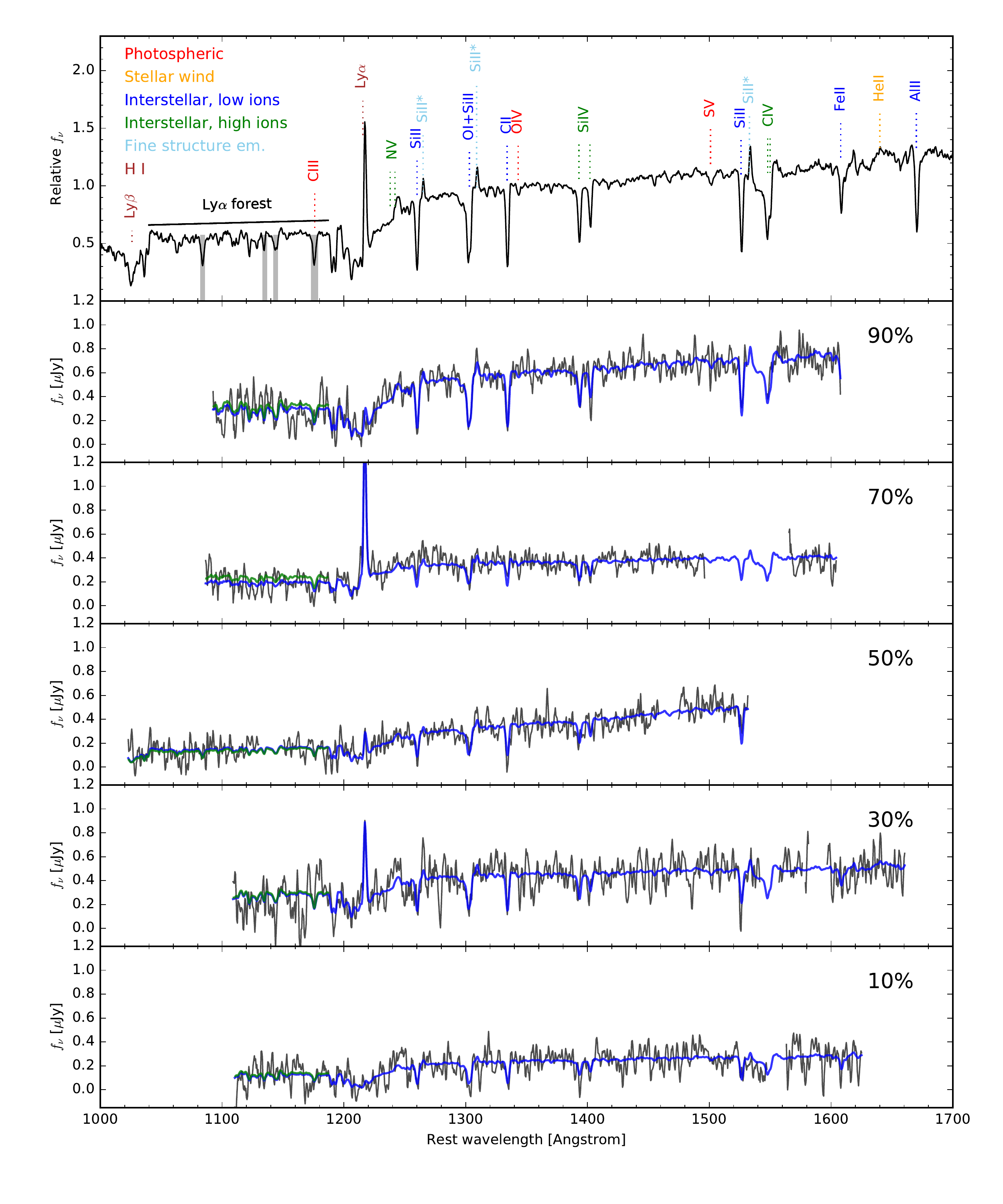}
\vspace{-10ex} 
\caption{\emph{Top panel:} Mean spectrum of $z > 2$ LBGs with confident redshifts. Several strong stellar and interstellar features are identified and colored according to their origin. Shaded boxes in the Ly$\alpha$ forest indicate regions that are masked for our tomography analysis. \emph{Lower five panels:} Representative spectra with {\tt zqual} $=3$ or 4 at the 10th, 30th, 50th, 70th, and 90th percentiles of the signal-to-noise distribution (from bottom to top) are plotted in gray after smoothing with a 3 pixel boxcar. Colored curves show the models described in Section~\ref{sec:lbgmodeling}. In the Ly$\alpha$ forest region, the blue curves have been adjusted using MFR while the green ones have not; note the very small differences. The models include the mean absorption $\langle F \rangle (z)$.
\label{fig:specexamples}}
\end{figure*}

\begin{figure}
\centering
\includegraphics[width=\linewidth]{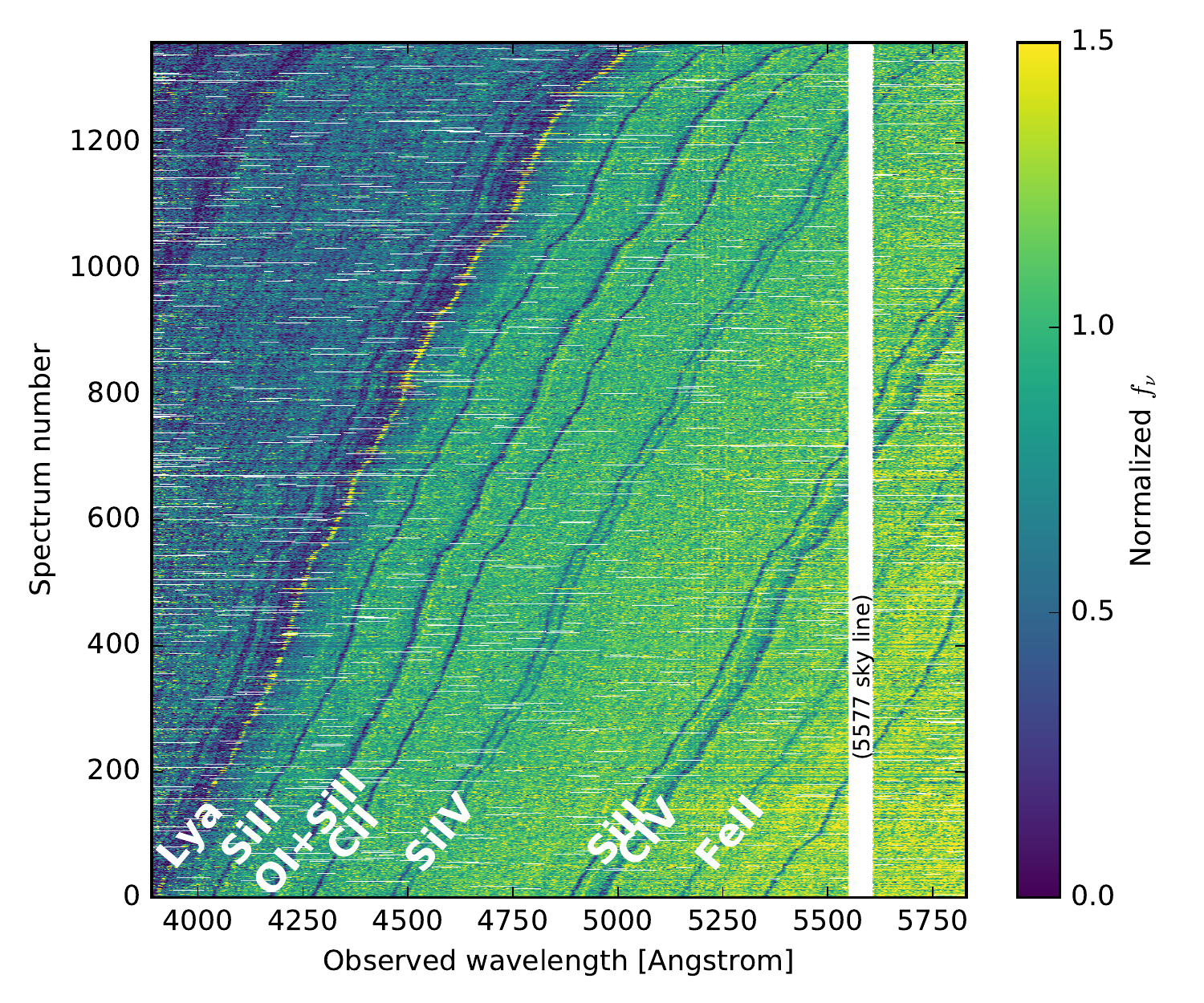}
\caption{Visualization of the 1360 LBG spectra with high-confidence redshifts at $z=2.2$-3.2. Color encodes the relative intensity, and spectra are ordered in increasing redshift from bottom to top. Many redshifting spectral features are clearly seen as labeled at bottom. White bands represent masked regions of spectra, which most commonly occur at chip gaps or zero-th order spectra\label{fig:specgraphic}}
\end{figure}

\begin{figure*}
\centering
\includegraphics[width=\linewidth]{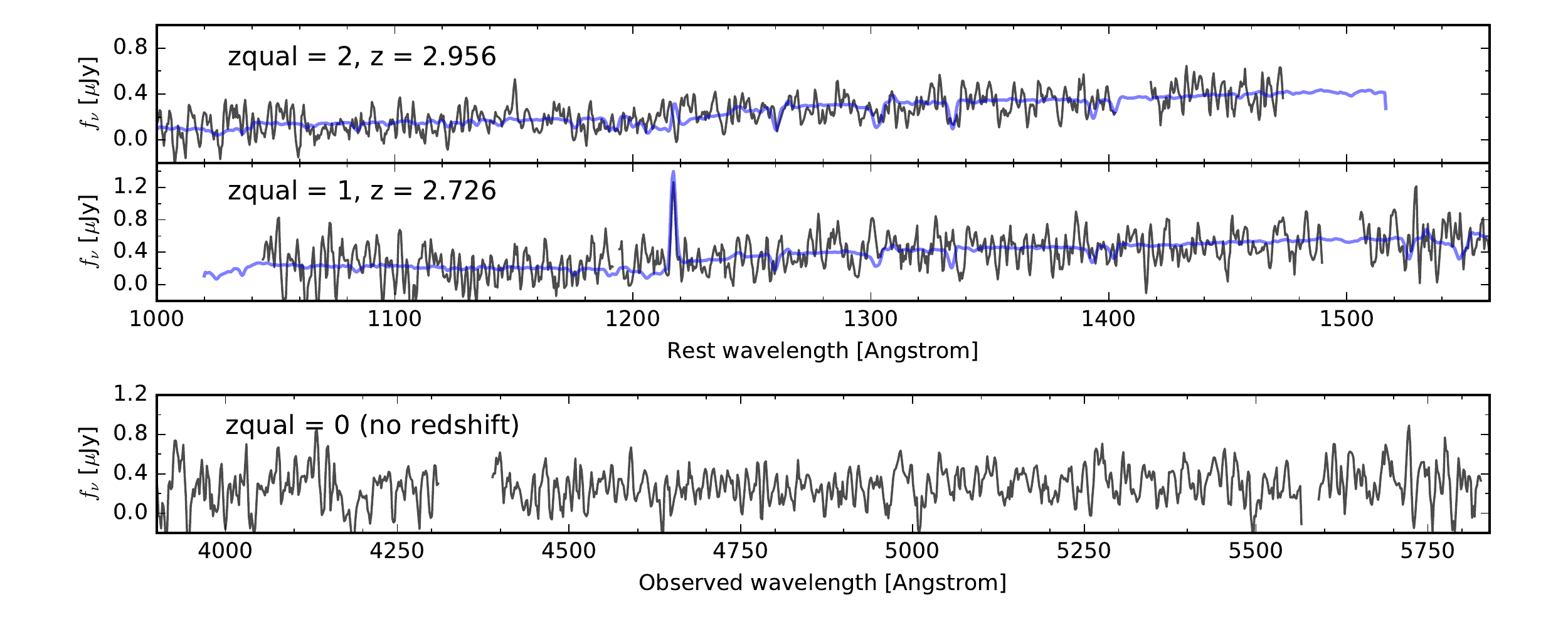}
\caption{Representative spectra with a lower confidence redshift ({\tt zqual} $=1$ and 2) or no redshift estimate ({\tt zqual} $=0$). Such spectra are, conservatively, not used in the remainder of this paper. As in Figure~\ref{fig:specexamples}, the spectra are smoothed with a 3 pixel boxcar, and blue curves show the models from which redshifts are derived.\label{fig:z012spec}}
\end{figure*}

Figure~\ref{fig:specexamples} gives an idea of the data quality by displaying a set of example spectra spanning the 10th-90th percentiles of the signal-to-noise distribution. All of these galaxies have high-confidence redshifts $z > 2.2$ and show clear evidence of multiple interstellar transitions indicated in the top panel. Figure~\ref{fig:specgraphic} graphically presents the full set of 1360 LBG spectra with high-confidence redshifts in the targeted range $z=2.2$-3.2. The Ly$\alpha$ forest region is colored blue. For completeness, in Figure~\ref{fig:z012spec} we show representative spectra with {\tt zqual} $=0$ (no redshift), 1 (single emission line), and 2 (low confidence). Although redshifts with ${\tt zqual}$ $=1$ or 2 are likely to be correct in most cases, we do not use them for the analyses in this paper.

We compared the redshifts of galaxies in common with the full VUDS \citep{LeFevre15,LeFevre19}, zCOSMOS-Bright \citep{Lilly09}, and zCOSMOS-Deep (S.~J.~Lilly et al.~in prep) surveys, which were not used to inform targeting. Galaxies were matched to the nearest source in our photometric catalogs within 1 arcsec. Throughout this paper, we only consider VUDS and zCOSMOS redshifts with quality flags of 3 or 4, corresponding to the most secure redshifts. There are 333 galaxies with high-confidence redshifts in both LATIS and one of these surveys. Among these, we identify 12 $5\sigma$ outliers. One is a QSO with an uncertain velocity, and 2 are blended systems where the target is uncertain. After reviewing the LATIS spectra of the remaining 9, we find that 3 support the LATIS redshift, 2 support the literature redshift, and 4 cases are ambiguous. We conclude that this external comparison supports our redshift identifications, with $\lesssim 2\%$ of the high-confidence LATIS redshifts called into question, all of which were graded with {\tt zqual} $ = 3$.

\begin{figure*}
\centering
\includegraphics[width=0.8\linewidth]{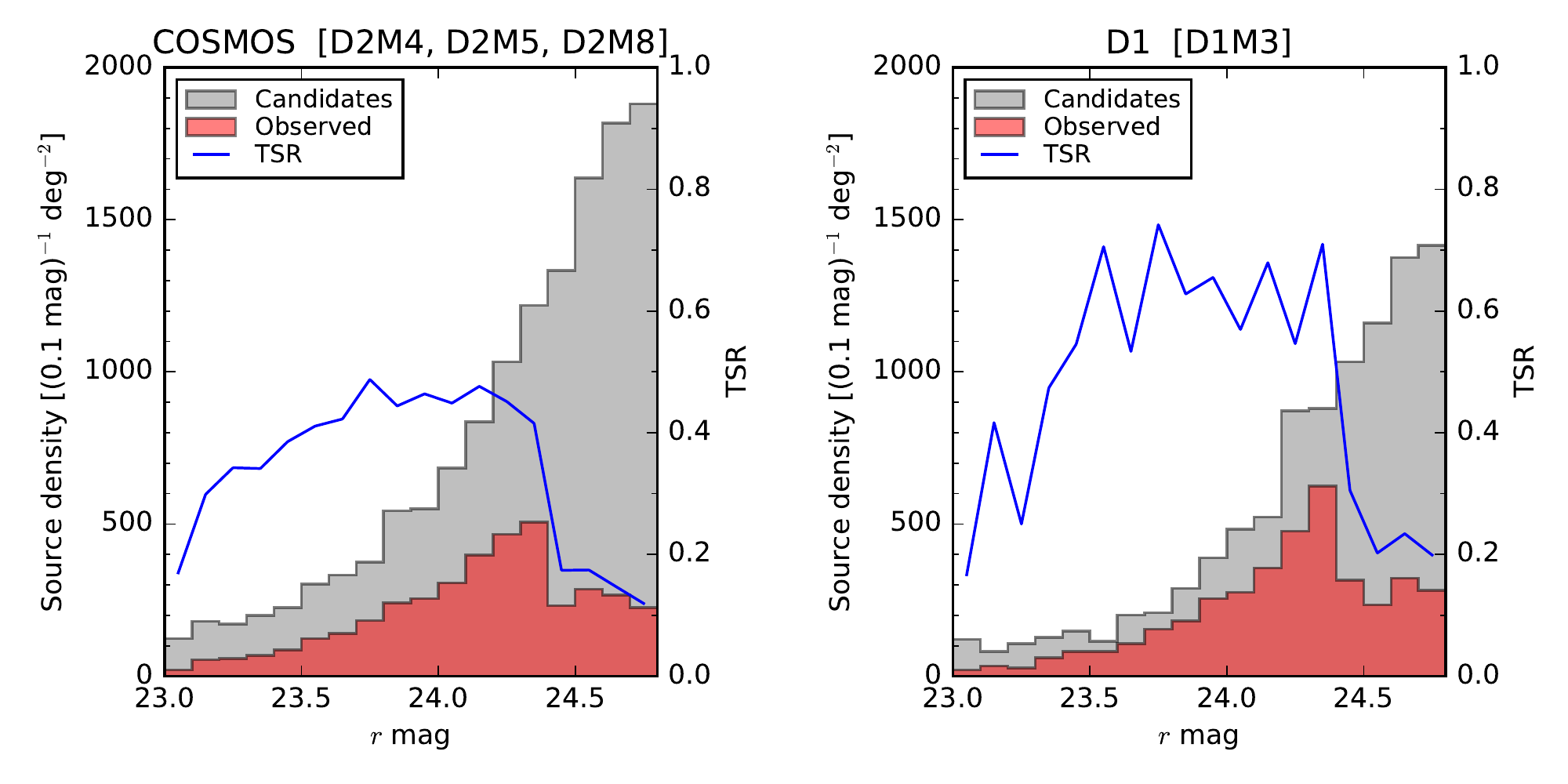}
\caption{Targeting rate as a function of $r$ magnitude. The left and right panels show rates for footprints in the COSMOS and D1 fields, respectively, where both of the main target sets have been observed. Histograms show the surface density of candidate targets (gray) and those actually observed (red). Their ratio is the target sampling rate (TSR), shown by the blue curves (right axes).\label{fig:targetsampling}}
\end{figure*}

\begin{figure*}
\centering
\includegraphics[width=0.8\linewidth]{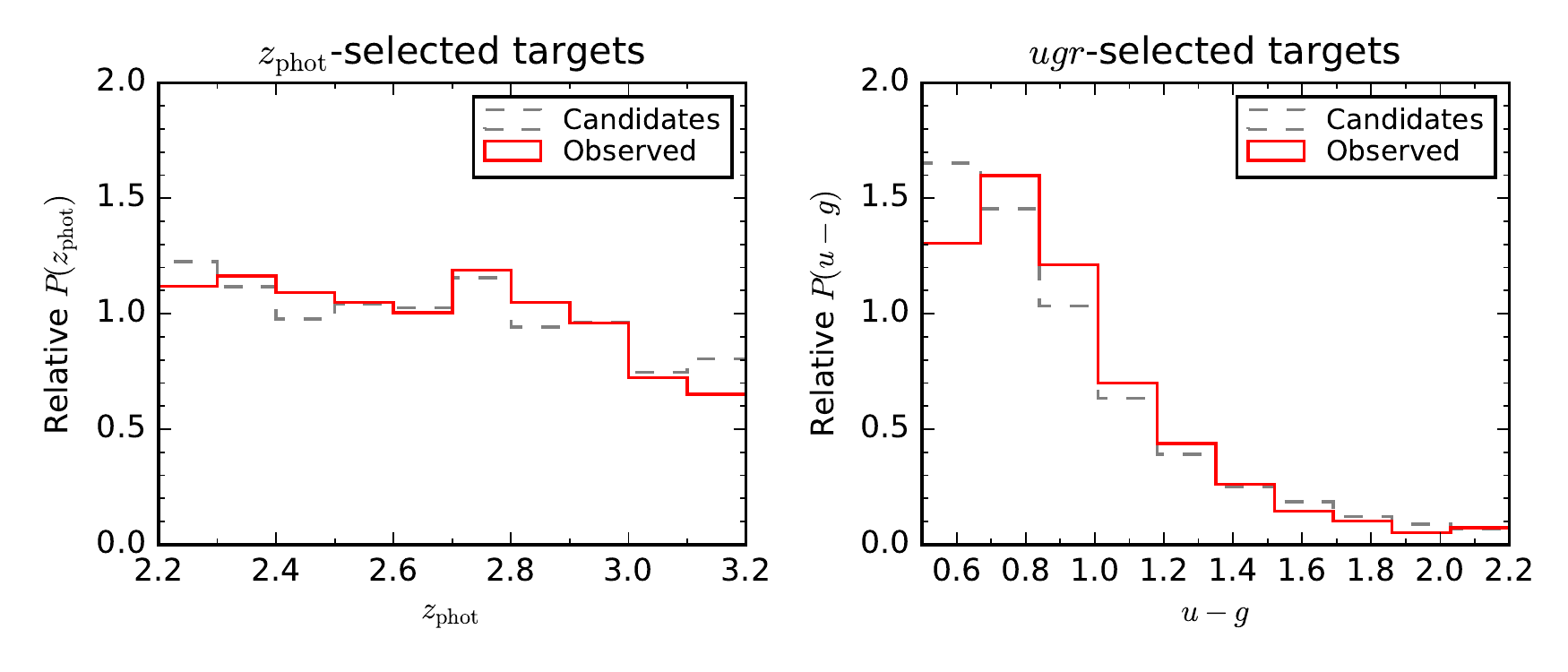}
\caption{\emph{Left:} Relative distribution of photometric redshifts $z_{\rm phot}$ for candidate targets (gray) and observed targets (red) that are $z_{\rm phot}$-selected, i.e., that have $2.2 < z_{\rm phot} < 3.2$. \emph{Right:} Relative distribution of $u-g$ colors for candidate and observed targets that are $ugr$-selected.\label{fig:target_sampling_colors}}
\end{figure*}

\begin{figure*}
\centering
\includegraphics[width=0.8\linewidth]{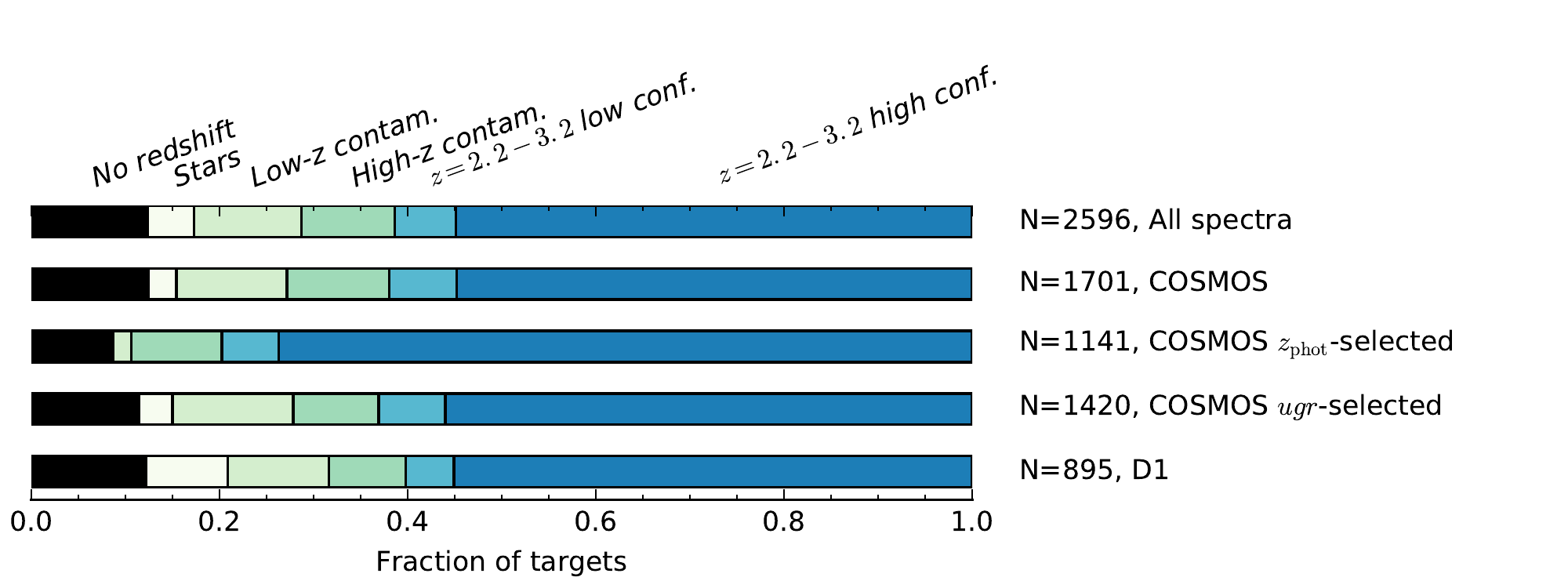}
\caption{The nature of targeted sources. The 5 rows consider different subsamples as indicated on the right. The fraction of targets is indicated which have no measured redshift ({\tt zqual} $=0$), stars, low-$z$ contaminants ($z < 0.5$), high-$z$ contaminants ($0.5 < z < 2.2$), and galaxies in the target volume $z=2.2$-3.2 at low confidence ({\tt zqual}$=1$ or 2) or high confidence ({\tt zqual} $=3$ or 4). For the survey as a whole, 55\% of targets have confident redshifts $z=2.2$-3.2.
\label{fig:purity}}
\end{figure*}

\subsection{Target Sampling Rate and Purity}
\label{sec:purity}

Figure~\ref{fig:targetsampling} shows the rate at which candidate LBGs and QSOs were targeted as a function of $r$-band magnitude. For this figure, we consider only those footprints in which both main target sets have been fully observed (D1M3, D2M4, D2M5, D2M8). In COSMOS, $\simeq 46\%$ of candidates have been observed near $r \simeq 24$, the highest-priority magnitudes, while in D1 the fraction is 64\%. The higher target sampling rate (TSR) in D1 is due to a lower number of candidates in the D1M3 footprint, which in turn seems to arise from cosmic variance; the higher TSR will likely not apply to the D1 field as a whole. Thus, overall, LATIS targets around half of the $\gtrsim L^*$ LBG candidates.

In both fields, there is a sharp decline in TSR at $r > 24.4$ reflecting the lower prioritization of these faint sources. In D1 there is also a decline at $r < 23.5$, since bright $ugr$-selected targets have lower priority, whereas in COSMOS this decline is more gradual since $z_{\rm phot}$-selected targets with $r=23$-23.5 are not deprioritized (Section~\ref{sec:priorities}). The colors and photometric redshifts of candidate and observed targets are compared in Figure~\ref{fig:target_sampling_colors}. The distributions are quite similar; galaxies at the edges of the $z_{\rm phot}$ range and those with the bluest $u-g$ colors are only slightly under-represented in the observed targets, reflecting the prioritization scheme discussed in Section~\ref{sec:priorities}. 

The overall purity of our targeting is illustrated in Figure~\ref{fig:purity}, which breaks down the targets according to their redshift and confidence. Overall 55\% of sources in our main target sets have confident redshifts in the desired range $z=2.2$-3.2, which we define as the purity. A further 7\% have redshifts in this range at lower confidence. The purity is similar between the COSMOS and D1 fields. Although Figure~\ref{fig:purity} shows that the purity is higher for $z_{\rm phot}$-selected galaxies in COSMOS than for $ugr$-selected galaxies, which are the only type available in D1, this does not translate to a large difference in the overall sample purity (compare second and fifth rows). The reason is that the surface density of $z_{\rm phot}$-selected sources only permits 2/3 of the slits to be filled, and a similar proportion (61\%) of the more abundant $ugr$-selected sources are also $z_{\rm phot}$-selected anyway. We note that 14\% of targets had a prior $z_{\rm spec}$ in the literature; excluding these would lower the purities discussed here at the 5\% level.

\begin{figure}
\centering
\includegraphics[width=\linewidth]{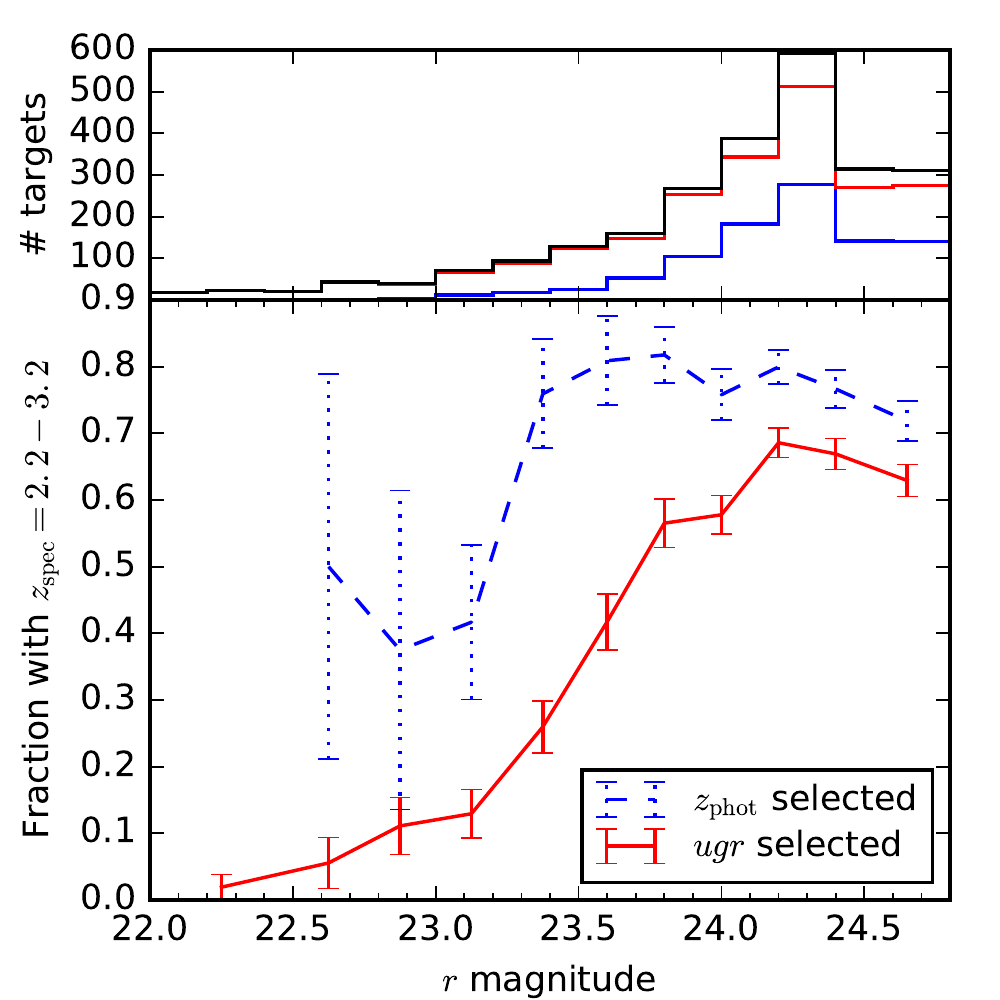}
\caption{Fraction of $z_{\rm phot}$- or $ugr$-selected LBGs candidates with LATIS redshifts in the target range $z=2.2-3.2$, with any confidence level, as a function of $r$ band magnitude. Sources with prior known $z_{\rm spec}$ are excluded. The histogram at top displays the magnitude distribution of the two selection methods and the total target set (in black). \label{fig:purity_rmag}}
\end{figure}

While Figure~\ref{fig:purity} encapsulates the overall statistics of our target selection, the purity is a strong function of magnitude. Figure~\ref{fig:purity_rmag} shows the fraction of targets for which we measured $z=2.2$-3.2 (at any {\tt zqual}). Here we consider only the LBG candidates without a prior $z_{\rm spec}$ and include targets from both our main and bright target sets. For candidates with $r \gtrsim 23.7$, the purity of the $z_{\rm phot}$ and $ugr$ selections is actually fairly similar. But for brighter galaxies, the $ugr$ selection is significantly less pure. This motivates our decision to deprioritize $ugr$-selected galaxies with $r=23$-23.5 in our main target sets, and to reserve $r < 23$ targets for the backup bright target masks.

In addition to the LBGs, we have observed 53 QSO candidates. The majority of these (44) were known as QSOs with spectroscopic redshifts in the literature. Among the additional 9 photometric candidates, 2 were confirmed as QSOs (1 at $z=2.2$-3.2) with most of the rest being stars. The low success rate among new candidates is likely due to the fact that 2/3 of the current data are in COSMOS, where the quasar population has already been well observed (Section~\ref{sec:selecting_qsos}).

\subsection{Bright Target Statistics}

The sample statistics for the bright target masks (Section~\ref{sec:brighttargetmasks}) are quite different. These masks consist of $ugr$- and $z_{\rm phot}$-selected targets with $r=22$-23.5, which are usually not high-redshift galaxies, as Figure~\ref{fig:purity_rmag} shows. Furthermore, the masks are observed in substandard conditions. Among the 243 targets with $r < 23.5$ that were observed only on the two bright target masks, 10\% were confirmed to be $z = 2.2$-3.2 galaxies with any {\tt zqual}. Most are stars or low-redshift galaxies. The yield is much higher (42\%) among the $z_{\rm phot}$-selected targets, but there are only 12 of these. These backup masks therefore do not contribute appreciable to the sightline density, but they do allow poorer conditions to be productively used to map the locations of very luminous galaxies up to $\simeq (4-5) L^*$.

\section{Spectrum Modeling and Redshift Measurements}
\label{sec:specmodeling}

With the spectra now classified and with preliminary measurements of redshifts, we now describe the techniques we use to model the LBG and QSO spectra. Modeling the spectra is needed to best estimate the intrinsic galaxy spectrum in the Ly$\alpha$ forest, the ``continuum'' against which foreground absorption will be measured. It also allows us to refine our initial redshift measurements. 

\subsection{LBG Spectrum Modeling}
\label{sec:lbgmodeling}

Although the Ly$\alpha$ forest is relatively flat in LBG spectra, it is not a featureless continuum. Furthermore, as we will show, the strength of the absorption features in the forest is correlated with the interstellar absorption features redward of Ly$\alpha$. Therefore, in order to make the best estimate of an LBG's intrinsic spectrum in the Ly$\alpha$ forest, it is best to model the entire spectrum. 

We do this by constructing a set of galaxy spectral templates from the LATIS data set. The templates and model fits were constructed iteratively. We divided the observed LBGs with high-confidence redshifts $z > 2.28$ (ensuring that part of the Ly$\alpha$ forest is included) into 5 bins of Ly$\alpha$ equivalent width (EW). We initially shifted these into the rest frame using the redshifts determined from manual inspection and comparison to the \citet{Shapley03} composite spectrum (Section~\ref{sec:specclass}). For each spectrum, we divided out the mean transmission $F(z)$ of the Ly$\alpha$ forest, as measured by \citet{FG08}. We then fit a power law to the spectrum redward of 1250~\AA, masking the strong interstellar absorption lines, and divided it from the entire spectrum to remove the continuum slope. All such spectra in a given bin of Ly$\alpha$ EW were then averaged, excluding a small fraction of sources with spectroscopic evidence of an AGN.

\begin{figure*}
\centering
\includegraphics[width=\linewidth]{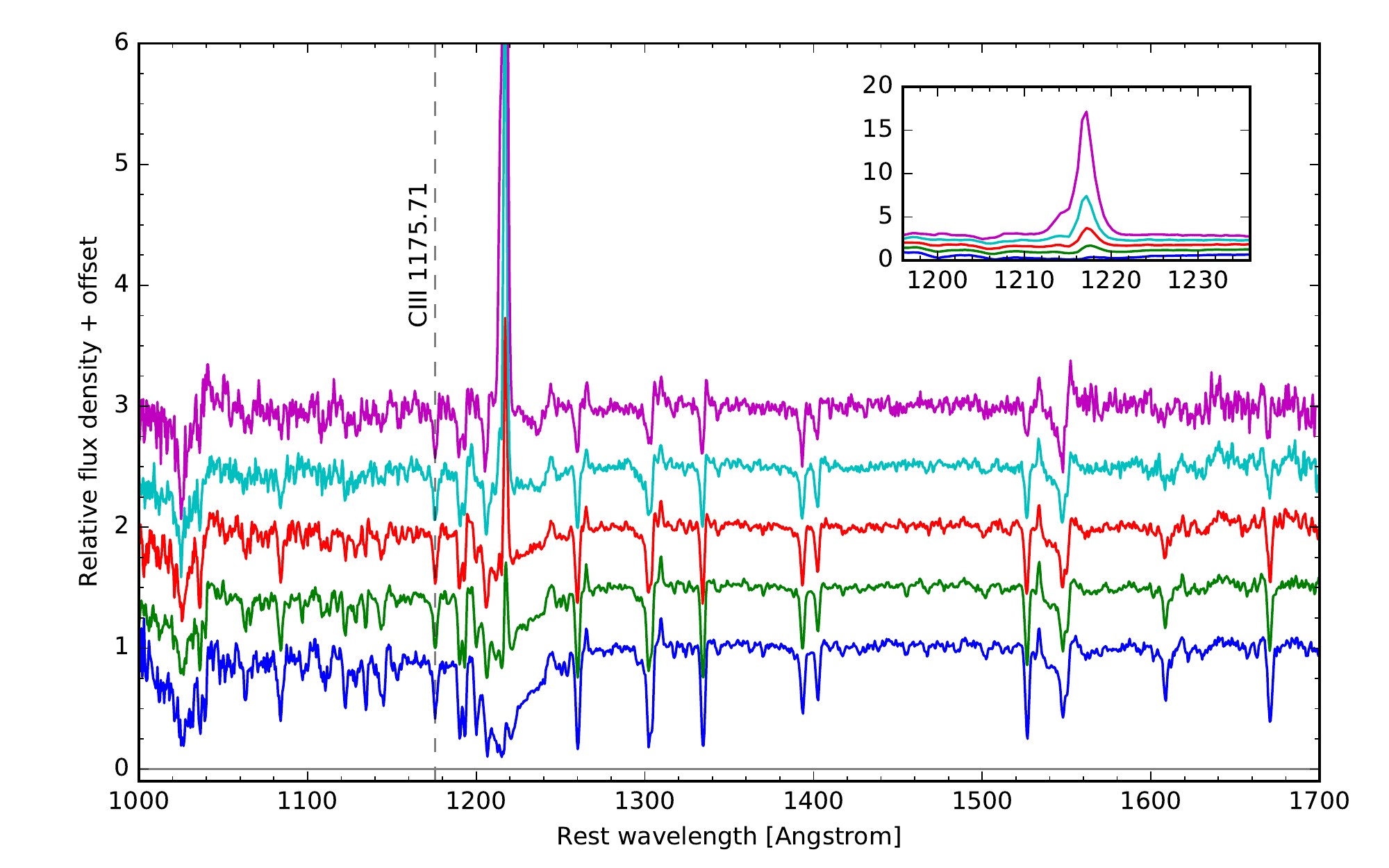}
\caption{LBG template spectra constructed by dividing the sample into 5 bins of Ly$\alpha$ EW. Before averaging, the spectra in each bin are divided by a power law fit to $\lambda_{\rm rest} > 1250$~\AA~and by the mean Ly$\alpha$ forest transmission $\langle F(z) \rangle$. Spectra are offset vertically by 0.5 for clarity. The inset shows the region around Ly$\alpha$ on an enlarged scale.
\label{fig:lbg_templates}}
\end{figure*}

The templates were then offset in velocity to $v_{\rm stars} = 0$ using the \ion{C}{3} 1175.7~\AA~line. (Consistent results are obtained using other photospheric lines, but this line is the strongest and the most robustly detected in all of the templates.) We then modeled each LBG as a non-negative linear combination $T(\lambda; z)$ of these 5 templates, redshifted and multiplied by $\langle F \rangle (z_{\rm Ly\alpha})$ (where $1+z_{\rm Ly\alpha}(\lambda) = \lambda / 1215.67$~\AA) and a power law continuum $C(\lambda)$. We included the Ly$\alpha$ forest in the fit so that it can contribute to the determination of the continuum slope. The product $T \times \langle F \rangle \times C$ was fit to the observed spectrum using a standard non-linear least squares method. The resulting redshifts should be more accurate than those based on the \citet{Shapley03} composite spectrum, since the templates are better matched to the spectral properties of each galaxy. Since the initial templates were constructed by stacking spectra with approximate redshifts, we then constructed an improved set by shifting each galaxy into its rest frame, now using our refined redshifts, and generating the templates again as just described. This procedure was then iterated a second time; by this point, changes in the redshifts and templates were quite minimal, indicating we had reached convergence.

\begin{figure}
\centering
\includegraphics[width=\linewidth]{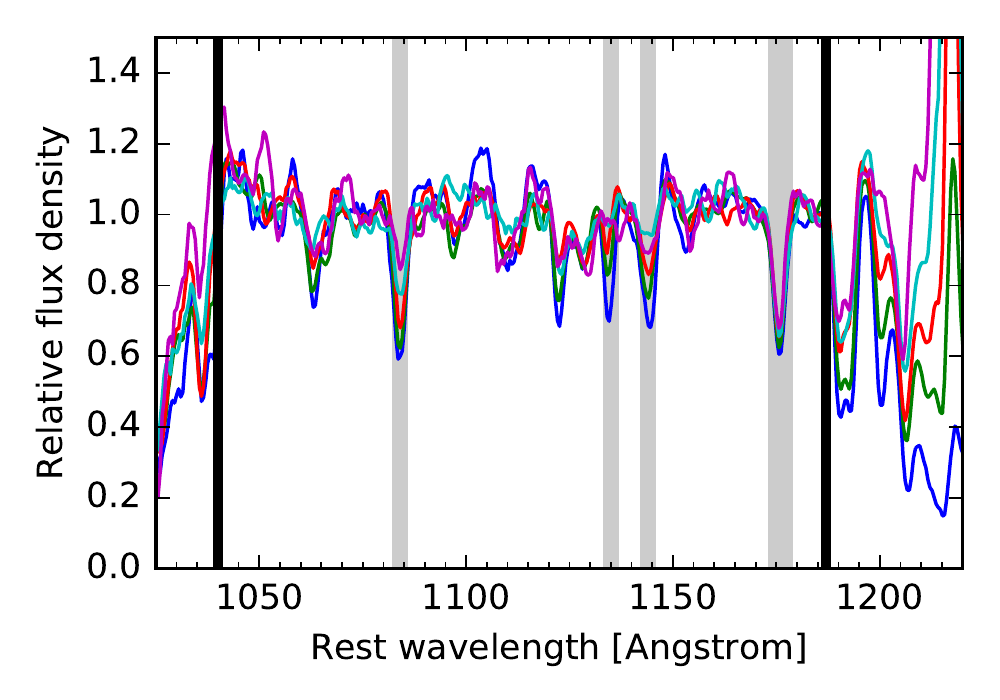}
\caption{LBG template spectra around the Ly$\alpha$ forest region, which is enclosed by the vertical black lines. Gray regions are masked in our Ly$\alpha$ forest analysis due to the strong and/or variable absorption features.
\label{fig:forest_masks}}
\end{figure}

Figure~\ref{fig:lbg_templates} shows the resulting 5 template spectra, which span a wide range in Ly$\alpha$ emission and absorption and in the strength of the interstellar lines. Figure~\ref{fig:forest_masks} compares the templates in the Ly$\alpha$ forest region, which we define to be 1040-1187~\AA~in order to exclude the Ly$\beta$ forest and to provide adequate separation from Ly$\alpha$ and \ion{Si}{2} $\lambda\lambda$ 1190, 1193 absorption. In our Ly$\alpha$ forest analysis, we exclude data in the 4 shaded regions, selected to include the two strongest lines and the two that show the largest variation among the templates (1084~\AA, 1135~\AA, 1144~\AA, and 1176~\AA). We mask a $\pm 2$~\AA~window in the rest-frame around each line ($\pm 3$~\AA~for 1176~\AA), amounting to 6\% of the forest length. Although the exact choice of mask is somewhat arbitrary, we found that adding the next two strongest lines (1063~\AA~and 1123~\AA) ultimately had a negligible effect on the tomographic maps.

This procedure generally does a good job at matching the continuum shape and the main absorption lines in the individual spectra, which Figure~\ref{fig:specexamples} demonstrates. The median reduced $\chi^2$ is 1.13 redward of Ly$\alpha$ (to exclude IGM fluctuations), indicating the models are generally sufficient and that our noise spectra are realistic. However, in order to ensure that the continuum is adequately modeled in the Ly$\alpha$ forest region, we employ the mean flux regularization (MFR) technique introduced by \cite{Lee12}. In this method, the Ly$\alpha$ forest region of a galaxy spectrum is multiplied by a low-order polynomial that best matches the spectrum to the mean flux $\langle F \rangle(z)$ determined from quasar measurements \citep{FG08}. This suppresses power on very large scales $\Delta z \simeq 0.2$ ($\sim 170$~\cMpch) while mitigating continuum errors that could adversely affect the smaller scales of interest. We set the order of the polynomial by the length of the Ly$\alpha$ forest contained in the spectrum. When $\Delta z_{\rm Ly\alpha} < 0.15$, we do not perform MFR. When $0.15 < \Delta z_{\rm Ly\alpha} < 0.3$, we fit and divide by a constant. When $\Delta z_{\rm Ly\alpha} > 0.3$, we fit and divide by a line. MFR typically makes only modest continuum adjustments by a factor of $0.99 \pm 0.12$ (median and rms; see Figure~\ref{fig:specexamples}).

\subsection{LBG Redshift Comparison}

As discussed in Section~\ref{sec:specclass}, we matched our redshifts to the full zCOSMOS and VUDS data sets, finding a small fraction of catastrophic outliers. Among $z > 2$ galaxies with high-confidence redshifts in LATIS and one of these surveys, we find a median offset of $c(z_{\rm LATIS} - z_{\rm zCOSMOS})/(1+z) = 118$~km~s${}^{-1}$ and $c(z_{\rm LATIS} - z_{\rm VUDS})/(1+z) = 187$~km~s${}^{-1}$, or about half of our instrumental resolution. Given the range of velocities that different features in the UV spectrum present, it is understandable that different measurement procedures could lead to systematically different redshifts. When we combine LATIS with the zCOSMOS and VUDS redshifts to plot the locations of galaxies in our tomographic maps (Section~\ref{sec:maps}), for consistency we adjust the zCOSMOS and VUDS redshifts onto the LATIS system using these offsets.

Our LBG spectral modeling is designed to produce redshifts that are, on average, the systemic redshift $z_{\rm sys}$, since the templates are shifted to $v_{\rm stars} = 0$. We assessed this by comparing LATIS redshifts to nebular redshifts from the MOSDEF survey. For 24 galaxies with high-confidence redshifts (excluding AGN), we find a median offset $c (z_{\rm LATIS} - z_{\rm MOSDEF}) / (1+z) = -92$~km~s${}^{-1}$, with a standard deviation of 100~km~s${}^{-1}$. This scatter is equal to that obtained when an optimal combination of Ly$\alpha$ and interstellar line redshifts is used to estimate $z_{\rm sys}$ \citep{Steidel18}, which indicates that the precision of the LATIS redshifts is good. The origin of the $-92$~km~s${}^{-1}$ offset is unclear. We apply a global shift to the LATIS redshifts, as well as the adjusted zCOSMOS and VUDS redshifts, to place them on the MOSDEF system when we compute the positions of galaxies in our tomographic maps. However, this amounts to a small correction of 0.9 $h^{-1}$ cMpc, well below the map resolution.

\subsection{QSO Spectrum Modeling}

Although free of narrow absorption features, the QSO continuum in the Ly$\alpha$ forest is complicated by the broad wings of the Ly$\alpha$ and Ly$\beta$ emission lines and the presence of metal emission lines. We obtain a first estimate of the intrinsic QSO spectrum (absent foreground absorption) using the suite of principal components determined by \citet{Suzuki05}. We use the first 10 eigenspectra, following the recommendation of Suzuki et al., and perform a least-squares fit to the spectrum redward of Ly$\alpha$. Metal lines from foreground absorbers are then identified and masked, and the fit is repeated. In a few cases, the model flux density became negative within the observed wavelength range; we then decrease the number of eigenspectra used in the fit until the model is everywhere positive. This procedure produces a predicted QSO continuum in the Ly$\alpha$ forest, in which the emission lines are predicted via their correlations with the emission lines redward of Ly$\alpha$.

\begin{figure}
\centering
\includegraphics[width=\linewidth]{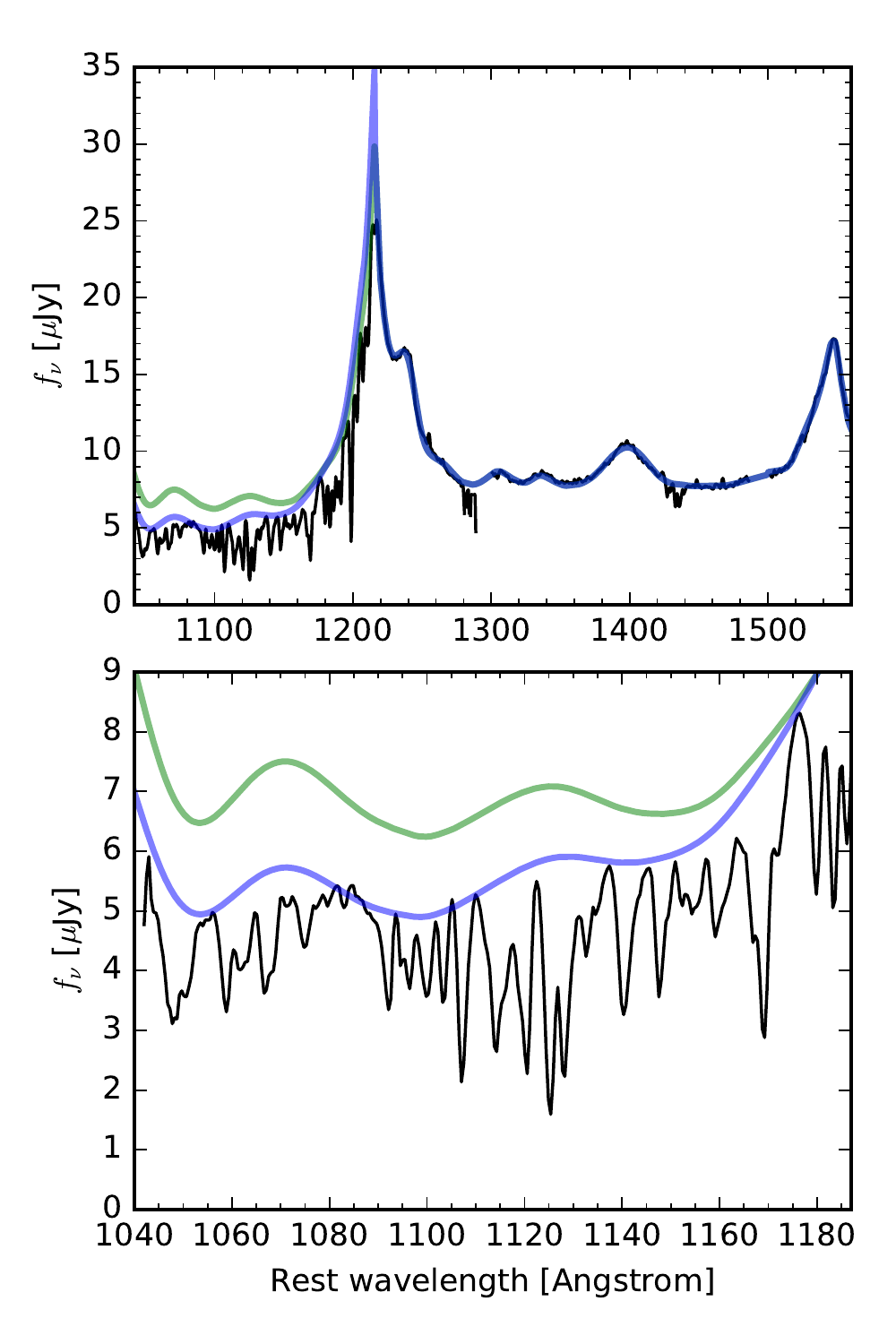}
\caption{Example spectrum of a bright QSO at $z=2.735$ with $r=20.9$ (black curve). The green and blue curves show the fitted model before and after mean flux regularization, respectively. The lower panel isolates the Ly$\alpha$ forest region.
\label{fig:demo_qso}}
\end{figure}

The model for one quasar is shown by the blue curve in Figure~\ref{fig:demo_qso}. As noted by \citet{Suzuki05}, the slope of the Ly$\alpha$ forest continuum is not always accurately predicted from the red part of the spectrum. Therefore, as for the LBGs, we use mean flux regularization to correct the forest continuum shape. Due to the more complex QSO continuum, we use polynomials of order 1 or 2 when the observed length of forest is $\Delta z = 0.15$-0.3 or $\Delta z > 0.3$, respectively. The blue curve in Figure~\ref{fig:demo_qso} shows that this procedure produces an accurate continuum model, both in terms of the continuum slope (due to MFR) and the higher frequency features (due to the principal components analysis). 

Our current data set contains 47 broad-line QSOs at $z=2.2$-3.2. 
Of the 44 sources targeted on the basis of a literature classification, the redshifts are almost always confirmed, but 8 turned out to be unsuitable for the Suzuki et al.~templates to model. Among the 47 broad-line QSOs in LATIS, 16 were excluded from the Ly$\alpha$ forest analysis either because they have broad absorption lines (3 QSOs); the length of the forest contained in our spectrum was too short to permit MFR ($\Delta z_{\rm Ly\alpha} < 0.15$), which in contrast to the LBGs seems to be necessary in most cases (7 QSOs); or visual inspection of the spectrum showed that it was otherwise not accurately modeled using the Suzuki et al.~templates (6 QSOs).

\subsection{Continuum Uncertainties}
\label{sec:continuumerrors}

Errors in the continuum placement directly propagate to the transmitted flux $F = S / C$, where $S$ and $C$ are the spectrum and continuum model, respectively. To estimate the continuum uncertainties, we measured the dispersion in $\langle F \rangle$ averaged over 3 pMpc ($\Delta z \approx 0.01$) segments of the Ly$\alpha$ forest. (We note that this scale is much smaller than the $\Delta z \gtrsim 0.1$ scales that are suppressed by the mean flux regularization.) \citet{FG08} measured the rms dispersion to be 0.11 at $z = 2.4$ based on high-resolution quasar spectra and found no large redshift dependence over the range relevant for LATIS. We first consider the LBGs and split these Ly$\alpha$ forest segments into bins of continuum-to-noise ratio, ${\rm CNR} = C / \sigma_{\rm noise}$, where $\sigma_{\rm noise}$ is the random noise. In each bin, we compute the rms $\sigma_{\rm obs}$ of $\langle F \rangle$ among the segments. We consider this dispersion to be composed of three components: $\sigma_{\rm obs}^2 = \sigma_{\rm noise}^2 + \sigma_{\rm IGM}^2 + \sigma_{\rm cont}^2$, where $\sigma_{\rm IGM} = 0.11$ represents the intrinsic IGM fluctuations and $\sigma_{\rm cont}$ incorporates any additional scatter. We think that continuum errors are likely the dominant contributor to $\sigma_{\rm cont}$, but this term also includes any additional noise beyond that propagated during the data reduction. 

\begin{figure}
    \centering
    \includegraphics[width=\linewidth]{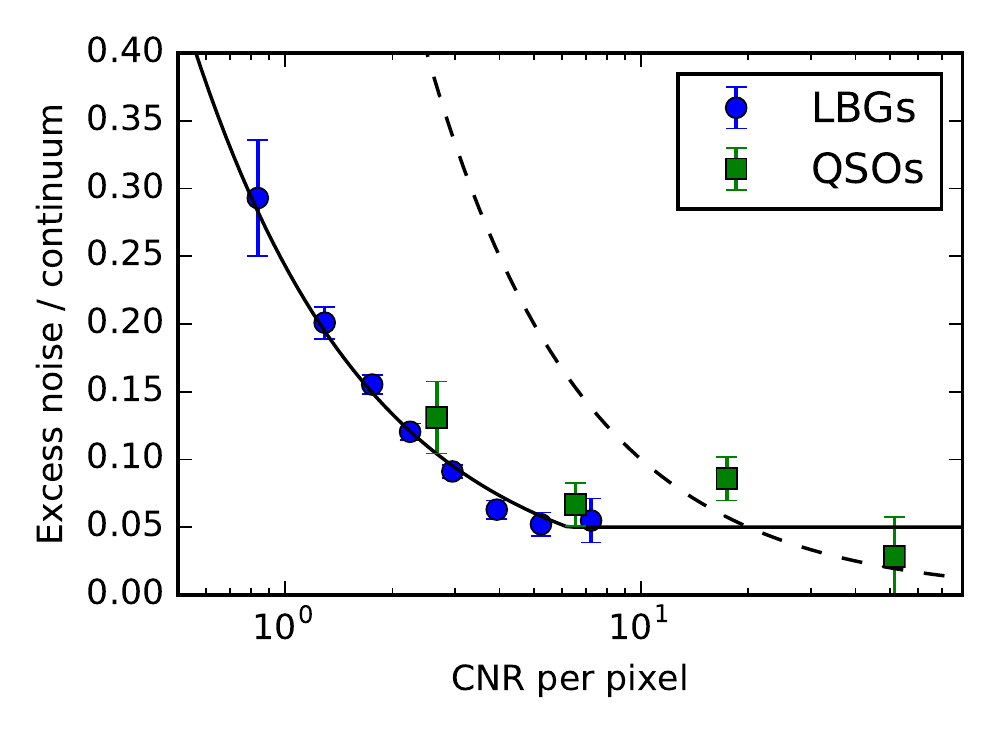}
    \caption{Excess noise in the Ly$\alpha$ forest, presumably due to continuum errors, is estimated as a function of the continuum-to-noise ratio as described in Section~\ref{sec:continuumerrors}. The solid line represents a fit to the LBG data (see text). The dashed line represents the random noise, 1/CNR.\label{fig:excess_noise}}
\end{figure}

The excess noise $\sigma_{\rm cont}$ relative to the continuum is shown in Figure~\ref{fig:excess_noise}. The solid line shows a simple fit $0.24 \times {\rm CNR}^{-0.86}$. We conservatively place a lower limit of 0.05. Repeating this procedure for the QSOs yields consistent but less precise estimates of the continuum errors, so we adopt the same continuum errors for LBGs and QSOs. The dashed line in Figure~\ref{fig:excess_noise}, 1/CNR, demonstrates that the random noise dominates when ${\rm CNR} < 20$, i.e., virtually always. We will incorporate this estimate of the continuum uncertainty when calculating uncertainties in the Ly$\alpha$ forest fluctuations (Section~\ref{sec:mapconst}) and when simulating LATIS (Section~\ref{sec:mocks}). Our estimates of the LBG continuum uncertainty are compatible with the CLAMATO survey \citep{Lee16}, and our QSO continuum uncertainties at high CNR are similar to the 4-7\% estimated by other authors using different techniques \citep[e.g.,][]{Lee12,Eilers17}.

We repeated this analysis for the subset of LBGs where the continuum is not adjusted using MFR because of the short length of the Ly$\alpha$ forest that is observed ($\Delta z_{\rm Ly\alpha} < 0.15$). The excess noise in these spectra is very similar to that in the full sample, giving us confidence that these spectra can be used for tomography. In contrast, QSO models produced without MFR were generally not usable. 

\subsection{Damped Absorbers}
\label{sec:damped}

Damping wings from high-column-density systems produce absorption over a wide wavelength range. When this range significantly exceeds our spectral resolution, it violates the mapping between wavelength and velocity that underlies tomographic reconstruction. We therefore use an automated procedure to mask these lines. An absorber with a column density of $N_{\rm HI} \gtrsim 10^{19.7}$~cm${}^{-2}$ and an equivalent width $W_0 = 5$~\AA~absorbs approximately half of the flux in the adjacent resolution elements of our spectra. We mask absorption lines that are detected at $>5\sigma$ and have an equivalent width $>5$~\AA~in the absorber frame. We find a total of 102 such absorbers over a total path length of $\Delta z = 384$. Roughly interpolating between prior measurements of the number density of sub-DLA \citep{Zafar13} and DLA \citep{Peroux03} systems at $z \approx 2.5$ indicates $dn/dz \approx 0.3$, so the expected $\sim115$ absorbers is in good agreement with the number we find, particularly since some damped systems may be present at a detection significance below our threshold.

\section{Tomographic Reconstruction} \label{sec:mapmethod}

With the spectra reduced, modeled, and characterized, we can now measure the Ly$\alpha$ forest fluctuations in each sightline and generate three-dimensional tomographic maps. We will construct maps over the 4 footprints where observations are complete for both of the main target sets: 3 footprints in COSMOS (D2M4, D2M5, D2M8) covering 0.43~deg${}^2$ and 1 footprint in the D1 field (D1M3) covering 0.15~deg${}^2$. The total  volume enclosed from $z=2.2$-2.8 is $1.4 \times 10^6$~$h^{-3}$~cMpc${}^{-3}$, one-third of the ultimate LATIS survey.

\subsection{Sightline Density and Continuum-to-Noise Ratio}
\label{sec:sldens}

\begin{figure*}
\centering
\includegraphics[width=0.48\linewidth]{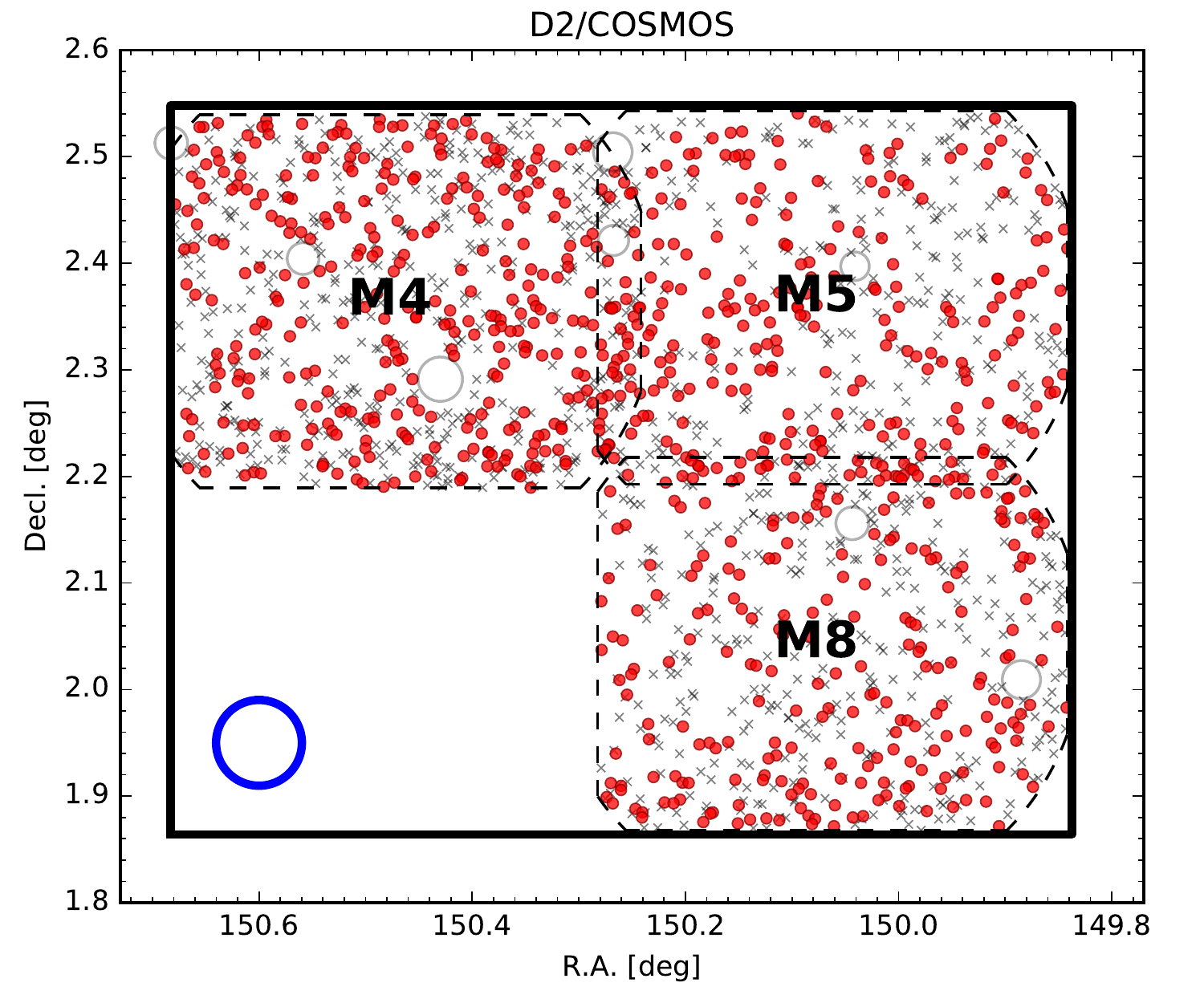} \hfill
\includegraphics[width=0.48\linewidth]{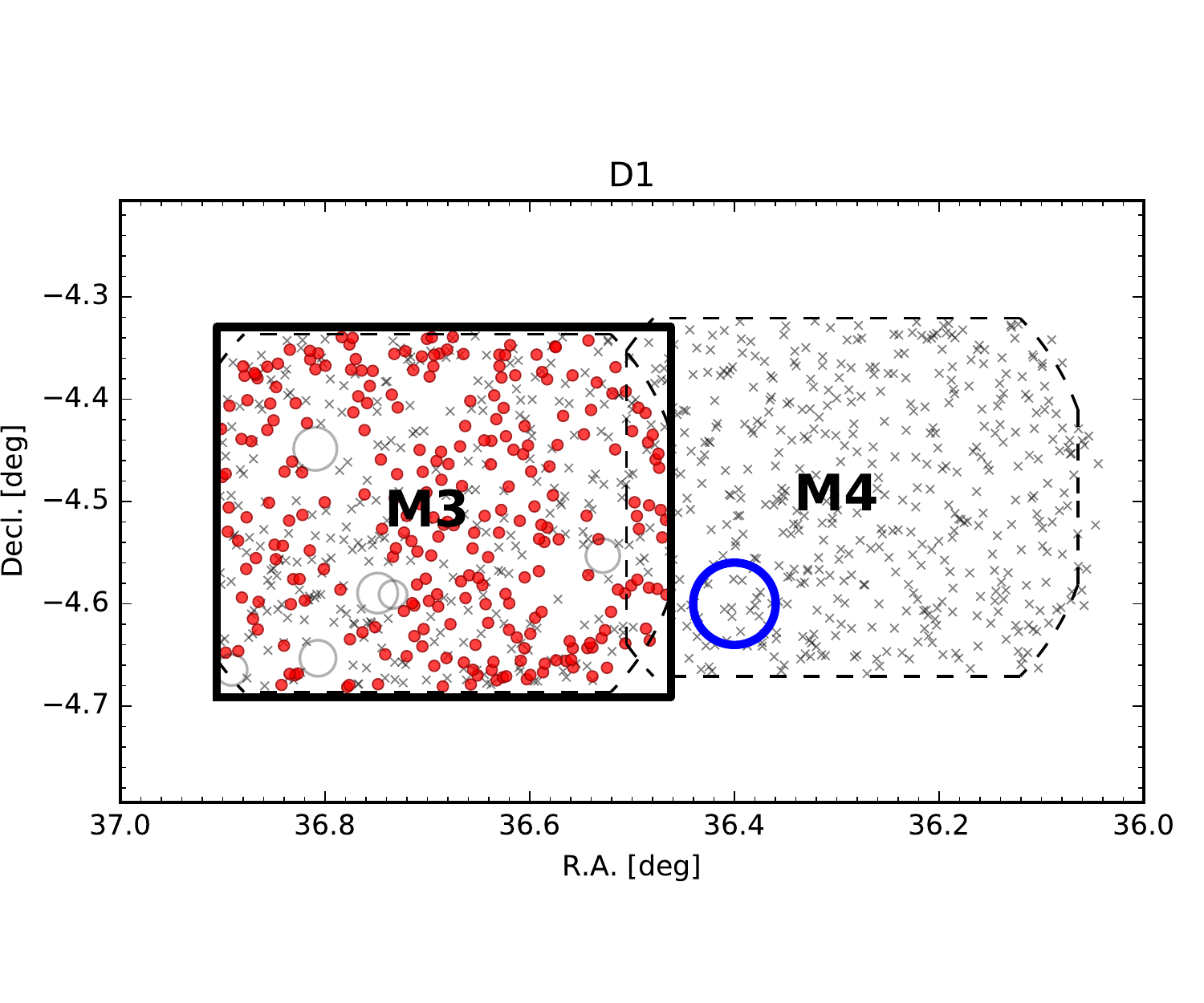}
\caption{The positions of targeted galaxies in the COSMOS (left panel) and D1 (right) fields. Red circles represent targets used for Ly$\alpha$ forest measurements, while gray crosses indicate targets that were not used to construct our maps (usually foreground galaxies). Dashed lines outline the individual IMACS footprints. The D1M4 field is not used for tomography in this paper since observations of both target sets are not yet complete. Thick black lines indicate the border of the tomographic maps. Blue circles have a radius of 3~$ h^{-1}$~cMpc and so approximate the map resolution element. Light gray circles enclose the largest regions (radius $> 45''$) from which targets are excluded due to a bright star.  
\label{fig:sightlinedist}}
\end{figure*}

Our tomographic reconstruction incorporates all sightlines contained within the 4 footprints listed above whose Ly$\alpha$ forest overlaps the range $z_{\rm Ly\alpha} = 2.2$-2.8, that have high-confidence redshifts, and that were not manually excluded due to reduction defects. These total 1071 sightlines (98\% LBGs, 2\% QSOs) with an areal density of 1850 deg${}^{-2}$. Figure~\ref{fig:sightlinedist} shows the positions of these sightlines on the sky (red circles). Although the targeted galaxies (red circles and gray crosses) are reasonably uniformly distributed, there are some areas with few or no sightlines usable for tomography. This is expected given the clustered nature of luminous galaxies. The consequences of a variable sightline density for map quality will be assessed using simulated mock surveys (Section~\ref{sec:mocks}). 

\begin{figure*}
\centering
\includegraphics[width=0.47\linewidth]{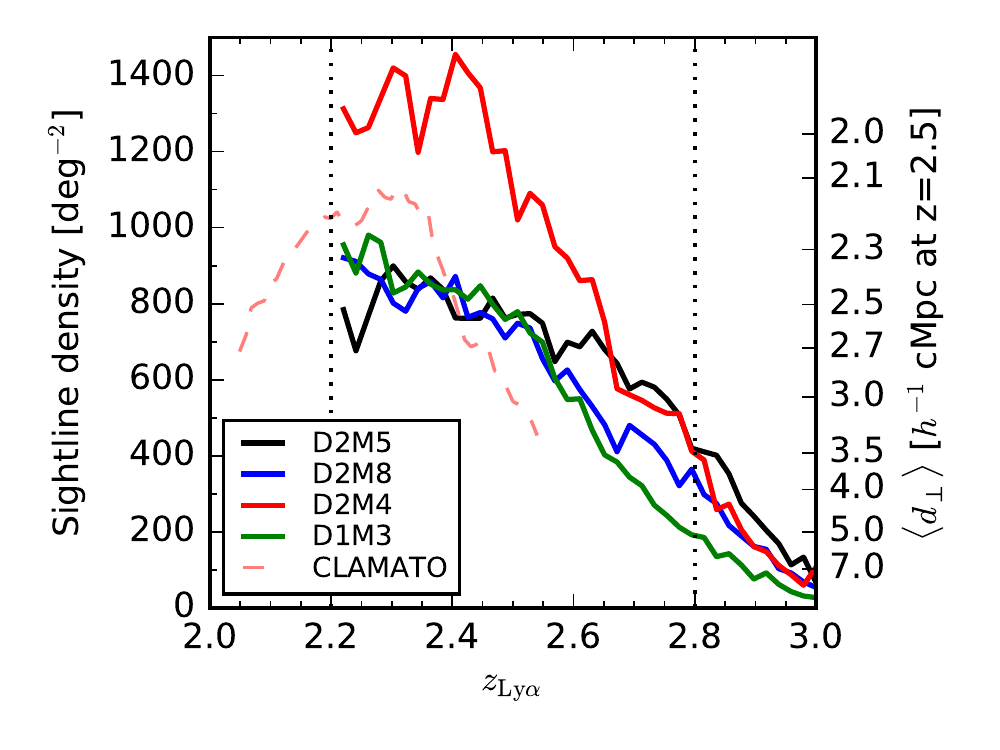}
\hfill
\includegraphics[width=0.47\linewidth]{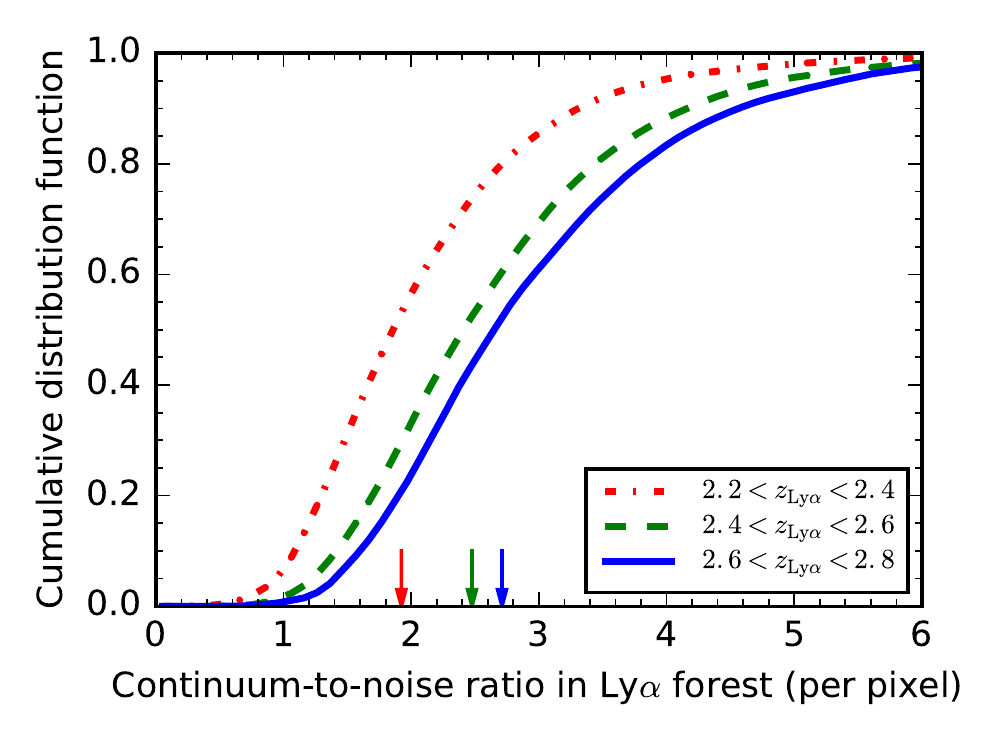}
\caption{\emph{Left:} The density of sightlines piercing a given $z_{\rm Ly\alpha}$, averaged over each of 4 footprints in the D2/COSMOS and D1 fields. The right axis shows the mean transverse separation $\dperp$ as described in the text. Dotted lines show the boundaries of our tomographic map. The CLAMATO DR1 \citep{Lee18} results are shown for reference as the dashed line. \emph{Right:} The cumulative distribution of CNR in the Ly$\alpha$ forest in 3 bins of $z_{\rm Ly\alpha}$. Arrows indicate the median in each bin.
\label{fig:sld}}
\end{figure*}

The sightline density and the continuum-to-noise ratio are key metrics determining the quality of a tomographic reconstruction. The left panel of Figure~\ref{fig:sld} shows the areal density $n$ of sightlines piercing a given $z_{\rm Ly\alpha}$, averaged over each footprint, and the corresponding mean transverse sightline separation at $z=2.5$, $\dperp = 70.6~n^{-1/2}~h^{-1}$~cMpc. In most fields, the sightline density is relatively constant at $\dperp \simeq 2.5$~\cMpch~over $z=2.2$-2.6, meeting the design goal of the survey. At $z > 2.6$ the sightline density declines. We limit the tomographic reconstruction to $z < 2.8$, where the sightline separation falls to $\dperp \approx 4$~\cMpch, which we take as the maximum useful value based on the simulations by \citet{Stark15a}. The distribution of CNR is shown in the right panel of Figure~\ref{fig:sld}. The median CNR varies with redshift due to the wavelength-dependent sensitivity of IMACS (Figure~\ref{fig:filter_grism}), ranging from 1.7-2.7 per pixel. These values are roughly consistent with the target ${\rm CNR} = 2$ that set our exposure times.

Interesting, in the D2M4 footprint we achieved a far higher sightline density than typical. Since targeting procedures and the depth of the spectra were not different, we conclude an overdensity of galaxies allows us to reach $n = 1300$~deg${}^{-2}$, the highest density yet employed for Ly$\alpha$ tomography.

\subsection{Tomographic Map Construction}
\label{sec:mapconst}

We are now ready to transform the LATIS spectra into three-dimensional maps of the IGM opacity. We reconstruct the transmitted flux $F$, rather than attempting to recover the underlying density field \citep{Pichon01,Gallerani11,Horowitz19}. We use Wiener filtering, a method that has widely been used in the mapping of large-scale structure, to invert the sightline data. The Wiener filter incorporates noise weighting and regularizes the output map, as described below. Its utility for IGM tomography was investigated theoretically by \citet{Pichon01} and \citet{Caucci08}, and more recently by \citet{Stark15b,Stark15a} in the context of the CLAMATO survey, which also employs a Wiener filter \citep{Lee14,Lee18}. For LATIS we specifically use the efficient {\tt dachshund} code developed by \citet{Stark15a}. 

The input data consist of measurements of flux contrasts $\delta_F$ and associated uncertainties $\sigma_{\delta}$ at a series of positions $(x,y,z)$ within the volume to be reconstructed. Each such measurement is one pixel in the Ly$\alpha$ forest of a background source. The flux contrasts are defined as fractional variations around the mean, the fundamental metric in which the spectra and the maps are expressed:
\begin{equation}
    \delta_F = \frac{F}{\langle F(z)\rangle} - 1,
\end{equation}
where $F = S/C$ is the continuum-normalized spectrum and $\langle F(z) \rangle$ is the mean flux transmission derived from quasar observations \citep{FG08}. The uncertainty $\sigma_{\delta}$ includes both random noise and the continuum uncertainty (Section~\ref{sec:continuumerrors}) added in quadrature. We have carefully assessed the accuracy of these noise estimates using multiple techniques, as described in the Appendix. To avoid placing excess weight on a few quasar sightlines with very high signal-to-noise ratios, we impose a floor of $\sigma_{\delta} > 0.2$. The coordinates $(x,y,z)$ are expressed in $h^{-1}$ cMpc and are aligned with the R.A., Decl., and redshift axes, respectively. We convert sky coordinates and redshifts to $(x,y,z)$ coordinates using redshift-dependent radial and transverse comoving distances. Although we express the line-of-sight coordinate as a distance, our method does not attempt to correct for peculiar velocities, so the maps are made in velocity space. This is all that is needed to compare to the galaxy distribution, our main concern in this paper. 

The 173,185 data points are used to reconstruct the IGM opacity in two volumes with dimensions $64 \times 51 \times 483$~$h^{-3}$~cMpc${}^3$ in D2/COSMOS and $33 \times 27 \times 483$~$h^{-3}$~cMpc${}^3$ in the D1 field. One quadrant of the COSMOS volume has no sightlines yet (see Figure~\ref{fig:sightlinedist}); we exclude this region with $x < 30$~\cMpch~and $y < 24$~\cMpch~from our analysis. This leaves a total volume of $1.7 \times 10^6$~$h^{-3}$~cMpc${}^3$ in the maps.\footnote{The map volume is sized to enclose all of the sightlines at redshifts $z=2.2$-2.8. It is slightly larger than the volume within the projected mask footprints, $1.4 \times 10^6$~$h^{-3}$~cMpc${}^3$, because of their non-rectangular shape (Figure~\ref{fig:sightlinedist}) and the flared geometry of the sightlines.} Each voxel in the maps occupies (1~$h^{-1}$~cMpc)${}^3$.

Wiener filtering interpolates between the sightlines to estimate $\delta_F$ in each voxel. When the underlying field and the noise are Gaussian, Wiener filtering is the optimal linear operator and can be shown to correspond to the maximum \emph{a posteriori} estimate in certain Bayesian approaches \citep{Pichon01}. Interpolation requires a statistical description of the underlying field. Specifically, Wiener filtering requires the covariance matrix between input data and map voxels, ${\rm C}_{\rm MD}$, as well as the covariance among the input data points, ${\rm C}_{\rm DD} + {\rm N}$. We assume independent Gaussian measurement errors, so that ${\rm N}$ is a diagonal matrix, and we follow the usual \emph{ad hoc} assumption that ${\rm C}_{\rm DD} = {\rm C}_{\rm MD} = {\rm C}(\vec{r}_1, \vec{r}_2)$ is Gaussian:
\begin{equation}
    C(\vec{r}_1, \vec{r}_2) = \sigma_F^2 \exp\left[ -\frac{\Delta r_{\parallel}^2}{2 \sigma_{\parallel}^2} - \frac{\Delta r_{\perp}^2}{2\sigma_{\perp}^2} \right],
\end{equation}
where $\Delta r_{\parallel}$ and $\Delta r_{\perp}$ are the components of $\vec{r}_1 - \vec{r}_2$ along and perpendicular to the line of sight, respectively (see, e.g., \citealt{Lee18}, Equation 3; \citealt{Caucci08}, Equation 10). As discussed by \citet{Caucci08}, choosing $\sigma_{\parallel} \sim \sigma_{\perp} \sim \dperp$ regularizes the output map by suppressing structure on scales smaller than the mean sightline separation. The amplitude $\sigma_F^2$ represents the \emph{a priori} expected variance in a volume of order $\sigma_{\perp}^2 \sigma_{\parallel}$. Where observational errors are much larger than this, Wiener filtering suppresses the signal in favor of the prior $\delta_F = 0$.

Since one of our goals is to compare the LATIS and CLAMATO maps where they overlap, and the surveys' sightline separations are comparable, we choose $\sigma_{\perp} = 2.5$~\cMpch~and $\sigma_F^2 = 0.05$ following \citet{Lee18}, who in turn relied on simulations by \citet{Stark15b} that showed these parameters to be nearly optimal. To account for smoothing of the spectra along the line of sight, we take $\sigma_{\parallel}^2 = \sigma_{\perp}^2 - \sigma_{\rm inst}^2$, where $\sigma_{\rm inst} = 1.4$~\cMpch~is the instrumental resolution expressed in line-of-sight distance at $z = 2.5$.

For most applications, we then smooth the Wiener-filtered maps using an isotropic Gaussian kernel. Smoothing reduces noise at the expense of resolution, and it must be tailored to the requirements of each application. For display purposes, we use $\sigma_{\rm kern} = 2$~\cMpch, while for some of the quantitative applications described in the rest of the paper we will we use a broader kernel with $\sigma_{\rm kern} = 4$~\cMpch. Finally the maps are multiplied by a calibration factor described in the next section.

\begin{figure}
\centering
\includegraphics[width=0.75\linewidth]{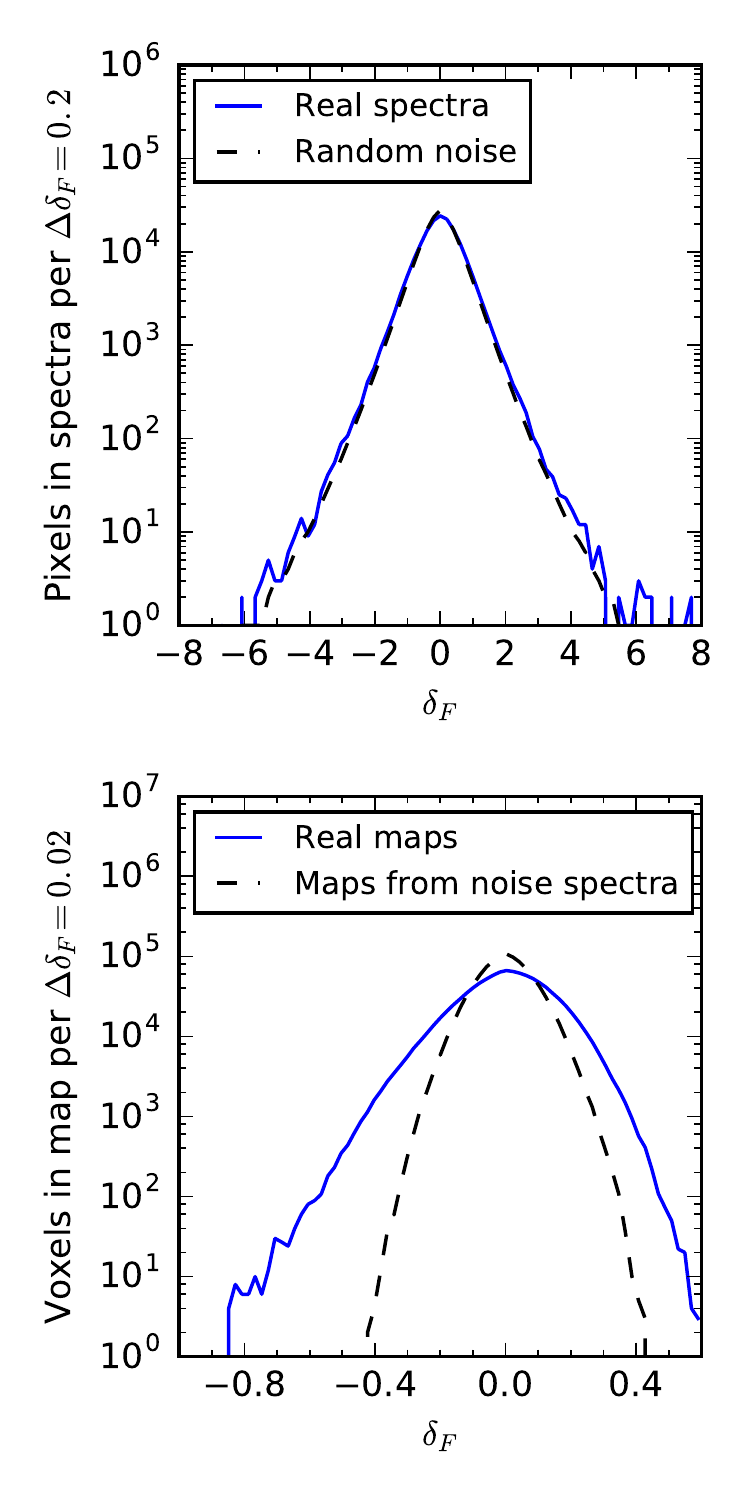}
\caption{\emph{Top:} The distribution of flux contrasts $\delta_F$ in the Ly$\alpha$ forest spectra is compared to that of random Gaussian noise, including continuum uncertainties. \emph{Bottom:} The distribution of $\delta_F$ in the Wiener-filtered maps, without any additional smoothing applied, is compared to that in maps constructed from spectra of independent Gaussian random noise. Although individual spectral pixels are noise-dominated (top panel), the maps recover significant structures (bottom) from spatially and spectrally coherent absorption.\label{fig:fluctuations}} 
\end{figure}

Figure~\ref{fig:fluctuations} shows that although the individual Ly$\alpha$ forest spectra are noisy and are, as an ensemble, consistent with Gaussian random noise at the level of individual pixels (top panel), the maps do contain significant structure. This can be demonstrated by comparing the fluctuations $\delta_F$ in the actual maps (bottom panel, solid line) with those in maps that are constructed from pure noise realizations (dashed line), i.e., from spectra composed of independent Gaussian random deviates with an rms of $\sigma_{\delta}$. The range of $\delta_F$ in the actual maps is considerably broader, indicating that LATIS recovers spatially and spectrally coherent fluctuations from spectra that individually are noisy.

\subsection{Mock Surveys}
\label{sec:mocks}

Before we display the LATIS maps, we first would like to estimate their uncertainties and assess the fidelity of a LATIS-like survey for mapping the underlying flux field. We do this by performing 90 mock LATIS surveys in a large $N$-body simulation.

Briefly, we use the particle data from the MultiDark Planck 2 (MDPL2) simulation \citep{Klypin16} recorded at $z=2.535$, near the midpoint of the LATIS redshift range. The density field in a grid with 0.25 \cMpch~cubic voxels is estimated using cloud-in-cell interpolation. We then use the fluctuating \citet{Gunn65} approximation (FGPA; e.g., \citealt{Weinberg98}) to estimate the Ly$\alpha$ flux field. This method assumes that the gas density follows the dark matter density and that there is a one-to-one mapping between density and temperature; we use the relation measured by \citet{Rudie12}. It therefore ignores astrophysical sources of scatter and breaks down on small scales where the gas is pressure supported. However, when the FGPA is applied to $N$-body simulations with a similar inter-particle spacing to MDPL2, it does produce estimates of the flux field that are fairly accurate on the large scales relevant to LATIS \citep{Sorini16}. Confirming this, the simulated one-dimensional flux power spectrum matches BOSS measurements \citep{BOSSPS} well on velocity scales larger than $k^{-1} \sim 50$~km~s${}^{-1}$, which is smaller than our instrumental resolution by a factor of 3 and so more than adequate for our purposes.

In each of 90 non-overlapping sub-volumes, we impose a Hubble flow to convert coordinates along one dimension into velocities. We construct mock spectra with the same relative $(x,y,z)$ coordinates as the LATIS data, i.e., matching the exact sightline distribution. The spectra are smoothed and sampled like the observations, and Gaussian random noise is added to match each sightline's noise properties. We also simulate continuum errors. For each sightline, we take the median CNR in the forest, determine the corresponding continuum uncertainty from Figure~\ref{fig:excess_noise}, draw a Gaussian random deviate with this dispersion, and modify the mock observed spectrum accordingly (see \citealt{Krolewski18}, Equation 5). We then feed these mock data to {\tt dachshund} to reconstruct the flux field using the same parameters applied to the real data.

\begin{figure*}
\centering
\includegraphics[width=0.31\linewidth]{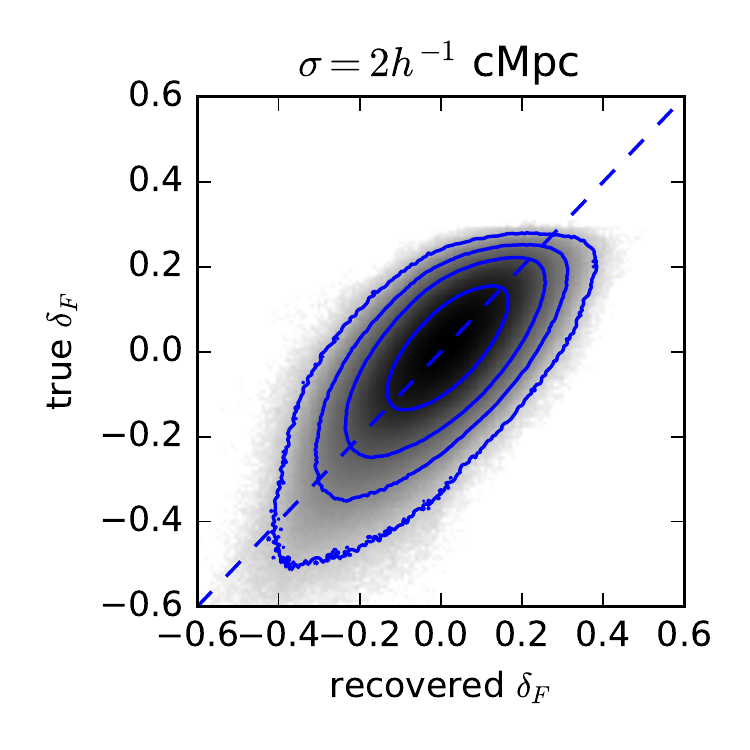}
\includegraphics[width=0.31\linewidth]{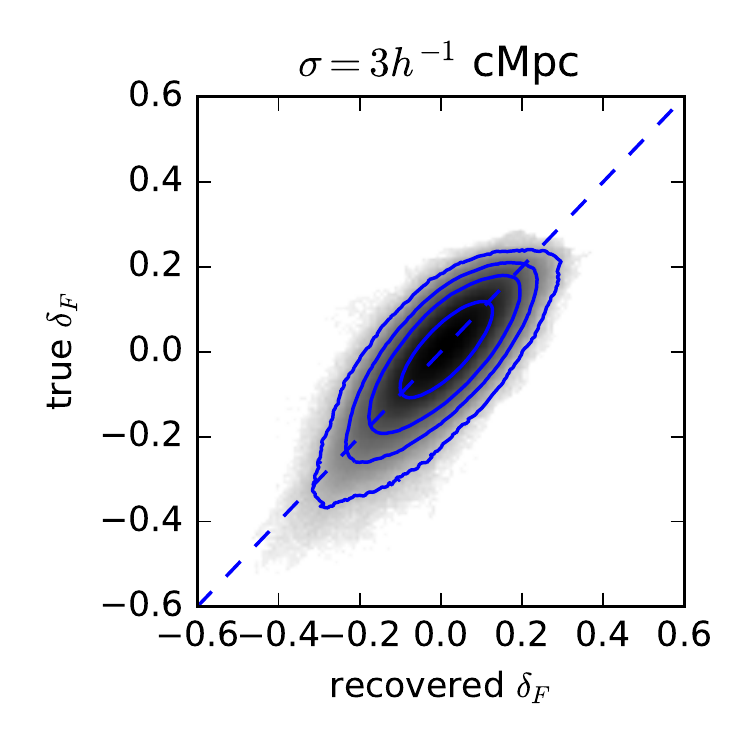} \hspace{0.5cm}
\includegraphics[width=0.31\linewidth]{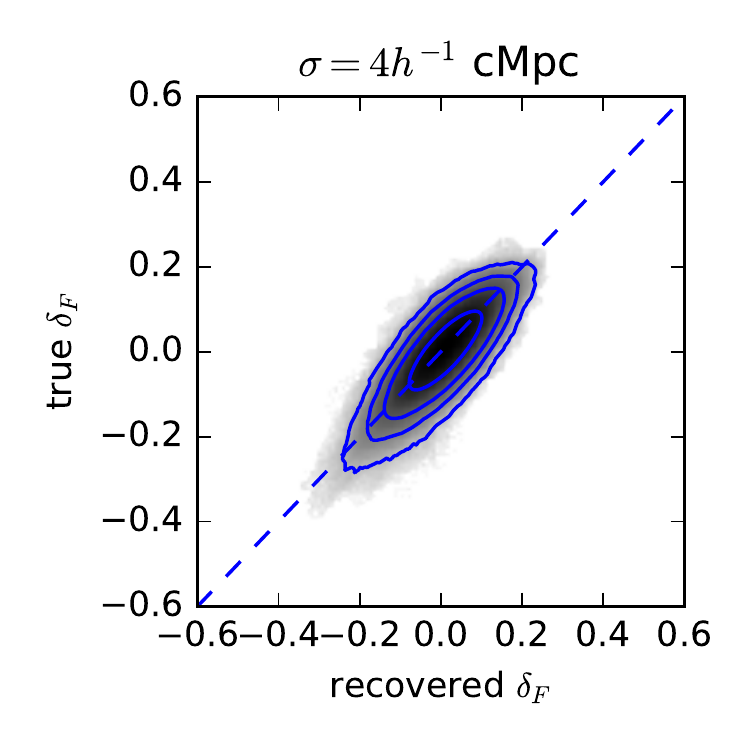} \hspace{0.5cm}\vspace{-6ex}
\caption{Comparison between the true and recovered flux fields in mock surveys of the LATIS COSMOS map. Both fields are smoothed by a Gaussian kernel with $\sigma = 2$, 3, or 4 \cMpch~(left to right). Shading shows the logarithm of the density of voxels. Contours enclose 68, 95, 99, and 99.9 percent of the voxels. Voxels close to the map boundary (within 4 \cMpch) are excluded. The blue line shows the 1:1 relation.
\label{fig:calibration}}
\end{figure*}

The relationships between the true $\delta_F^{\rm true}$ and recovered $\delta_F^{\rm rec}$ flux fields, smoothed on several scales, are shown in Figure~\ref{fig:calibration}. The mock surveys are clearly able to recover fluctuations with a meaningful precision relative to the range present in the simulated volumes. For larger smoothing kernels, this relationship tightens, as expected. We fit lines to the relations in Figure~\ref{fig:calibration} and determine slopes of 0.69, 0.77, and 0.85 for kernels with $\sigma = 2$, 3, and 4 \cMpch, respectively.\footnote{These slopes are derived from the COSMOS mock surveys. In the D1 mocks, the slopes are slight different: 0.73, 0.86, and 0.98.} The slopes are shallower than unity primarily because of reconstruction errors that scatter $\delta_F^{\rm rec}$ away from the peak of the distribution at $\delta_F^{\rm true} \approx 0$. A fitting method that attempts to measure the relation between $\delta_F^{\rm rec}$ and $\delta_F^{\rm true}$ in the absence of noise would likely yield a steeper slope. However, our main purpose is to minimize the squared error ${\rm Var}(\delta_F^{\rm rec} - \delta_F^{\rm true})$ for a given value of $\delta_F^{\rm rec}$ in the maps, which we will use when calculating the map signal-to-noise ratio below. To a first approximation, this is achieved by multiplying the maps by a calibration factor equal to the ordinary least squares slope. Ultimately this is relevant only for the signal-to-noise ratio, since for other applications we will normalize each map by its standard deviation, which we denote $\sigma_{\rm map}$,\footnote{Voxels within 4~\cMpch~of the map boundary are excluded when calculating $\sigma_{\rm map}$.} and any global calibration factor thus cancels out.

\begin{figure}
\centering
\includegraphics[width=\linewidth]{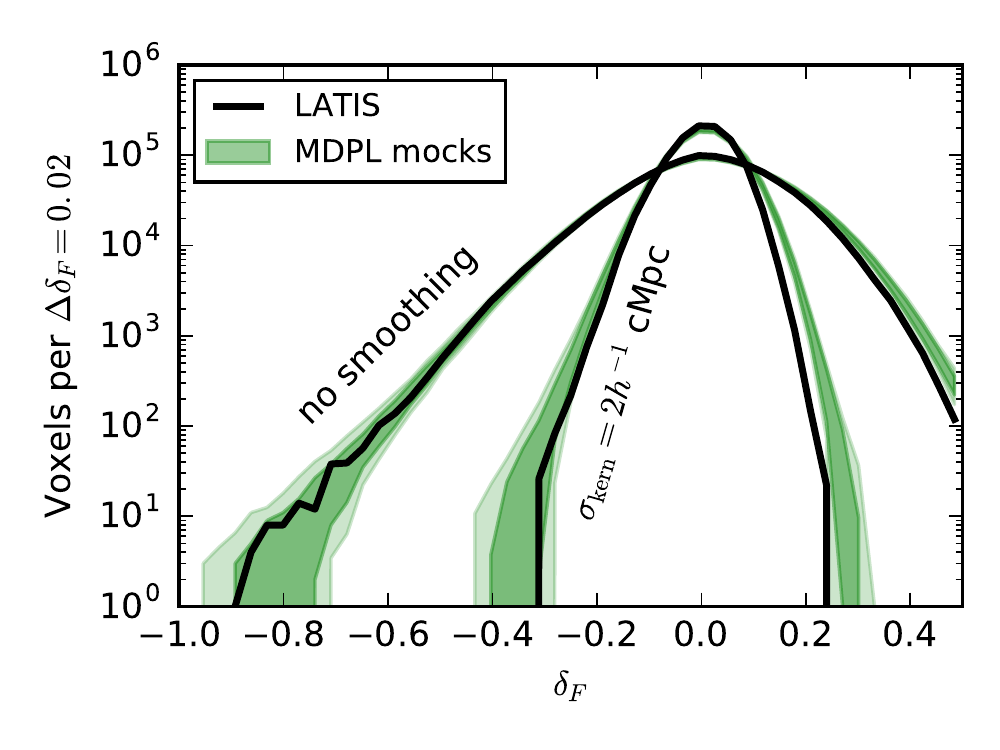}
\caption{Histograms of flux contrasts $\delta_F$ in the LATIS maps (black curves) and in mock surveys of the MultiDark MDPL2 simulation (green bands enclosing 68\% and 90\% of simulations). The broader curves show the Wiener-filtered maps without any further smoothing; the narrower curves show maps smoothed with a $\sigma_{\rm kern} = 2$~\cMpch~kernel. The broad level of agreement indicates that the LATIS maps have structure consistent with $\Lambda$CDM expectations.
\label{fig:deltaFdist}}
\end{figure}

\begin{figure*}
    \centering
        \includegraphics[width=0.3\linewidth]{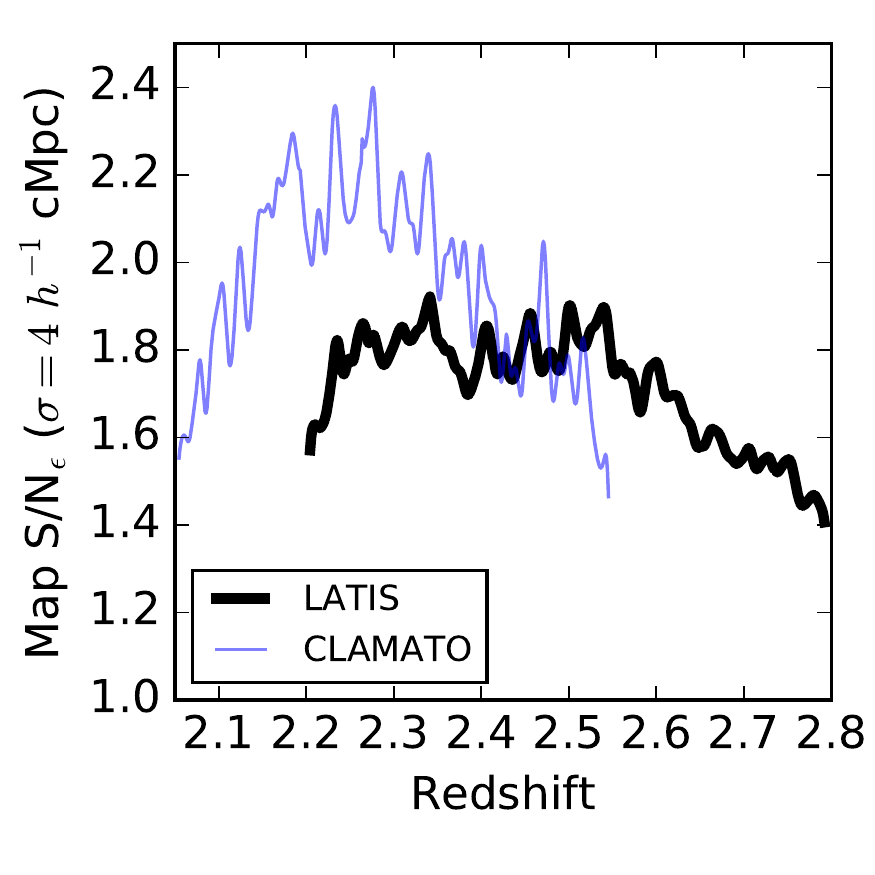} \hspace{0.5cm}
        \includegraphics[width=0.64\linewidth]{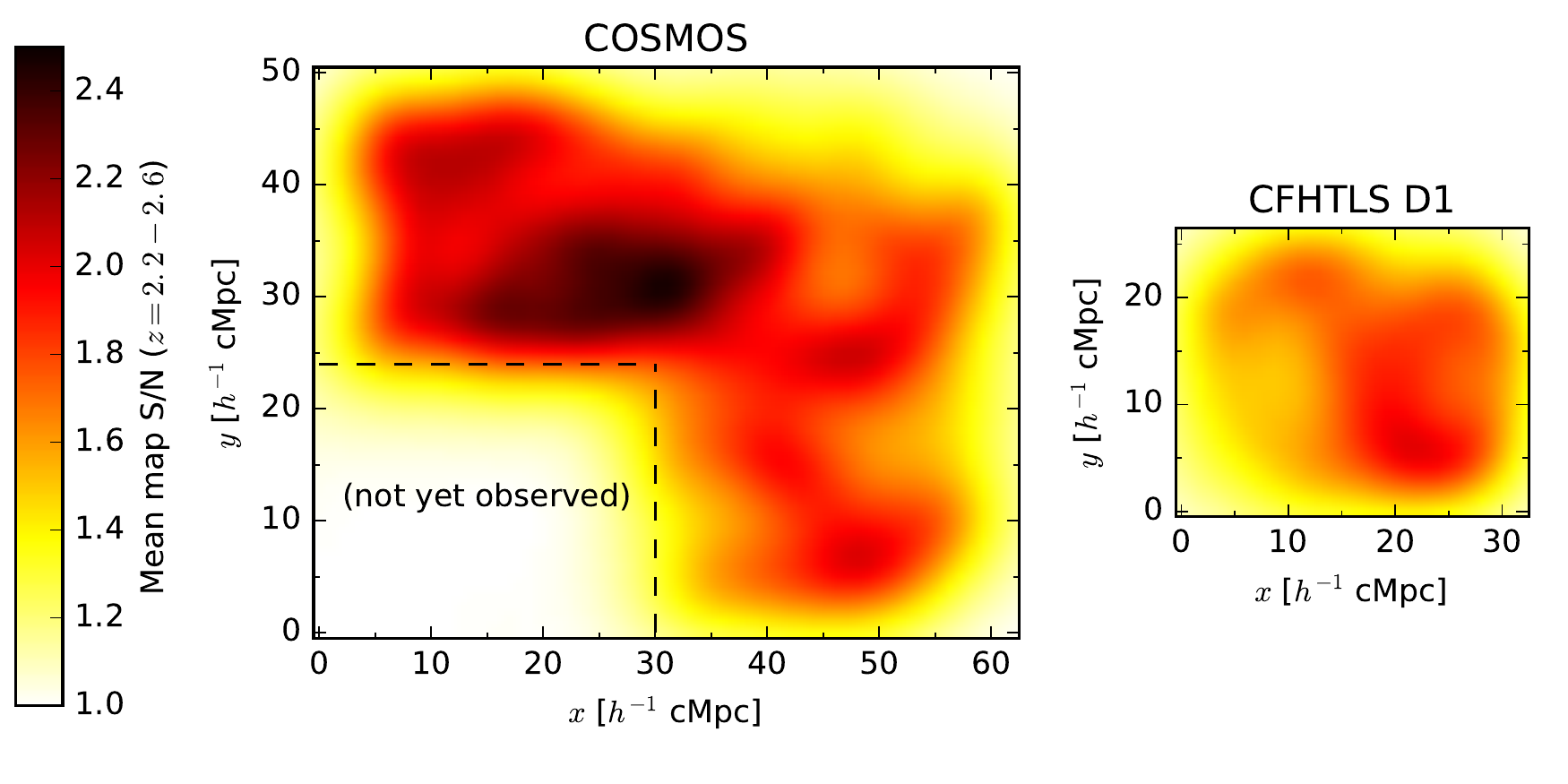}
    \caption{{\emph Left:} Mean signal-to-noise ratio ${\rm S/N}_{\epsilon}$ (Equation~\ref{eqn:SN}) of the LATIS maps, after smoothing with Gaussian $\sigma = 4$~\cMpch~kernel, as a function of redshift. Regions within 4~\cMpch~of the volume edge are excluded. For comparison we show the same calculation for CLAMATO data release 1 (see text; \citealt{Lee18}). \emph{Middle and right:} Mean ${\rm S/N}_{\epsilon}$ over the redshift range $z=2.2$-2.6 as a function of sky position in the COSMOS and D1 fields. There is little trend with redshift over this range (left panel), but a significant dispersion depending on the density and S/N of nearby sightlines.\label{fig:sne}}
\end{figure*}

The distribution of fluctuations in the LATIS maps is compared to the mock surveys in Figure~\ref{fig:deltaFdist}. (Throughout this section, we exclude voxels within 4~\cMpch~of the map edge where boundary effects are strong.) The broad curves show the raw Wiener filter output, while the narrow curves show maps after smoothing by $\sigma_{\rm kern} = 2$~\cMpch. Although there is a slight deficit of voxels with high $\delta_F$ (matter underdensities) in the real maps relative to the simulations, overall the agreement is strikingly good. We do not expect a perfect agreement for several reasons, including our approximate treatment of the IGM and the fact that the simulated maps are fixed at $z=2.5$. Nonetheless, this comparison confirms that at the level of the one-point statistic, the LATIS maps are compatible with expectations for $\Lambda$CDM in the Planck cosmology.

The mock surveys also allow us to estimate the noise in the maps. We follow \citet{Lee14,Lee18} and use the metric
\begin{equation}
    {\rm S/N}_{\epsilon}^2 = \frac{{\rm Var}(\delta_F^{\rm true})}{{\rm Var}(\delta_F^{\rm true} - \delta_F^{\rm rec})},\label{eqn:SN}
\end{equation}
where $\delta_F^{\rm true}$ and $\delta_F^{\rm rec}$ are the true and recovered flux fields, both smoothed by $\sigma = 4$~\cMpch. We evaluate the denominator at each voxel in the map, measuring the variance over the 90 mock surveys, which allows us to measure the variation in ${\rm S/N}_{\epsilon}$ throughout the volume. The left panel of Figure~\ref{fig:sne} shows the mean ${\rm S/N}_{\epsilon}$ as a function of redshift, which is fairly constant at ${\rm S/N}_{\epsilon} \simeq 1.8$ over the range $z=2.2$-2.6 and then declines toward the back of the volume due to the falling sightline density. In other words, the noise in the reconstruction is about half of the intrinsic IGM fluctuations on 4~\cMpch~scales.

\citet{Lee14} considered a good map construction to have ${\rm S/N}_{\epsilon} \approx 2$-2.5. LATIS falls slightly short of this range, but only by 10\%. We also show ${\rm S/N}_{\epsilon}$ for the CLAMATO first data release \citep{Lee18}, which we have calculated using the same methods as applied to LATIS. Note that we have adjusted the noise properties of the CLAMATO spectra based on a close analysis discussed in the Appendix, which was also applied to the LATIS data, and this lowers ${\rm S/N}_{\epsilon}$ by $\simeq20$\% from the value quoted by \citet{Lee18}. Compared to CLAMATO, the current LATIS map has a $\simeq10$\% lower ${\rm S/N}_{\epsilon}$ on average, but covers a $4.4\times$ larger volume.

\section{LATIS IGM Maps: Visualization and Characterization of Structures}
\label{sec:maps}

\begin{figure*}
    \centering
    \includegraphics[width=\linewidth]{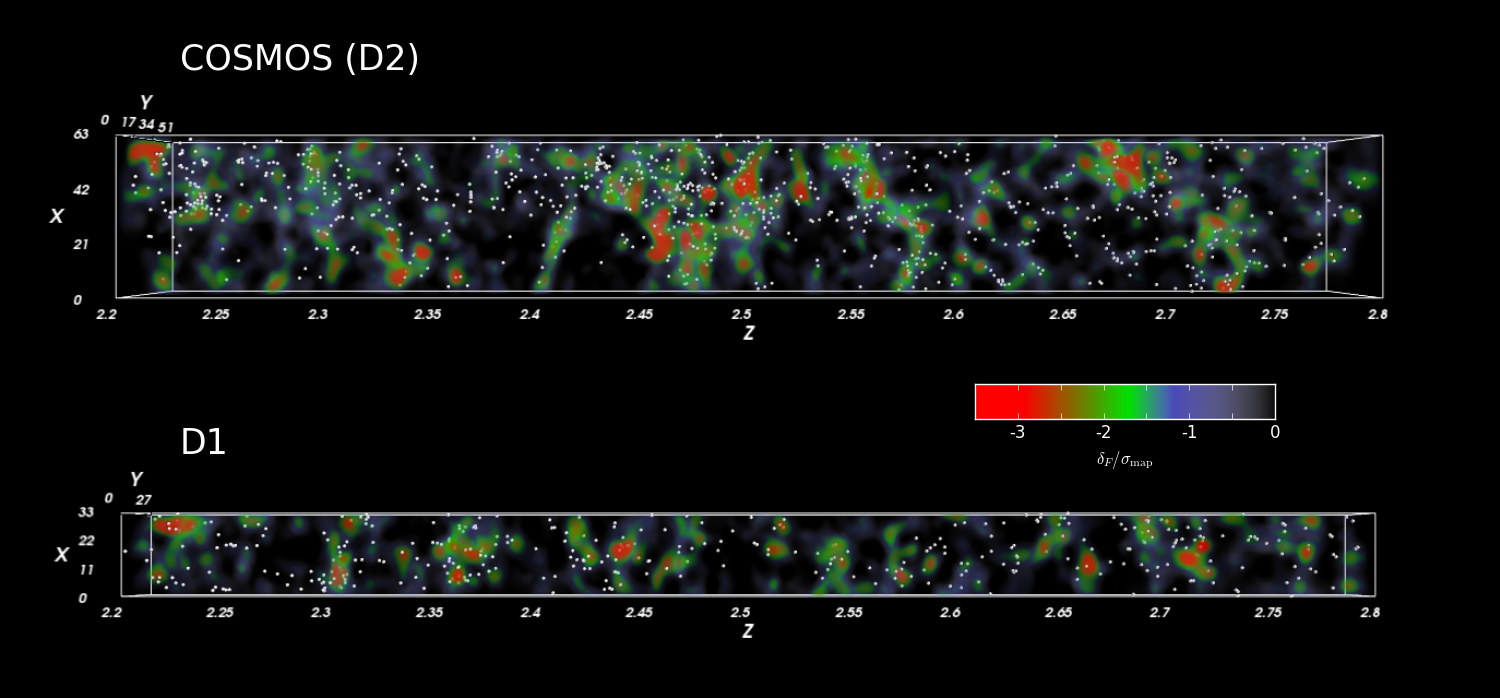}
    \caption{Renderings of the IGM opacity in the LATIS fields observed to date. The COSMOS (top) and D1 (bottom) maps, smoothed by $\sigma_{\rm kern} = 2$~\cMpch, are each viewed in a side-on projection. The $z$ axis shows the redshift while the $x$ and $y$ axes are in $h^{-1}$ cMpc. Red colors correspond to more negative $\delta_F$, i.e., lower transmitted flux and higher matter densities, while bluer colors represent the reverse. Regions with $\delta_F > 0$, i.e., with higher than mean transmission, are completely transparent in these renderings. The positions of galaxies from the LATIS, VUDS, and zCOSMOS surveys are overlaid. (Note that one quadrant of the COSMOS volume is not yet observed; see Figure~\ref{fig:sightlinedist}.)  \label{fig:3dmaps}}
\end{figure*}

\begin{figure*}
    \centering
  \includegraphics[width=0.8\linewidth]{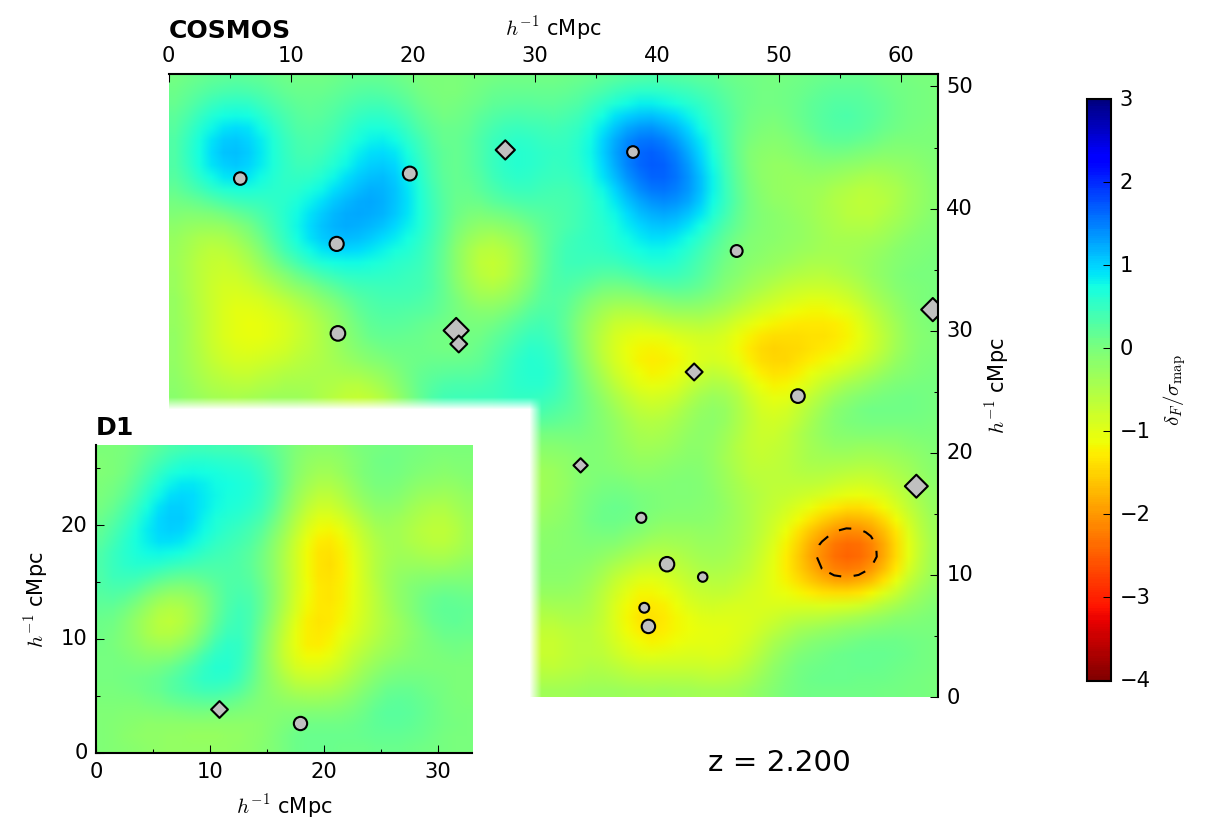}
    \caption{Animated rendering of the IGM opacity in the LATIS maps. Each frame shows a cross section of the two tomographic maps, smoothed by $\sigma = 2$~\cMpch. The L-shaped large region is the COSMOS field, while the smaller disconnected region in the lower-left corner is the D1 field. Points show the positions of galaxies within $\pm 4$~\cMpch~of the plotted redshift, as measured in LATIS (circles) and in the VUDS and zCOSMOS surveys (diamonds). Larger symbols denote brighter galaxies. Redder colors encode more negative $\delta_F$, i.e., lower transmission and higher matter densities, while blue colors show the reverse, as indicated by the color bar. Dashed black and solid white contours enclose $\delta_F / \sigma_{\rm map} = -2, -3, \ldots$ and $\delta_F / \sigma_{\rm map} = 2, 3, \ldots$, respectively. Note that $\delta_F$ is normalized by the dispersion $\sigma_{\rm map}$ of the map, not the noise. {\bf The animation can be viewed at \url{https://youtu.be/AAKO1oA1Ghw} or downloaded from the ancillary material on arXiv.}
    \label{fig:movie}}
\end{figure*}

Figure~\ref{fig:3dmaps} shows three-dimensional renderings of the IGM opacity in the LATIS survey volume to date. Since 4 of 12 footprints are included in these maps, the final LATIS maps will be $3\times$ larger. The maps already show a rich suite of structures. To aid in visualizing our maps and their correlations with the galaxy distribution, we provide a movie in Figure~\ref{fig:movie} that scans through the COSMOS and D1 maps in redshift.

In this section, we will identify a set of secure matter over- and underdensities in the IGM tomographic maps. We will then demonstrate their reality by comparing the IGM maps to the galaxy distribution, to structures previously detected via other methods, and to the CLAMATO IGM maps where they overlap LATIS.

Structures are present with a range of opacities and detection significances. For this paper, we will concentrate primarily on a set of securely detected large-scale overdensities. We identify flux minima within Wiener-filtered maps smoothed with a $\sigma_{\rm kern} = 4$~\cMpch~kernel, and we normalize each map by its own standard deviation $\sigma_{\rm map}$. \citet{Stark15a} have shown that $\sigma_{\rm kern} = 4$~\cMpch~is well matched to the signal expected for protoclusters and so acts like a matched filter. We set a detection threshold using the mock surveys described in Section~\ref{sec:mocks}. Comparing values in the recovered maps to the actual flux distribution, we find that voxels with $\delta_F / \sigma_{\rm map} < -2.35$ have a 95\% probability of lying in the bottom 10\% of the actual flux distribution. In the median, they are in the bottom 1\%. (Recall that low fluxes correspond to high matter densities.) 

This defines one reasonable threshold, specific to LATIS, for a securely detected overdensity. We then locate minima in the $\delta_F$ maps that satisfy the $\delta_F / \sigma_{\rm map} < -2.35$ threshold. Since the edges of the maps are noisier (see Figure~\ref{fig:sne}) and can suffer from edge effects, we exclude voxels within 4~\cMpch~of the map boundary, i.e., 1 $\sigma_{\rm kern}$. We will call these flux minima ``peaks'' since they are expected to correspond to matter overdensities. Although it is beyond the scope of this paper to fully analyze the topology of the maps and identify which peaks may be part of common structures, we make a simple attempt to avoid selecting multiple blended peaks by requiring that a peak be the minimum $\delta_F$ within a 12~\cMpch~sphere, i.e., $3\sigma_{\rm kern}$.

With these criteria, we find 18 peaks in the COSMOS map and 7 peaks in the D1 map. Applying the same peak-finding method to the suite of mock survey maps, which naturally incorporates both noise in LATIS and cosmic variance, we find $14 \pm 3$ peaks in COSMOS mocks and $5 \pm 2$ in D1 mocks. Thus the number of detected peaks is fully consistent with cosmological expectations. Among the 25 LATIS peaks, we note that 12 have $\delta_F  / \sigma_{\rm map} < -3$, the criterion suggested by \citet{Lee16} to identify likely protoclusters, defined as progenitors of $M > 10^{14} h^{-1} \msol$ halos. The average detection significance of the peaks is $3.6\sigma_{\rm resid}$, where $\sigma_{\rm resid} = [{\rm Var}(\delta_F^{\rm true} - \delta_F^{\rm rec})]^{1/2}$ (see Equation~\ref{eqn:SN}) represents the rms error in the mock survey maps at the location of a given peak.

\begin{figure*}
    \centering
    \includegraphics[width=0.72\linewidth]{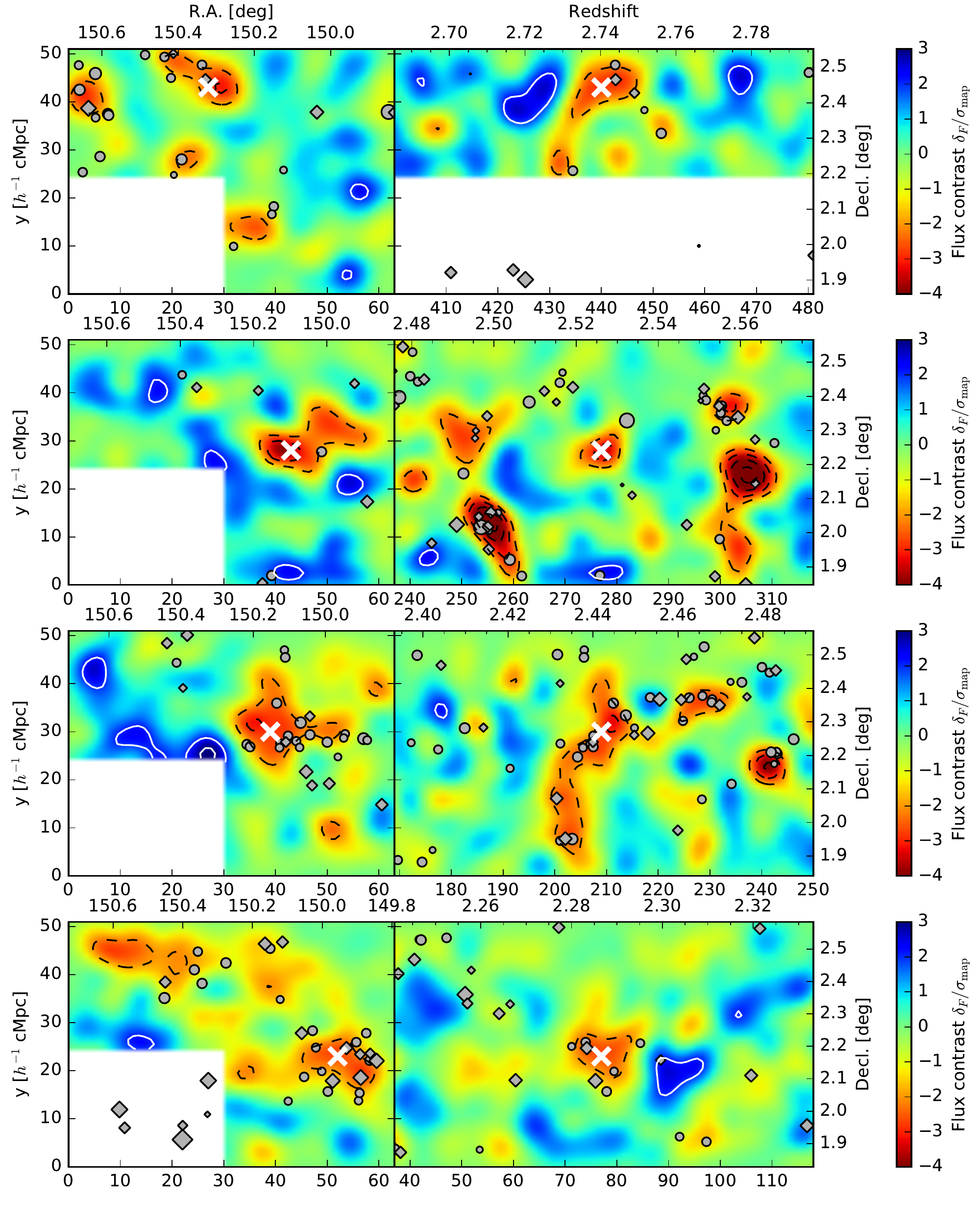} \\
    \vspace{-4ex}
    \includegraphics[width=0.72\linewidth]{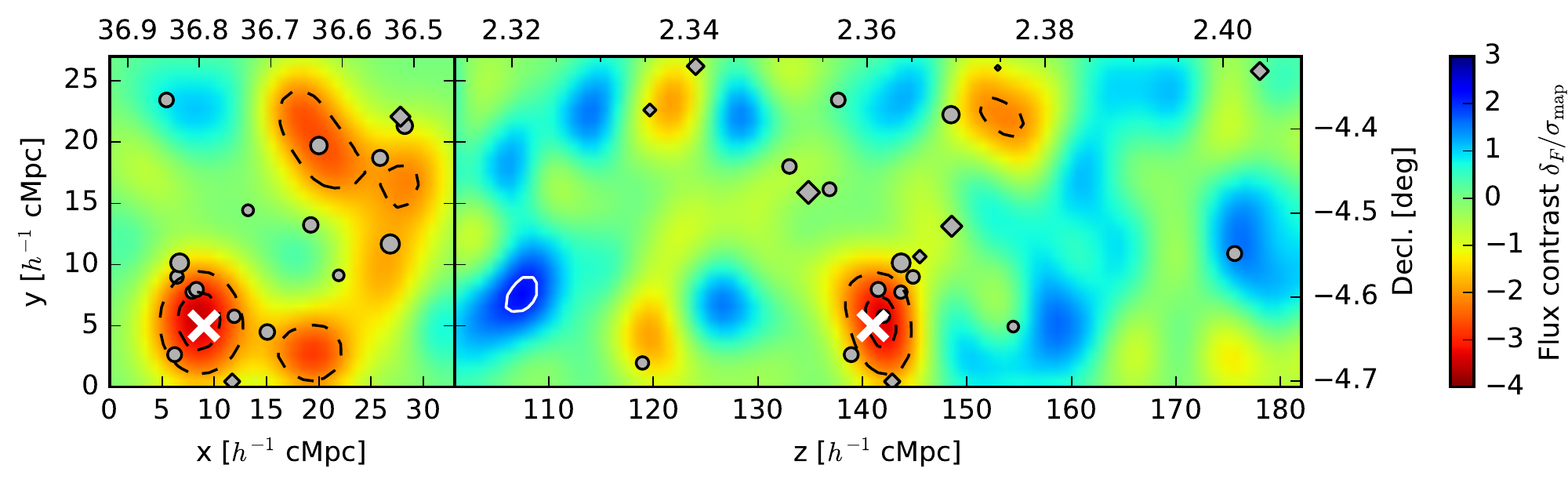}
    \caption{Maps of the IGM opacity in the vicinity of 5 representative matter overdensities detected in the LATIS maps. In each row, the left panel shows an $xy$ cross section of the tomographic map, smoothed by $\sigma = 2$~\cMpch, at the redshift of the overdensity. The right panel shows a $zy$ cross section around the $x$ position of the overdensity, whose position is indicated by a cross. Redder colors encode more negative $\delta_F$, i.e., lower transmission and higher matter densities, while blue colors show the reverse, as indicated by the color bar. Points (circles and diamonds) show the locations of galaxies. The meaning of the points and contours follows that in Figure~\ref{fig:movie}. The lower panel shows the D1 field while the others show COSMOS.\label{fig:slices}}
\end{figure*}

Maps of a representative set of 5 structures are shown in Figure~\ref{fig:slices}. Each row shows two projections of an IGM-selected overdensity. Overlaid are the positions of galaxies from LATIS (circles) along with the full zCOSMOS and VUDS data sets (diamonds).  The IGM overdensities are clearly rich in galaxies compared to random locations, and galaxies often trace the map features with a remarkable level of detail (e.g., the filamentary structure in the middle row, right panel). The sizes and morphologies of IGM structures are often resolved. If we consider the $-2\sigma_{\rm map}$ contour surrounding each of the 25 peaks as an {\it ad hoc} measure of their extent (outer dashed contours in Figure~\ref{fig:slices}), then the median enclosed volume is equal to a sphere with diameter 12~\cMpch, which is similar to the half-mass sizes predicted for massive protoclusters \citep{Chiang13}.

\begin{figure*}
    \centering
    \includegraphics[width=0.7\linewidth]{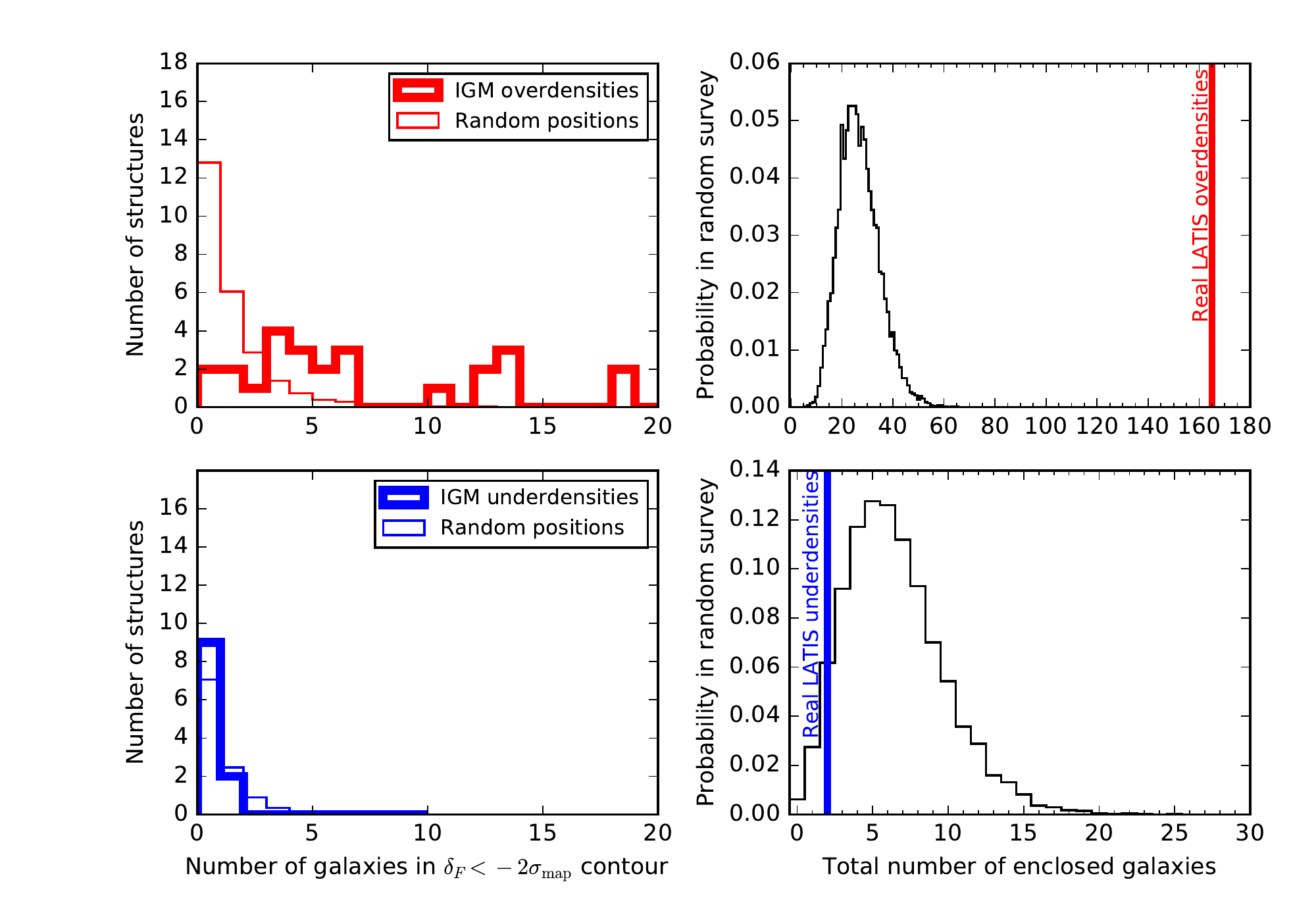}
    \caption{\emph{Left column:} The distribution of galaxy counts within IGM-selected overdensities (thick lines, top panel) and underdensities (bottom panel) is compared with the distribution at random locations (thin lines) within the same volumes. Galaxies are counted within a $\delta_F < -2\sigma_{\rm map}$ contour in the Wiener-filtered maps after smoothing by $\sigma_{\rm kern} = 2$~\cMpch. \emph{Right column:} The total number of galaxies in all IGM-selected overdensities (vertical line, top panel) and underdensities (bottom) is compared to the distribution of counts in random surveys. In each random survey, all observed structures are shifted to random positions as discussed in the text. As an ensemble, we detect a $17 \sigma$ enhancement and a $2.1 \sigma$ deficit of galaxies in over- and underdensities, respectively.
    \label{fig:gals_in_overdensities}}
\end{figure*}

We can now quantify the galaxy richness of the LATIS overdensities. This is an essential validation test of the maps, although we note that the presence of a correlation between IGM opacity and galaxy density on Mpc scales is not a new result \citep[e.g.,][]{Adelberger03,Lee14b}. 
Figure~\ref{fig:gals_in_overdensities} shows the number of galaxies within a 3D contour defined by $\delta_F < -2 \sigma_{\rm map}$ that surrounds each overdensity peak. This is an arbitrary threshold that does not necessarily include all associated galaxies, but it is adequate for our purposes of comparing galaxy richness. We count galaxies in LATIS as well as the VUDS and zCOSMOS surveys with confident redshifts (see Section~\ref{sec:specclass}). There are 6.6 galaxies, on average, in the 25 IGM-selected peaks. At  random locations, created by shifting and reflecting the observed structures to preserve their volume, there is only 1.0 galaxy on average. This difference is significant for most of the individual structures (top-left panel) and extremely significant for the ensemble: the top-right panel compares the total number of galaxies in the IGM-selected overdensities with random surveys, in which the position of each of the 25 structures is randomized. The ensemble of IGM-selected overdensities is enriched in galaxies at a confidence of $17 \sigma$.\footnote{This requires approximating the distribution of galaxy counts as Gaussian, which is not correct in detail but is adequate to demonstrate a very significant detection.}

A possible concern is that a relatively rare absorption line with a high column density of \ion{H}{1}, often found close to galaxies \citep{Rudie12}, might substantially influence its vicinity in our maps and masquerade as a large-scale overdensity, even though we mask the strongest absorbers (Section~\ref{sec:damped}). We assessed this possibility for each overdensity by removing each individual sightline with an impact parameter $< 4$~\cMpch~$= \sigma_{\rm kern}$ and reconstructing the map. On average, such a resolution element is pierced by 10 sightlines, and removing a single sightline rarely has an appreciable effect. Only in 2 of our 25 overdensities (8\%) can one sightline perturb the map by more than the $1\sigma$ uncertainty estimated from our mock surveys. Even in these cases, absorption is present in the other sightlines, so removing a sightline does not erase the flux decrement $\delta_F < 0$, but it can reduce its amplitude by $\simeq 40\%$. We conclude that our detection of overdensities is not very sensitive to any individual absorber. For the maps smoothed by $\sigma_{\rm kern} = 2$~\cMpch~that we use for display purposes, only a few sightlines pierce a resolution element, and the amplitude of map features can be more sensitive to individual sightlines.\footnote{In 20\% of overdensities, the $\delta_F$ peak smoothed on $\sigma_{\rm kern} = 2$~\cMpch~scales can change by more than the $1\sigma$ uncertainty from the mock surveys. Again, omitting a single sightline can reduce the amplitude of a peak by 40\%, but it never removes the flux decrement entirely.}

The LATIS maps also contain underdensities, visible in Figure~\ref{fig:slices} as the large blue regions. We identify these using a similar criterion to that applied to find matter overdensities. Using our mock surveys, we find that flux maxima in maps smoothed by $\sigma_{\rm kern} = 4$~\cMpch~that have $\delta_F > 2.54 \sigma_{\rm map}$ have a 90\% chance of being in the top 10\% of the true flux distribution. The current LATIS maps have 11 such matter underdensities or ``voids.'' (Applying the stricter 95\% significance threshold that we used to select matter overdensities would have resulted in 3 underdensities; we chose a slightly looser cut to generate a larger sample for exploration in this paper.) These voids never contain more than one galaxy from the aforementioned surveys (Figure~\ref{fig:gals_in_overdensities}, bottom-left panel); on average, they contain 0.18 versus 0.59 at random locations. Even at mean density, the joint LATIS-VUDS-zCOSMOS sample usually contains no galaxies within the volume of a typical void, highlighting the difficulty of mapping mean-to-underdense environments using galaxies as tracers. Since the absence of galaxies in individual underdensities is not a powerful test, we consider the ensemble of IGM-selected underdensities (Figure~\ref{fig:gals_in_overdensities}, bottom-right panel). Only 3\% of random surveys have fewer galaxies than the actual LATIS voids.\footnote{For our purposes, we use an \emph{ad hoc} criterion for identifying a set of extended and secure voids. \citet{Krolewski18} identify a larger sample of voids in CLAMATO using looser criteria, which they calibrated based on simulations, and find the voids have $2\times$ fewer galaxies than random locations. This is slightly less density contrast than we find here, as expected, but it corresponds to a higher statistical significance because of their larger sample.} Therefore the IGM-selected underdensities as a group are depleted in galaxies at $2.1\sigma$ confidence, but we emphasize that they could not readily be identified in redshift surveys.

\subsection{Comparison to CLAMATO Map}

\begin{figure*}
    \centering
    \includegraphics[width=0.48\linewidth]{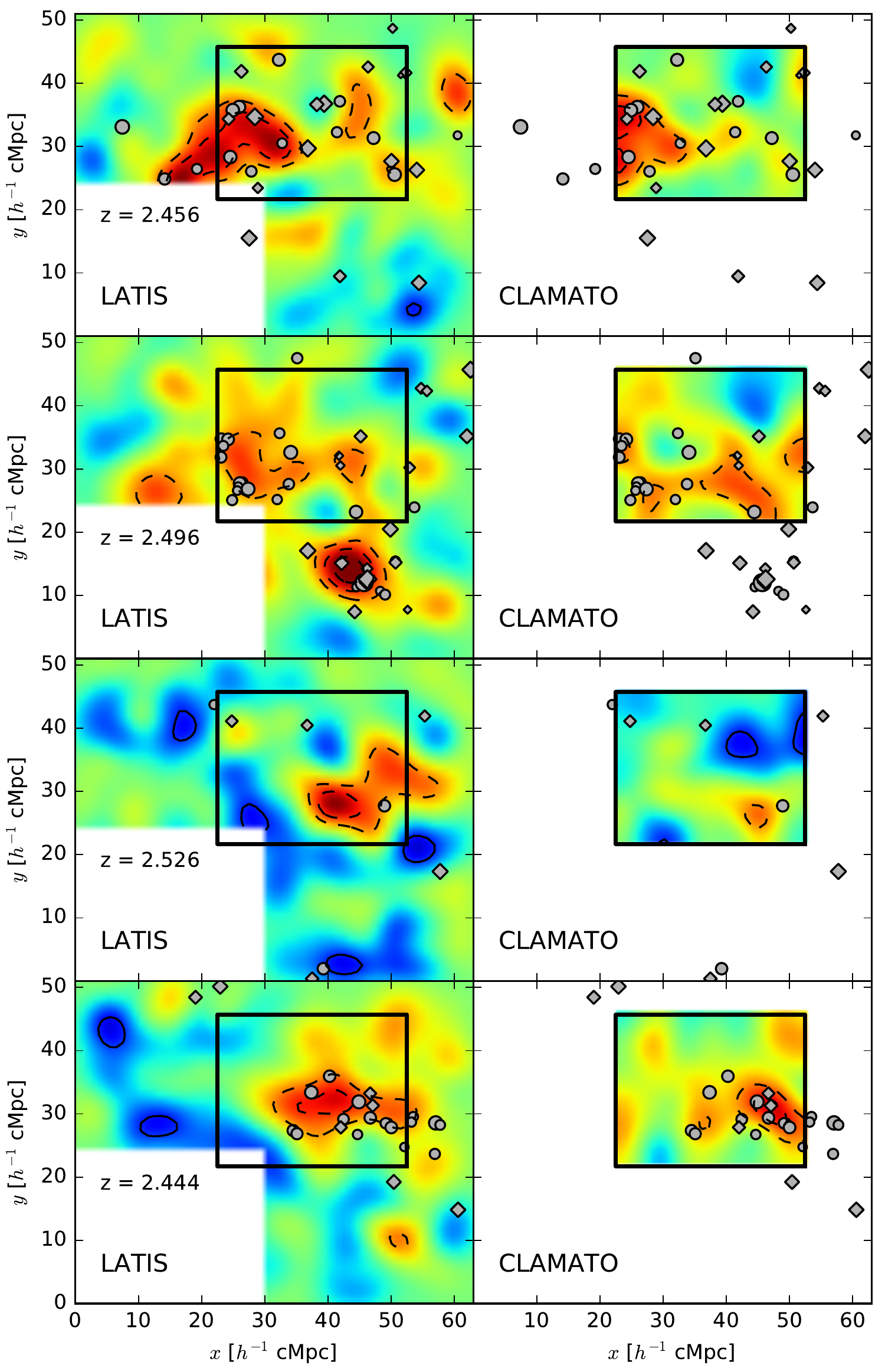} \hfill
    \includegraphics[width=0.48\linewidth]{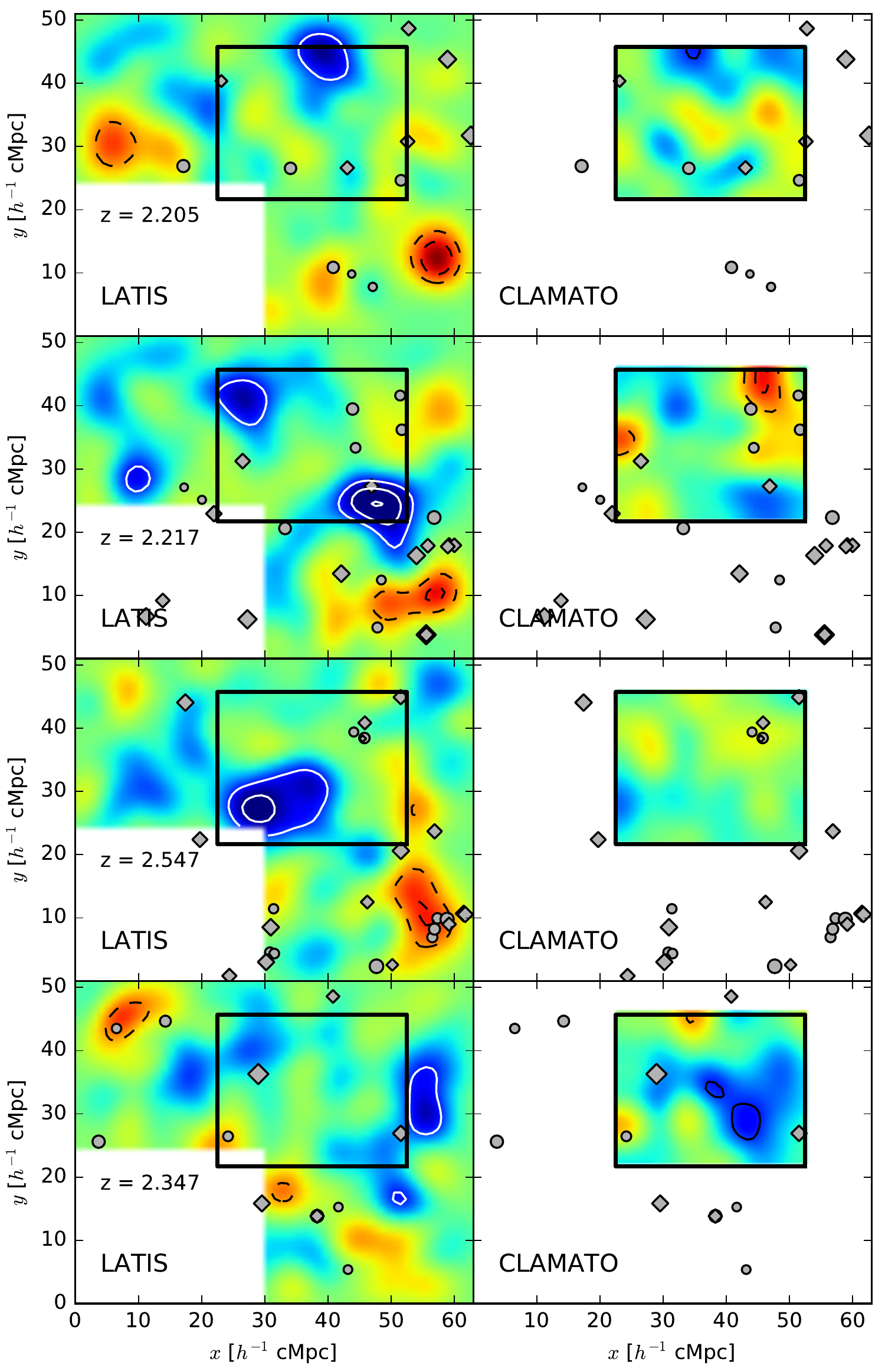}
    \caption{\emph{Left:} The structures of four overdensities, discussed in the text, are compared in the LATIS (left subpanels) and CLAMATO (right) tomographic maps. Colors and symbols match those in Figure~\ref{fig:slices}. The CLAMATO footprint is outlined in all panels. \emph{Right:} As in the left panel, but showing matter underdensities. \label{fig:compare_clamato}}
\end{figure*}

Nearly the entire footprint of the CLAMATO first data release \citep{Lee18} is already contained within LATIS. The two surveys overlap in the redshift range $z=2.2$-2.55. Using the criteria described above, we identify 4 overdensities in either map that fall within the region of overlap. (We exclude peaks close to the CLAMATO map boundary, since edge effects are likely significant.) The LATIS and CLAMATO maps around these peaks are compared in the left panel of Figure~\ref{fig:compare_clamato}. In the first, second, and fourth rows, there is a reasonably good agreement in the positions and morphologies of the overdensities in the two maps. The agreement is less good in the third row, where a strong LATIS peak is weaker and less extended in the CLAMATO map. Examining the LATIS sightlines near this position, we find that all show some absorption, but the strength of the map feature enhanced by one sightline that is nearly opaque. 
There are also 4 underdensities identified in either map within their common volume, which are shown in the right panel of Figure~\ref{fig:compare_clamato}. The first three rows show LATIS-detected voids; among these, the first two are seen in the CLAMATO maps, although at reduced significance. The third is not evident in CLAMATO, but it falls within just 5~\cMpch~of the back of the CLAMATO volume, where edge effects and the declining sightline density (Figure~\ref{fig:sne}) might explain the difference. The fourth row shows a CLAMATO-detected void that is not present in the LATIS map. Since the ${\rm S/N}_e$ is not particularly low at this location in our map, the reason for this discrepancy is not clear. 

This is a first comparison of Mpc-resolution tomographic maps produced by independent groups using different data sets, and we conclude that the strong overdensities discussed in this Section usually have similar morphologies. The reproducibility of matter underdensities may be somewhat worse, but a larger sample of structures that are common to two surveys is needed to confirm this possibility and better understand its origins.

\begin{figure*}
\centering
\includegraphics[width=0.43\linewidth]{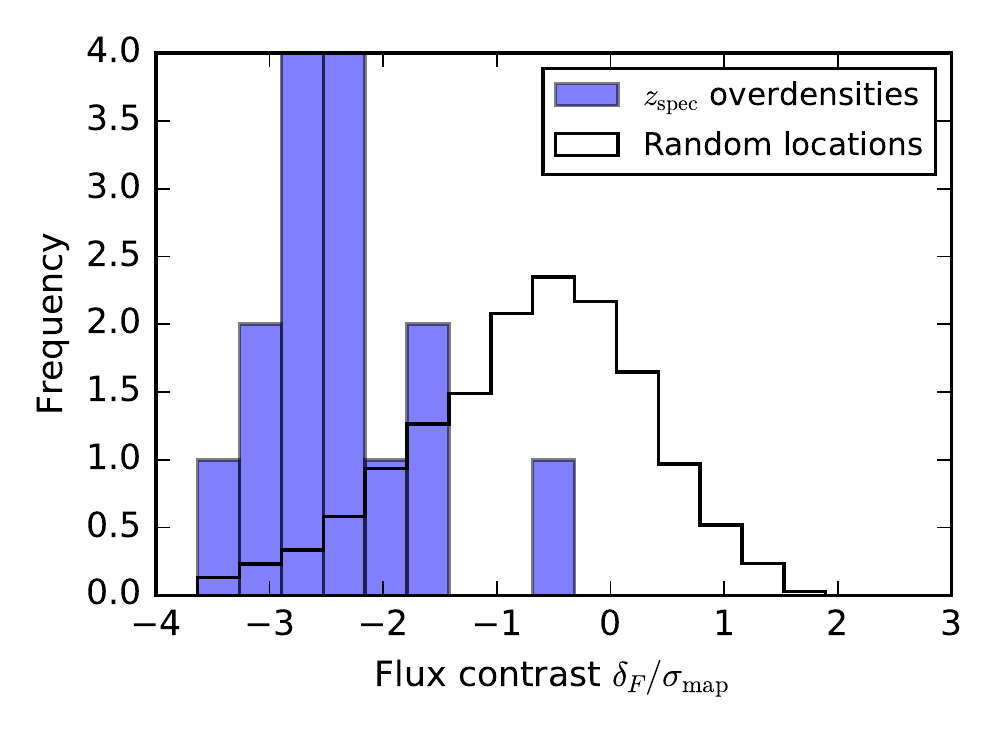}
\includegraphics[width=0.43\linewidth]{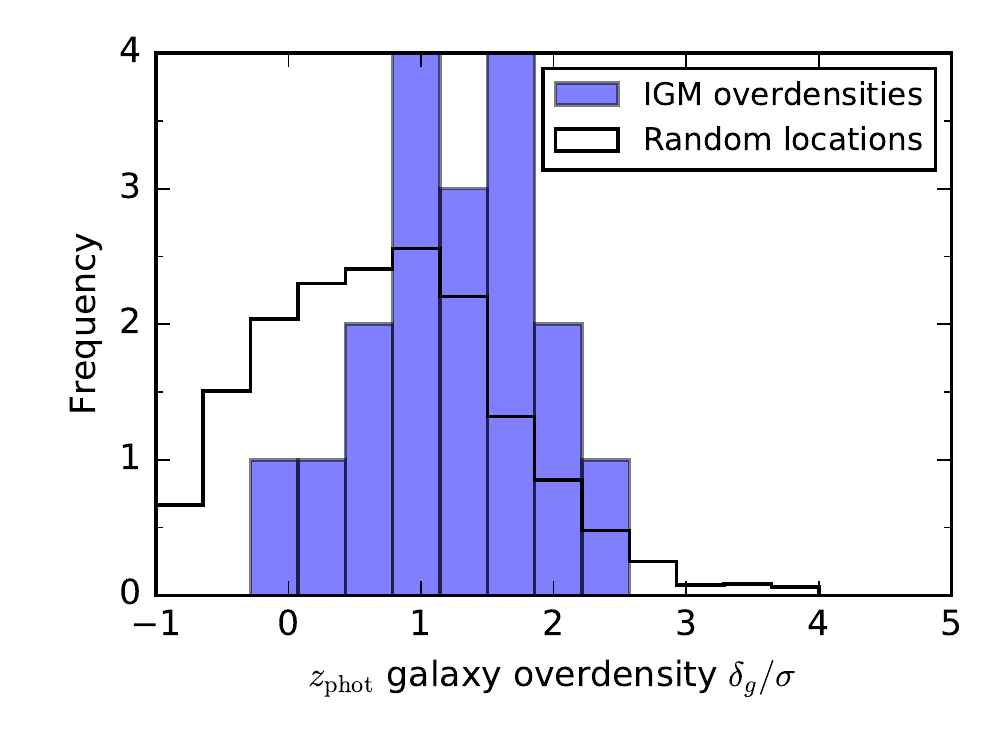}
\vspace{-3ex}
\caption{\emph{Left:} The IGM environments of protogroups spectroscopically identified by \citet{Diener13} using the zCOSMOS survey. For each candidate protogroup, we identify the minimum $\delta_F$ in the LATIS maps within $\pm \sigma_{\rm kern} = 4$~\cMpch~(in all dimensions) of the reported galaxy centroid (blue histogram). This procedure mitigates uncertainties in both the galaxy centroid and the location of the LATIS peak. The black histogram shows the same statistic evaluated at random locations, demonstrating that the spectroscopic overdensities are found in the high-density (low transmission) tail of the LATIS maps. \emph{Right:} The galaxy environments of IGM-selected overdensities as traced using photometric redshifts \citep{Scoville13}. The galaxy density map is smoothed by $\sigma_{\rm kern} = 4$~\cMpch. The maximum galaxy overdensity $\delta_g$ within $\pm 4$~\cMpch~of each IGM overdensity is normalized by the dispersion of $\delta_g$ and plotted in blue. Random locations are shown in black. Rare IGM fluctuations are almost always found to be overdense in galaxies ($\delta_g > 0$) but only loosely relate to the magnitude of the $z_{\rm phot}$-traced $\delta_g$.\label{fig:compare_structures}}
\end{figure*}

\section{Discussion} \label{sec:disc}

In this paper, we constructed the largest 3D maps of the IGM to date with a resolution $\simeq2.5$~\cMpch~(1 pMpc). Several tests validate the overall reliability of the LATIS maps for tracing large-scale structures. First, we performed mock observations of $N$-body simulations with an identical sightline distribution and noise to the real observations. The 3D Ly$\alpha$ flux PDF in the recovered maps is very similar in the real and mock data (Figure~\ref{fig:deltaFdist}), showing that structures in the LATIS maps are consistent with cosmological expectations. Second, we identified a set of 25 secure overdensities and show that these IGM-selected features are clearly enriched in galaxies as traced by the LATIS, VUDS, and zCOSMOS surveys (Figure~\ref{fig:gals_in_overdensities}). Large underdensities, or voids, detected by LATIS are found to be underdense in galaxies. Finally, we compared the morphologies of overdensities covered in both the LATIS and CLAMATO maps and found them to generally be in good agreement (Figure~\ref{fig:compare_clamato}). Although a more detailed and quantitative comparison awaits future work, this first cross-validation shows that the recovery of strong matter overdensities in Ly$\alpha$ tomographic maps is broadly reproducible. 

Many known structures that have been detected via other methods have counterparts in the LATIS maps. For example, the maps clearly contain the $z=2.47$ structure identified by \citet{Casey15} as an overdensity of sub-mm-bright starbursting galaxies, the $z=2.44$ overdensity traced by Ly$\alpha$ emitters that \citet{Chiang15} located in the HETDEX pilot survey \citep{Adams11,Blanc11}, and the starbursting cluster core with extended X-ray emission at $z=2.506$ discovered by \citet{Wang16}. These are all part of an enormous superstructure, named Hyperion by \citet{Cucciati18}, who identified 7 peaks in galaxy density spanning $z \simeq 2.4$-2.5. Part of this structure is contained in the CLAMATO maps and was investigated by \citet{Lee16}. The wider LATIS footprint now includes all of the 7 peaks discussed by Cucciati et al. Although a full investigation of this remarkable system is beyond the scope of this paper, 6 of these 7 peaks do appear to have counterparts in the LATIS map. Figure~\ref{fig:3dmaps} (top row) includes Hyperion and shows a clear correspondence between the galaxy distribution and IGM opacity on large scales, as well as some potentially interesting differences on smaller scales that will be investigated in future work. 

IGM overdensities are seen not only at the locations of very massive structures like Hyperion. We examined the environments of galaxy overdensities identified by \citet{Diener13} using the zCOSMOS spectroscopic redshifts. Diener et al.~argue that this sample represents a range of environments, with most systems being the progenitors of structures in the $M_{z=0} \approx 10^{13}-10^{14} \msol$ range, i.e., protogroups. Fifteen of these systems lie within the current LATIS map (excluding those within 4~\cMpch~of the edge). The left panel of Figure~\ref{fig:compare_structures} shows that the Diener et al.~protogroups are located in low transmission regions of the LATIS maps, with an average $\delta_F = -2.36 \sigma_{\rm map}$ that is very close to our threshold for identifying the most secure IGM structures. At the same time, the spectroscopically identified protogroups span a range of flux contrasts in our maps, which could help to quantify their masses.

An appealing feature of Ly$\alpha$ tomography is that, on the Mpc and larger scales probed by LATIS or CLAMATO, the signal is expected to be closely tied to the matter density. For the overdensities discussed in Section~\ref{sec:maps}, we can typically measure the peak $\delta_F$ smoothed on $\sigma_{\rm kern}=4$~\cMpch~scales to 28\% precision. Our mock surveys indicate that the recovered $\delta_F^{\rm rec}$, which includes realistic observational errors, predicts the smoothed matter overdensity $\delta_m = \rho_m / \langle \rho_m \rangle - 1$ with 39\% precision, although we caution that we have used a simplified treatment of the IGM physics. These are both impressive levels of precision, and integrating over larger volume like a protocluster would increase the precision further. To highlight the complementarity of Ly$\alpha$ tomography with spectroscopic redshift surveys, we consider the precision with which the galaxy overdensity $\delta_g$ can be measured. Since the density fields in this comparison were smoothed with a $\sigma_{\rm kern} = 4$~\cMpch~Gaussian kernel, we count the numbers of galaxies in the LATIS, VUDS, and zCOSMOS surveys within an $R = 6.2$~\cMpch~spherical top-hat kernel, which has the same volume, in order to compare overdensities at the same resolution. Even at the locations of LATIS-identified overdensities, there is an excess of only 3.1 galaxies. Poisson fluctuations thus impose an uncertainty of 63\% in $\delta_g$, even after combining data from 3 of the largest spectroscopic surveys. At mean density the average number of galaxies at this resolution is 0.7, showing again that $z=2$-3 redshift surveys cannot easily separate mean and underdense environments.

Another important strength of Ly$\alpha$ tomography is that structures are identified essentially independent of their galaxy populations, which mitigates concerns that using specific tracers, e.g., sub-mm sources, red sequence galaxies, or UV-bright galaxies, could bias the types of structures that are discovered. In principle, this concern could be partially addressed by using photometric redshifts to locate galaxy overdensities. We examined the \citet{Scoville13} galaxy overdensity maps, constructed using the photometric redshifts in the COSMOS field, at the positions of the LATIS-selected overdensities. We first smooth the Scoville et al.~maps with a $\sigma=4$~\cMpch~kernel so that galaxy and IGM overdensities are measured at the same resolution. We then compare the $z_{\rm phot}$-determined galaxy overdensity $\delta_g$ at the locations of the LATIS overdensities with random locations. The right panel of Figure~\ref{fig:compare_structures} shows that, reassuringly, there is a galaxy overdensity ($\delta_g > 0$) detected near 83\% of the IGM overdensities, but the strength is muted. The LATIS overdensities, which are rare 3$\sigma$ fluctuations in the IGM maps, are typically only 1.3$\sigma$ fluctuations in the galaxy density field as estimated using photometric redshifts. This suggests that photometric redshift catalogs can identify large-scale overdensities, but the magnitude of the signal may relate rather loosely to a galaxy's environment. This is particularly true for complex structures like Hyperion that contain multiple peaks that overlap in projection and are separated in redshift by less than the $z_{\rm phot}$ uncertainty.

Although opacity fluctuations in the IGM arise mainly from density fluctuations, they are also affected by inhomogeneities in the ionizing radiation field, which are most profound in the vicinity of quasars. \citet{Schmidt19} showed that radiation from hyperluminous quasars might erase the flux deficit that otherwise would be produced by the surrounding matter overdensity. IGM tomography can then be used to trace quasar light echoes, providing a novel way to age date individual systems. More typical quasars will have a less profound but potentially significant effect on our tomographic maps, which can be investigated in future work through detailed cross-comparisons with spectroscopic and photometric galaxy density fields.

These comparisons (see also the Introduction) show that Ly$\alpha$ tomography is a very promising tool for detecting and characterizing structures in the ``cosmic noon'' era that is quite complementary to other techniques. In the near term, we expect to complete LATIS at the end of 2020. We began observing in December 2017 and by April 2019 had fully mapped one-third of the survey area and completed half of the total survey exposure time. We are therefore on track to finish LATIS within our three-year schedule. At that point, LATIS will be not only the largest Ly$\alpha$ tomographic survey with comparable resolution, but also one of the largest spectroscopic surveys at $z\sim2.5$. We expect 3900 high-confidence redshifts in the $z=2.2$-3.2 range, similar to the total of the VUDS and zCOSMOS surveys in this same range\footnote{Here we count VUDS and zCOSMOS galaxies with flags of 2, 3, 4, and 9 following \citet{LeFevre15}.} and double the number with similar redshift quality flags. (These surveys probe a much wider redshift range and include fainter galaxies than LATIS, but their spectral resolution is lower than optimal for Ly$\alpha$ tomography.) After the survey is complete, we intend to provide a public data release that we expect will enable a variety of novel studies of the galaxy-IGM connection.

In the more distant future, many planned and proposed facilities could greatly expand the possibilities for Ly$\alpha$ tomography. Highly multiplexed spectrographs with 4000-20000 fibers on large, wide-field telescopes \citep{Ellis19,Marshall19,Schlegel19} will be able to survey far larger volumes than is possible with current instruments. Extremely large telescopes (ELTs) will have the sensitivity to observe fainter background galaxies, dramatically increasing the density of sightlines: reaching 1 mag fainter gives an order-of-magnitude gain in density \citep{Lee14}. ELTs will enable very high-fidelity mapping of \ion{H}{1} on IGM scales, but also will have the power to resolve the distribution and kinematics of \ion{H}{1} and metals within the circumgalactic medium surrounding individual galaxies \citep{Newman19,Rudie19}.

\acknowledgments
We appreciate the thoughtful report provided by the anonymous referee. S.~R.~was supported by a grant from the Rose Hills Foundation. We thank the staff at Las Campanas Observatory for their dedicated support that was essential for this project.
 Based on observations obtained with MegaPrime / MegaCam, a joint project of CFHT and CEA/IRFU, at the Canada-France-Hawaii Telescope (CFHT) which is operated by the National Research Council (NRC) of Canada, the Institut National des Science de l'Univers of the Centre National de la Recherche Scientifique (CNRS) of France, and the University of Hawaii. This work is based in part on data products produced at Terapix available at the Canadian Astronomy Data Centre as part of the Canada-France-Hawaii Telescope Legacy Survey, a collaborative project of NRC and CNRS. 
This research made use of Astropy,\footnote{\url{http://www.astropy.org}} a community-developed core Python package for Astronomy \citep{astropy13, astropy18}. 

\appendix

As described in Section~\ref{sec:mapconst}, we estimate the noise $\sigma_{\delta}$ in our measurements of $\delta_F$ by adding in quadrature the error spectrum propagated during the data reduction and the continuum uncertainties estimated in Section~\ref{sec:continuumerrors}. We now check the accuracy of our noise estimates using two methods. In the first, we consider the rms fluctuations in the $\delta_F$ sightline data (i.e., pixels in the observed Ly$\alpha$ forest spectra). From the simulations described in Section~\ref{sec:mocks}, we expect the intrinsic rms fluctuations in $\delta_F$ at the LATIS spectral resolution to be $\sigma_{\rm IGM} = 0.19$. This is much smaller than the median $\sigma_{\delta} = 0.57$, indicating that the dispersion in $\delta_F$ is dominated by measurement errors. The normalized median absolute deviation (NMAD) of $\delta_F$, a robust measure of the standard deviation, is 0.60. This compares very well to our expectation of 0.58 based on our $\sigma_{\delta}$ estimates and $\sigma_{\rm IGM}$. This agreement is not very surprising, since we estimated the continuum noise based on the excess noise in the Ly$\alpha$ forest spectra. In the second method, we consider an independent set of data: the spectra redward of Ly$\alpha$. In the range $1260~\textup{\AA} < \lambda_{\rm rest} < 1600~\textup{\AA}$, we first smooth each LATIS spectrum of a high-redshift galaxy with an 11 pixel boxcar and subtract this to remove the continuum. We then divide by the error spectrum and measure the NMAD of the normalized residuals, excluding pixels within $\pm 1000$~km~s${}^{-1}$ of strong interstellar absorption lines to isolate the relatively featureless part of the spectrum. We increase this NMAD by a small factor $1.05\times$ to account for the suppression of the variance caused by subtracting the smoothed spectrum. The median value among all the observed spectra is 1.02, which further supports the validity of our noise estimates.

We then applied the same tests to the CLAMATO first data release \citep{Lee18}. In the first test, we find that the NMAD of the $\delta_F$ pixel data is 0.72, which is significantly higher than the expected 0.58 based on the reported errors $\sigma_{\delta}$ and our estimated $\sigma_{\rm IGM} = 0.21$ at CLAMATO's spectral resolution. This disagreement can be reconciled if the errors $\sigma_{\delta}$ are increased by $1.27\times$. We then examined the noise in the galaxy spectra redward of Ly$\alpha$. We considered only the spectra from the blue arm of LRIS, which also observes the Ly$\alpha$ forest, and we excluded data at $\lambda > 4850$~\AA~near the dichroic transition. Following the same procedure applied to LATIS, we find that the NMAD of the normalized residuals is 1.23. Both methods indicate that the CLAMATO noise is underestimated by around 25\%. (We note that correlations introduced by resampling during data reduction would tend to reduce the variance rather than increase it.) Investigating the source of this discrepancy is beyond the scope of this paper, but the underestimation of the errors seems robust. Therefore we simply increase all of the reported $\sigma_{\delta}$ by $1.25\times$ when performing mock CLAMATO surveys to evaluate ${\rm S/N}_{\epsilon}$ in Figure~\ref{fig:sne}.

\bibliographystyle{aasjournal}
\bibliography{ms}

\end{document}